\documentclass[twocolumn]{aastex701}
\usepackage{comment}
\usepackage{empheq}
\usepackage{rotating}
\usepackage{amsmath}

\newcommand{\Add}[1]{\textcolor{black}{#1}}

\shorttitle{}
\shortauthors{Namekata et al.}

\graphicspath{{./}{../analysis-2022/figures/}{figures/}}

\begin{document}

\title{Do Young Suns Produce Frequent, Massive CMEs? Results from Five-Year Dedicated Optical Observations of EK Draconis and V889 Hercules}


\author[0000-0002-1297-9485]{Kosuke Namekata}
\affil{Heliophysics Science Division, NASA Goddard Space Flight Center, 8800 Greenbelt Road, Greenbelt, MD 20771, USA}
\affiliation{The Catholic University of America, 620 Michigan Avenue, N.E. Washington, DC 20064, USA}
\affiliation{The Hakubi Center for Advanced Research, Kyoto University, Yoshida-Honmachi, Sakyo-ku, Kyoto 606-8501, Japan}
\affiliation{Department of Physics, Kyoto University, Kitashirakawa-Oiwake-cho, Sakyo-ku, Kyoto, 606-8502, Japan}
\email{namekata@kusastro.kyoto-u.ac.jp}

\author[0000-0003-0332-0811]{Hiroyuki Maehara}
\affil{Okayama Branch Office, Subaru Telescope, National Astronomical Observatory of Japan, NINS, Kamogata, Asakuchi, Okayama 719-0232, Japan}
\email{hiroyuki.maehara@nao.ac.jp}

\author[0000-0002-0412-0849]{Yuta Notsu}
\affil{Laboratory for Atmospheric and Space Physics, University of Colorado Boulder, 3665 Discovery Drive, Boulder, CO 80303, USA}
\affil{National Solar Observatory, 3665 Discovery Drive, Boulder, CO 80303, USA}
\email{Yuta.Notsu@colorado.edu}

\author[0000-0001-6653-8741]{Satoshi Honda}
\affil{Nishi-Harima Astronomical Observatory, Center for Astronomy, University of Hyogo, Sayo, Hyogo 679-5313, Japan}
\email{honda@nhao.jp}

\author[0000-0002-5978-057X]{Kai Ikuta}
\affil{Department of Social Data Science, Hitotsubashi University, 2-1 Naka, Kunitachi, Tokyo 186-8601, Japan}
\email{kaiikuta.astron@gmail.com}

\author[0000-0001-9588-1872]{Daisaku Nogami}
\affil{Department of Astronomy, Kyoto University, Kitashirakawa-Oiwake-cho, Sakyo, Kyoto 606-8502, Japan}
\email{nogami@kusastro.kyoto-u.ac.jp}

\author[0000-0003-1206-7889]{Kazunari Shibata}
\affil{Kwasan Observatory, Kyoto University, 17 Ohmine-cho, Kita-Kazan, Yamashina, Kyoto 607-8471, Japan}
\affil{Department of Environmental Systems Science, Doshisha University, 1-3 Tataramiyakodani, Kyotanabe, Kyoto 610-0394, Japan}
\email{shibata@kwasan.kyoto-u.ac.jp}
\begin{abstract}

We report results from a five-year (132-night) dedicated observational campaign targeting two nearby young solar-type stars, EK Draconis ($\sim$50-125 Myr age) and V889 Hercules ($\sim$30 Myr age), using the 3.8m Seimei Telescope and Transiting Exoplanet Survey Satellite. The aim is to observationally constrain statistical properties of flaring radiation/heating as well as coronal mass ejections (CMEs), through high time-cadence H$\alpha$ spectroscopy. We obtained an unprecedented sample of 15 H$\alpha$ superflares, including two blueshifted absorption, two blueshifted emission, one redshifted emission, and nine line broadening events. We obtain the following results: (1) Larger flares exhibit broader H$\alpha$ line widths, up to 14.1$_{\pm 2.4}$ {\AA}, indicating higher chromospheric heating than solar flares. (2) The long-lasting redshifted event at $\sim$100 km s$^{-1}$ may indicate dense post-flare loops. (3) H$\alpha$ blueshifted absorptions/emissions provide evidence of massive filament/prominence eruptions, the core structures of CMEs. One newly identified event showed an unexpected rapid decrease in velocity. (4) The lower limit of the CME/eruption association rate with superflares is 27$_{-16}^{+25}$\%, yielding occurrence rates of 0.21$_{\pm0.12}$ and $<$0.32$^{+0.46}_{-0.32}$ events per day for EK Draconis and V889 Hercules, respectively. (5) We derived the first direct estimate of the lower limit of the mass-loss rate driven by super-CMEs ($\gtrsim10^{33}$ erg) for EK Dra as $4 \times (10^{-13}$--$10^{-12})$ $M_{\odot}$ yr$^{-1}$, comparable to the stellar wind mass loss at a similar age. This study provides critical observational constraints on the radiation and plasma environment around young solar-type stars and the early Sun, which can drive planetary space weather and stellar mass/angular momentum loss.

\end{abstract}



\section{Introduction}\label{sec:1}

Solar and stellar flares are the largest explosions in atmospheres of cool stars, observable across a broad wavelength range from radio to X-rays \citep{2011LRSP....8....6S,2017LRSP...14....2B,2024LRSP...21....1K}. 
These events are driven by the conversion of magnetic energy into kinetic and thermal energy through magnetic reconnection.
On the Sun, major flares emit intense X-ray and ultraviolet (UV) radiation and are often accompanied by coronal mass ejections (CMEs), which can significantly affect planetary magnetospheres and ionospheres \citep[e.g.,][]{2021LRSP...18....4T}.
Thus, we have empirically learned that major eruptive flares on central stars play a crucial role in shaping the atmospheres, habitability, and potential civilizations on surrounding planets \citep[e.g.,][]{2022LRSP...19....2C}.

The most energetic solar flares observed on the modern Sun release the bolometric energy up to $10^{32}$ erg \citep[e.g.,][]{2012ApJ...759...71E,2013A&A...549A..66A}. However, astronomical observations suggest that the young Sun (age $\le 600$ Myr) produced more energetic flares with the energy exceeding $10^{33}$ erg, known as ``superflares". Optical photometry from the Kepler and and Transiting Exoplanet Survey Satellite (TESS) missions, along with three decades of X-ray observations, has shown that young, rapidly rotating solar-type stars (G dwarfs) frequently exhibit superflares, with rates of approximately one event per day \citep[age $\sim$100 Myr;][]{1999ApJ...513L..53A,2000ApJ...541..396A,2012Natur.485..478M,2013ApJS..209....5S,2019ApJ...876...58N,2020arXiv201102117O,2022PASJ...74.1295Y,2022A&A...661A.148C,2022ApJ...926L...5N}. These results imply that the young Sun's superflares, accompanied by intense X-ray/UV radiation and ``super-CMEs" (here defined as more massive CMEs than \Add{current} solar ones of $\sim10^{17}$ g), likely influenced the magnetospheric and atmospheric evolution of early Venus, Earth, and Mars over timescales of $\sim$300 Myr.
It is widely discussed that intense X-ray/UV radiation from young stars likely drove photoionization, promoting atmospheric escape on early Earth \citep{2017ApJ...836L...3A,2021MNRAS.505.2941Y}.
In addition, CMEs themselves may have significantly enhanced atmospheric escape by dynamically interacting with planetary magnetospheres and atmospheres. Observations from the MAVEN mission has shown that CME impacts can temporarily increase atmospheric escape from Mars by up to an order of magnitude \citep{2017JGRA..122.1714M,2018GeoRL..45.8871L}. 
Besides escape, flare- and CME-associated energetic particles may have initiated chemical reactions that produced greenhouse gases and prebiotic molecules \citep{2016NatGe...9..452A,2023Life...13.1103K}. 
However, compared to radiation-driven atmospheric responses, the long-term and cumulative effects of CMEs remain less well constrained due to limited observational evidence, especially for young solar-type stars or the young Sun. 
Reviews of present-day escape rates and extrapolations to the early Sun suggest that current estimates may only represent conservative lower limits \citep{2018Icar..315..146J}.
These highlight the importance of further observational constraints on super-CMEs from stellar astronomy.

Nearby young solar-type stars, such as V889 Hercules (V889 Her, G0V, $\sim$ 30 Myr), DS Tucanae A (DS Tuc A, G6V, $\sim$ 45 Myr old; \citealt{2022A&A...661A.148C}) and EK Draconis (EK Dra, G1.5V, $\sim$50--125 Myr), have been often observed as ideal proxies for the infant Sun at the near-zero-age main-sequence stage, when Earth's atmosphere transitioned from a primary to a secondary N- and CO$_2$-rich atmosphere  \citep[e.g.,][]{2007LRSP....4....3G}.
Over the past three decades, X-ray observations of these stars have provided detailed characterization of quiescent and flaring coronal activity in young solar-type stars \citep{1999ApJ...513L..53A,2000ApJ...541..396A,2022A&A...666A.198P,2023ApJ...944..163H,2024ApJ...961...23N}. 
Later, space-based optical photometry from the Kepler and TESS have greatly improved our understanding of the frequency and temporal evolution of optical white-light flares. \citep{2012Natur.485..478M,2013ApJS..209....5S,2019ApJ...876...58N,2020arXiv201102117O,2022PASJ...74.1295Y,2022A&A...661A.148C,2022ApJ...926L...5N}.
However, key pieces are still missing for evaluating its impact on planet: specifically, our understanding of the UV/optical flaring radiation mechanisms from the transition region and chromosphere, as well as direct measurements of ambient winds and CMEs.


H$\alpha$ spectroscopic observations are a powerful and classical tool for characterizing chromospheric radiation and heating during flares, as well as filament/prominence eruptions associated with CMEs.
Of course, in addition to this method, there have been a number of recent efforts, particularly focused on M/K dwarfs, to detect other signatures of CMEs, such as post-flare coronal dimming \citep{2021NatAs...5..697V,2022ApJ...936..170L} and type II and IV radio bursts \citep{2020ApJ...905...23Z,2024A&A...686A..51M,2018ApJ...856...39C,2018ApJ...862..113C,2019ApJ...871..214V}.
Among these, H$\alpha$ spectroscopy has the advantage of a long history in the study of solar and M-dwarf flares \citep[e.g.,][]{1962BAICz..13...37S,1984SoPh...93..105I,1987ApJ...322..999C,2013ApJS..207...15K,2020PASJ...72...68N,2020ApJ...895....6G,2022ApJ...933..209N,2022ApJ...939...98O,2023ApJ...945...61N}, as well as extensive temporal coverage enabled by ground-based observations.
However, despite these advantages, no H$\alpha$ flare spectra had been reported from G-type stars prior to 2020, primarily due to the lower occurrence rate of detectable flares compared to M dwarfs.
In this context, as part of the observational campaign of this study, \citet{2022NatAs...6..241N} reported the first detection of H$\alpha$ flaring spectra from young solar-type stars. 
This was followed by subsequent studies \citep{2022ApJ...926L...5N,2024ApJ...961...23N,2024MNRAS.532.1486L}.
By using the 3.8m Seimei Telescope and TESS, \citet{2022NatAs...6..241N} identified a blueshifted H$\alpha$ absorption feature at --510 km s$^{-1}$ associated with a $2\times10^{33}$ erg superflare on EK Dra (the late phase of the same flare event was also captured by \citealt{2024MNRAS.532.1486L}) which is interpreted as evidence of a stellar filament eruption, where cool eruptive material moves toward the observer along the line of sight.
In the standard solar picture, filaments become destabilized in association of flares, erupt, and evolve into CMEs, forming the core component of the CMEs. 
Comparisons with solar observations and theoretical models suggest that massive, high-velocity filament eruptions are likely associated with CMEs \citep{2022NatAs...6..241N,2024ApJ...961...23N,2024ApJ...976..255N,2024ApJ...963...50I}.
Later, \citet{2024ApJ...961...23N} detected two events with blueshifted emission profiles, indicating stellar prominence eruptions extended beyond the stellar limb. 
Filament and prominence eruptions are essentially the same physical phenomenon; the distinction arises from their observational appearance in absorption or emission, depending on whether they are seen against the stellar disk or off-limb along the line of sight. 
In contrast, \citet{2022ApJ...926L...5N} reported a long-lasting, $10^{34}$ erg-class stellar flare that did not exhibit significant asymmetry or line broadening. 
Later, \citet{2024MNRAS.532.1486L} showed that four flares on EK Dra exhibited redshifted emission or broadening, although their time evolutions were poorly characterized due to the 30-minute cadence (cf. time cadence in \citealt{2022NatAs...6..241N,2024ApJ...961...23N} were 1$\sim$2 min).
In summary, three clear cases of stellar filament/prominence eruptions have been successfully identified. However, the current sample size remains too limited to robustly constrain their frequency, and the temporal evolution of redshifted emission and asymmetry during flares remains poorly understood.

The frequency of CMEs is a critical input for estimating the production rates of greenhouse gases and prebiotic molecules \citep{2016NatGe...9..452A,2023Life...13.1103K} and atmospheric escape rates 
\citep{2017JGRA..122.1714M,2018GeoRL..45.8871L,2018Icar..315..146J} of young planets, as well as for evaluating the impact of stellar mass loss on planetary environments and stellar evolution.
In the case of the present-day Sun, the solar wind mass-loss rate, approximately $2 \times 10^{-14}$ $M_{\odot}$ yr$^{-1}$, dominates over the CME-driven mass-loss rate, which averages around $4 \times 10^{-16}$ $M_{\odot}$ yr$^{-1}$ \citep{2010ApJ...722.1522V,2021LRSP...18....3V}.
For active, young stars, both theoretical studies \citep{2011ApJ...741...54C,2023ApJ...957...71S} and observational studies \citep{2005ApJ...628L.143W,2021LRSP...18....3V} suggest that the ambient wind mass-loss rate is on the order of $10^{-12}$ $M_{\odot}$ yr$^{-1}$, $\sim$100 times higher than the solar value, at an age of around 100 Myr or for stars with rotation periods of a few days.
In contrast, \citet{2015ApJ...809...79O} estimated the CME-driven mass-loss rate for EK Dra based on the X-ray flare occurrence frequency \citep{1999ApJ...513L..53A} and the solar CME mass–flare energy relationship \citep{2012ApJ...760....9A,2013ApJ...764..170D}. 
Their estimated CME mass-loss rate ranges from $10^{-12}$ to $10^{-11}$ $M_{\odot}$ yr$^{-1}$, depending on the assumed lower-energy threshold for CME/flare occurrence.
Furthermore, \citet{2017ApJ...840..114C} empirically estimated the CME mass-loss rate using the relationship between surface-averaged magnetic flux and CME properties. Their results suggest a mass-loss rate of $10^{-11}$ to $10^{-10}$ $M_{\odot}$ yr$^{-1}$ at an age of approximately 100 Myr.
These studies imply that CME-driven mass loss can dominate over stellar wind in terms of total mass loss, with significant implications for planetary environments.
However, there have been no direct estimates of the CME-driven mass-loss rate for stars other than the Sun.

With the aim of characterizing the statistical properties of flaring radiation and heating, as well as CMEs and prominence eruptions from young solar-type stars and the young Sun, we conducted a five-year observational campaign targeting the nearby young solar-type stars EK Dra and V889 Her. 
This campaign utilized H$\alpha$ spectroscopy from the 3.8m Seimei Telescope and partial optical photometry from TESS.
This campaign includes our initial reports by \cite{2022NatAs...6..241N,2022ApJ...926L...5N,2024ApJ...961...23N}.
The total number of observing nights under clear-sky conditions was 96 for EK Dra and 36 for V889 Her, representing one of the most extensive optical spectroscopic campaigns using a 4m-class telescope for this purpose. 
Our aim was to detect at least ten flares, and we ultimately identified 15 superflares.
Section~\ref{sec:2} describes the targets, observations, and datasets; Section~\ref{sec:3} outlines the analysis methods; Section~\ref{sec:4} presents the results from our five-year campaign; and Section~\ref{sec:5} discusses the radiation and heating mechanisms, as well as the statistical properties of stellar CMEs and eruptions.

\section{Observations and Data}\label{sec:2}

\subsection{Strategy and Target Selection}\label{sec:2-1}

Our targets were selected with the aim of detecting flares within realistic observation times (e.g., a few days to ten days).
Based on the first two years result of TESS photometry \citep[e.g.,][]{2020ApJ...890...46T,2021ApJS..253...35T}, we selected target stars by the following strategy:
(1) Classified as solar-type stars (G dwarfs)\footnote{In this paper, we adopt the broad term ``solar-type stars," which we define as G-type main-sequence stars. While our target stars can also be described using more specific terms such as ``Sun-like," ``solar-like," or ``solar-analogue" stars based on their similarity to the Sun in spectral type, mass, and radius, we consistently use the broader terminology throughout this work for clarity and consistency with previous papers \citep{2022NatAs...6..241N,2022ApJ...926L...5N,2024ApJ...961...23N}.}, (2) Bright enough to be observed with high signal-to-noise ratio (S/N) and temporal resolution (i.e., $V_{\text{mag}}$ $\lesssim$ 8-9) with optical spectroscopy onboard the 3.8m Seimei Telescope. (3) Observable from the northern hemisphere. (4) Having planned observations by TESS. (5) High frequency of flares detectable within realistic observation periods  (e.g., a few days to ten days).
Given these criteria, the number of suitable candidates was relatively limited. 
As a result, we selected EK Dra and V889 Her as our primary targets. 

Table \ref{tab:targets} summarizes the parameters of the target stars.
EK Dra is a young solar-type main-sequence star with an age of 50–125 Myr \citep{2005AandA...435..215K}, and V889 Her is a similar star at $\sim$30 Myr \citep{2003AandA...411..595S}.
EK Dra and V889 Her are both rapidly rotating, with rotation periods of 2.7 and 1.3 days, respectively.
They exhibit dense, hot coronae, large starspots, and frequent superflares (EK Dra: \citealt{1999ApJ...513L..53A,2005AandA...432..671S,2017MNRAS.465.2076W,2020ApJ...890...46T,2021ApJS..253...35T,2022ApJ...926L...5N,2024ApJ...961...23N}, V889 Her: \citealt{2008A&A...488.1047J,2019A&A...622A.170W,2020ApJ...890...46T,2021ApJS..253...35T}).
V889 Her is believed to be a single star \citep{2005AandA...435..215K}, while EK Dra has a faint low-mass companion located 20 AU away \citep{2003AandA...411..595S}.
EK Dra A and B are not thought to be magnetically connected \citep{2017MNRAS.465.2076W}, and the G-dwarf (EK Dra A) is the primary source of stellar superflares.

\begin{deluxetable*}{ccc}
\tablecaption{Stellar Parameters.}
\tablewidth{0pt}
\tablehead{
\colhead{Parameters} & \colhead{EK Dra} & \colhead{V889 Her} \\
\colhead{} & \colhead{(TIC 159613900)} & \colhead{(TIC 471000657)} 
}
\startdata
Spectral Type & G1.5V$^{(1)}$ & G0V$^{(6)}$ \\
V$_{\rm mag}$ & $7.60\pm0.01^{(2)}$ & $7.45\pm0.04^{(7)}$ \\
Age & 50--125 Myr$^{(4)}$$^{\S}$ & 30 Myr$^{(6)}$ \\
$T_{\text{eff}}$ (K) & 5560--5750$^{(1,3,4)}$$^{\dagger}$ & $5830\pm50^{(6)}$ \\
Radius ($R_{\odot}$) & $0.94\pm0.07^{(1)}$ & $1.09\pm0.05^{(6)}$ \\
Mass ($M_{\odot}$) & $0.95\pm0.04^{(1)}$ & $1.06\pm0.02^{(6)}$ \\
Distance (pc) & $34.40\pm0.03^{(5)}$  & $35.36\pm0.02^{(5)}$ \\
$P_{\text{rot}}$ (d) & $2.766\pm0.002^{(1)}$ & $1.3371\pm0.0002^{(6)}$ \\
$v\sin i$ \Add{(km s$^{-1}$)} & $16.4\pm0.1^{(1)}$ & $39.0\pm0.5^{(6)}$ \\
RV (km s$^{-1}$) & $-20.687\pm 0.004^{(8)}$ & $-23.6\pm1.5^{(6)}$ \\
Inclination (deg) & $60\pm5^{(1)}$ & $\approx55^{(6)}$  \\
Binarity & low-mass companion$^{(4)}$ & single$^{(6)}$ \\
\enddata
\tablecomments{$^{(1)}$\cite{2017MNRAS.465.2076W}, $^{(2)}$\cite{2000AandA...355L..27H}, $^{(3)}$\cite{2018AandA...620A.162J}. $^{(4)}$\cite{2005AandA...435..215K}, $^{(5)}$Gaia Early Data Release 3 \citep{2021AandA...649A...1G}, $^{(6)}$From Table 2 of \cite{2003AandA...411..595S}. $^{(7)}$\cite{2012AcA....62...67K}, $^{(8)}$\cite{2018AandA...616A...2L}. $^{\S}$Note that different papers report different estimates of age ranging from 30 Myr to 125 Myr. $^{\dagger}$Here we assume the stellar temperature is $\approx$5700 K from \cite{2005AandA...435..215K} for flare energy calculation to hold a consistency with our previous studies.}
\label{tab:targets}
\end{deluxetable*}

\subsection{TESS}

TESS conducts photometric observations using four cameras equipped with the TESS filter, spanning the optical band from 6000 to 10000 {\AA}. Each sector is observed for approximately 27 days \citep{2015JATIS...1a4003R}. 
The data were obtained from the Multimission Archive at the Space Telescope (MAST) archive\footnote{\url{https://mast.stsci.edu/portal/Mashup/Clients/Mast/Portal.html}}.
To date, TESS has observed EK Dra across 13 sectors and V889 Her for 3 sectors (see, Table \ref{tab:obslog:tess} in Appendix \ref{app:obslog}). 
Both EK Dra and V889 Her have been observed in TESS's short cadence mode (2 minutes) for almost all the sectors.
However, EK Dra was not included in the target list for Sector 50 and was only observed using the 10-minute-cadence full frame image (FFI).
Also, data with the 20-sec cadence have been available since Sector 41.

For uniform data analysis, we decided to use the 2-minute cadence data for all sectors except Sector 50. 
Although the 20-second cadence provides higher temporal resolution, it also increases the photon noise, which can disadvantageously affect flare detection. 
For Sector 50, we utilized the FFI ``raw" flux in the output by \texttt{eleanor} pipeline \citep{2019PASP..131i4502F} for flare detection (see, \citealt{2024ApJ...961...23N}). 
For short cadence data, we primarily used the PDC-SAP data. However, from Sector 75 in 2024 onwards, the photometric errors in TESS seem to have increased to some extent, resulting in substantial gaps in the PDC-SAP data and reducing overlap time with H$\alpha$ observations. 
After comparing SAP and PDC-SAP data, we determined that SAP data are sufficiently reliable for flare science, and thus we opted to use SAP data from 2024 onwards.

\subsection{3.8-m Seimei Telescope}

We performed long-term spectroscopic observations with the Kyoto Okayama Optical Low-dispersion Spectrograph with optical-fiber Integral Field Unit (KOOLS-IFU) installed in the 3.8 m Seimei telescope at Okayama Observatory \citep{2019PASJ...71..102M, 2020PASJ...72...48K}. 
The KOOLS-IFU is a spectrograph covering a wavelength from 5800 to 8000 {\AA} with a spectral resolution of R $\sim$ 2000 (VPH-683 grism). 
Our observational campaign was conducted over the 5-yr period from February 14, 2020 to April 25, 2024.
In Table \ref{tab:obslog:1} and \ref{tab:obslog:2} in Appendix \ref{app:obslog}, we have compiled the information on observing dates, number of frames, exposure times, and readout times of KOOLS-IFU.
The exposure times can vary from 20 to 160 seconds, depending on different observers and/or weather conditions. 
Furthermore, since this telescope began operations in 2019, instrumental improvements have been sometimes made, resulting in variations in readout times on different days.

The data reduction process was carried out following established ways by \cite{2020PASJ...72...68N,2022NatAs...6..241N,2022ApJ...926L...5N}, employing both \texttt{IRAF} and \texttt{PyRAF} software tools. 
This process includes basic calibrations, such as bias subtraction, wavelength calibration, sky subtraction, cosmic-ray subtraction, heliocentric correction, and continuum fitting, providing wavelength-calibrated, continuum-normalized 1D spectrum for each frame. 
The radial velocity due to proper motion of EK Dra \citep[--20.7 km s$^{-1}$;][]{2018AandA...616A...2L} and V889 Her \citep[--23.6 km s$^{-1}$;][]{2003AandA...411..595S} are also calibrated.
As for the data of V889 Her on 25 July 2022, we carefully removed the water vapor line trends from data taken on different dates (see Appendix \ref{app:watervapor}).

We measured the H$\alpha$ equivalent width (hereafter referred to as ``EW"), integrated for the wavelength interval of 6562.8--10 {\AA} to 6562.8+10 {\AA}, and used the EW for H$\alpha$ light curves.
\Add{In this paper, we define the EW for an emission profile as positive value.
Also, we define the EW changes relative to pre-flare conditions (EW$_{\rm preflare}$) as $\Delta$EW = EW -- EW$_{\rm preflare}$.}

\section{Analysis}\label{sec:3}


\subsection{Detection and Characterization of TESS White-Light Flares}

\subsubsection{Detection of White-Light Flares}

We performed an automatic TESS's flare detection by following the traditional method \citep[cf.][]{2012Natur.485..478M,2022ApJ...926L...5N}. Here, we briefly introduce the technique employed. Initially, the detrending of the TESS light curve was carried out using the fast Fourier transformation \texttt{numpy.fft.fft}. The TESS light curves generally exhibit nearly even data sampling (here we use two minutes cadence data), but if large data gaps exceeding 0.2 days were present, detrending was conducted separately for each uninterrupted segment.
Following the detrending, if two or more consecutive points exceeded five times the flux error (1$\sigma$ error in TESS's fits file; \citealt{2015JATIS...1a4003R}), these points were identified as flare candidates. However, if a flare occurred too close to data gaps (within fewer than five data points), it was excluded. 
We opted not to use three times flux errors as a threshold because it was less than the actual noise level (three times standard deviations of the detrended light curves) likely due to stellar activities, which lead to a significant over-identification of noise as flares. 
For the flares detected, we then defined the start and end points of each flare as the points where the flux dropped back to the flux error level. Subsequently, the energy and duration of each flare were automatically calculated.

In the study, significant H$\alpha$ flares without its significant white-light emissions were primarily classified as ``non-white-light" flares (non-WLFs). However, the flare event (EK5, see the label definition in Section \ref{sec:4-1}), despite not reaching the threshold, exhibited slight enhancements in white light, which were visually confirmed concurrently with H$\alpha$ flares. 
For this flares, we manually applied smoothing techniques to remove the background trends and estimated the flare energy. 
These cases were not statistically classified as white-light flares (e.g., the flare frequency distribution in Figure \ref{fig:ffd} in Appendix \ref{app:tess}), but instead, these were categorically noted as white-light flares in parentheses across various sections (see Table \ref{tab:flare-basic}).

\begin{deluxetable*}{lccccccccc}
\tablecaption{The list of H$\alpha$ superflares on EK Dra and V889 Her in our 5-year campaign observations through 2020 to 2024.}
\tablewidth{0pt}
\tablehead{
\colhead{} & \colhead{ID} & \colhead{H$\alpha$ Asymmetry} & \colhead{$E_{\rm WL,bol}$} & \colhead{$E_{\rm H\alpha}$} & \colhead{$t_{\rm WL}$} & \colhead{$\tau_{\rm WL}$} & \colhead{$t_{\rm H\alpha}$} &  Ref.  \\
\colhead{}& \colhead{} & \colhead{} & \colhead{} & \colhead{} & \colhead{(FWHM)} & \colhead{(e-decay)} & \colhead{(FWHM)} & &   \\
\colhead{} & \colhead{} & \colhead{} & \colhead{[10$^{33}$ erg]} & \colhead{[10$^{31}$ erg]} & \colhead{[min]} & \colhead{[min]} & \colhead{[min]}  &   
}
\startdata
\textsf{EK Dra}\\
\hline
2020 Mar 14 & (EK1) & No$^{(R)}$/Broadening  & 26$_{\pm 3}$ & 40$_{\pm 4}$ & 130 & 26  & 130 & (1) \\
2020 Apr 05 & (EK2) & Blue absorption & 2.0$_{\pm 0.1}$ & 1.7$_{\pm 0.1}$$^{(\ddag)}$ & 6.0 & 5.4  & 7.8$^{(\ddag)}$ & (2) \\
2021 Apr 21 & (EK3) & No/Broadening & -- & 18.7$_{\pm 0.6}$ & -- & -- & 69.5 & (4) \\
2021 Apr 30 & (EK4) & No/Broadening & -- & 35.2$_{\pm 0.9}$ & -- & -- & 88.8 & (4) \\
2022 Feb 23 & (EK5) & No/Broadening & $(2.7_{\pm0.3}/1.9_{\pm0.2}/0.9_{\pm0.1})$ & $\geq27.3_{\pm 1.6}$ & (46/32/30) & (24/22/17) & $\geq$242 & (4) \\
2022 Apr 10 & (EK6) & Blue emission & 1.5$_{\pm 0.1}$ & 48.7$_{\pm 1.7}$(9.1$_{\pm 2.6}$)$^{\ast}$ & 20 & 18 & 48 & (3)  \\
2022 Apr 16 & (EK7) & Blue emission & 12.2$_{\pm 0.2}$ & 17.7$_{\pm 0.7}$ & 40 & 20 &  54 & (3) \\
2022 Apr 17 & (EK8) & No & 3.4$_{\pm 0.1}$  & 2.9$_{\pm 0.3}$ & 30  & 9.6 & 23 & (3) \\
2023 Mar 08 & (EK9) & Blue absorption  & -- & $1.1_{\pm 0.1}$$^{(\ddag)}$ & -- & -- & 11.5 & (4) \\
2024 Mar 20 & (EK10) & No/Broadening & 14.8$_{\pm0.2}$ & $\geq7.8_{\pm 0.5}$ & 8.0 & 13.6 & $\geq35.7$ & (4) \\
2024 Mar 29 & (EK11) & No & (Non-WLF) & 20.2$_{\pm 0.8}$ & (Non-WLF) &  (Non-WLF) & 136 & (4)  \\
2024 Apr 19 & (EK12) & Red emission & (Non-WLF) & 14.0$_{\pm 0.8}$ & (Non-WLF) & (Non-WLF) & 86.6 & (4)  \\
\hline
\textsf{V889 Her}\\
\hline
2021 Apr 24 & (V1) & No/Broadening & -- & \Add{$>23.2_{\pm 0.3}$} & -- & -- & $>$33 & (4) \\
2022 Jul 25 & (V2) & No/Broadening & -- & \Add{$\sim 32.5_{\pm 0.9}$} & -- & -- & $\sim$84 & (4) \\
2023 Mar 31 & (V3) & No/Broadening & -- & \Add{$\geq26.4_{\pm 0.6}$} & -- & -- & $\geq$110 & (4) \\
\enddata
\tablecomments{
$t_{\rm WL}$ and $t_{\rm H\alpha}$ is the FWHM duration of the WL and H$\alpha$ flare.  
$\tau_{\rm WL}$ is the e-folding decay time of the flare.
Regarding $t_{\rm WL}$ and $\tau_{\rm WL}$, EK5 and EK12 include multiple flares. For EK5, the white-light emission was treated separately. For EK12, each peak was treated as a single event, and the final peak was discarded due to poor data quality. 
$^{\ast}$The value in bracket is the radiation energy of the central component without the blueshifted component. 
$^{\ddag}$The flare emission on April 5 in \cite{2022NatAs...6..241N} is very short-lived and its light curve is a combination of blueshift absorption and flaring emission, so the H$\alpha$ flare duration is expected to be underestimated.
$^{(R)}$In \cite{2022ApJ...926L...5N}, we reported a possible redshifted emission with a velocity of $\sim$100 km s$^{-1}$, but it was derived from the red wing residual and single Gaussian fit produce up to 26.6 km s$^{-1}$ which is below the threshold in the paper.
References: (1) \cite{2022ApJ...926L...5N}, (2) \cite{2022NatAs...6..241N}, (3) \cite{2024ApJ...961...23N}, (4) This study.
}
\label{tab:flare-basic}
\end{deluxetable*}

\subsubsection{Flare Energy, Duration, e-folding Decay Time}

The bolometric white-light (WL) flare energy is calculated following the methodology described in \cite{2013ApJS..209....5S}. We assume a blackbody radiation spectrum at 10,000 K to compute the bolometric WL energy. Additionally, the full-width at half maximum (FWHM) duration ($t_{\rm WL}$) and the e-folding decay times of white-light flares ($\tau_{\rm WL}$) are derived. The e-folding decay times ($\tau_{\rm WL}$) are determined by fitting the decay phase of the flares using an exponential function and employing \texttt{scipy.optimize.curve\_fit} for the fitting process.

\subsection{Detection and Characterization of H$\alpha$ Flares}\label{sec:3-2}

\subsubsection{Definition and Threshold of H$\alpha$ Flare}

We define H$\alpha$ flares as events showing an increase in equivalent width ($\Delta$EW) that meets the following criteria:
(1) The event must exhibit a change in equivalent width of more than 0.1 {\AA} and display a light curve characteristic of flares, featuring an increase and decay within a timescale of hours ($<$timescale of background rotational modulations$\sim$ten hours to a day; \citealt{2022ApJ...926L...5N}). This is an empirical threshold based on the data noise level \citep{2024ApJ...961...23N}. Almost all events were selected based on this criterion.
(2) If the rising phase is not captured, the event can be judged as a flare if the H$\alpha$ peak EW level exceeds the normal H$\alpha$ EW level by more than 0.2 {\AA} (Only applied to the V1 event).
(3) If the H$\alpha$ EW does not show significant increase and decay but shows very unique light curve like post-flare absorption, we identified it as a flare if its total variation is over 0.1 {\AA} (Only applied to the EK9 event).
Based on these criteria, we detected 15 stellar flares, including 12 from EK Dra and 3 from V889 Her, as summarized in Table \ref{tab:flare-basic}.

\subsubsection{Calculating H$\alpha$ Flare Energy and Duration}

The H$\alpha$ radiation energy is estimated from the $\Delta$EW light curve, which measures the emission. We set the threshold for integrating the flux at one-fifth of the peak values. Due to the high noise levels that often lead to under- or overestimation of the total duration, a smoothed light curve is employed to derive the energy and duration. The smoothing parameter in the Python function \texttt{numpy.convolve} was manually adjusted based on the flare's light curve shape and duration, as well as the noise level on the day of observation. The energy is calculated by multiplying the enhanced H$\alpha$ $\Delta$EW by the continuum flux and integrating over time.

The continuum flux of EK Dra around H$\alpha$ is calculated as 1.57 W m$^{-2}$ nm$^{-1}$ at 1 AU, derived from a flux-calibrated spectrum of EK Dra and the stellar distance provided by Gaia Data Release 2 \citep{2018AandA...616A...2L}. 
\Add{This value is commonly used across our series of papers \citep{2022NatAs...6..241N,2022ApJ...926L...5N,2024ApJ...961...23N} and consistent with the value estimated from Gaia Data Release 3 \citep{2023A&A...674A...1G}, 1.39$_{\pm0.03}$ W m$^{-2}$ nm$^{-1}$ at 1 AU at 650 nm (i.e., 11\% uncertainty is estimated).
For V889 Her, a flux-calibrated spectrum \Add{from our ground-based observation} was not available, so we decided to use the value from Gaia Data Release 3 \citep{2023A&A...674A...1G}, 1.66$_{\pm0.03}$ W m$^{-2}$ nm$^{-1}$ at 1 AU at 650 nm.}


\subsubsection{Identifying Line Asymmetry and Broadening}\label{sec:3-2-3}

In most studies, line asymmetry and broadening are visually identified from observed spectral profiles \citep{2016A&A...590A..11V,2018PASJ...70...62H,2020PASJ..tmp..253M,2020ApJ...891..103N,Notsu2023}.
In this study, we primarily rely on visual inspection of the line profiles rather than applying the automatic identification criteria (e.g., BIC method in \citealt{2025ApJ...979...93K}).
In addition to visual inspection, we also evaluate asymmetry based on the relative shifts of the line center components. The process is as follows.

First, in dynamic spectra, if blueshifted or redshifted absorption and emission components are visually evident separated from the line center, 
the flare is identified as an asymmetry flare (EK2, EK6, EK7, and EK9 are identified in Table \ref{tab:flare-basic}). 
For these flares, line broadening is not characterized because of the contamination between asymmetry and broadening. 

Otherwise, to determine if the line center component itself exhibits blue/redshift, the following approach is adopted: (1) Spectra integrated over $\sim$10 minutes are generated to enhance the S/N. (2) The spectrum is fitted with a single Gaussian component to check whether the central velocity (i) exceeds the half of the velocity resolution, i.e., 75 km s$^{-1}$, and (ii) exceed the line center above the error bars. 
In the fitting, spectra were integrated over 10-minute intervals to produce better S/N spectra and the integrated spectra with $\Delta$EW$>$0.05 {\AA} (a relatively conservative threshold intended to exclude extreme outliers) were fitted with a single Gaussian. Fits with the Gaussian amplitudes below 0.01 were also discarded, as this criterion is roughly equivalent to a 3-sigma level. (With this process, EK12 is identified.) 

Finally, for no line asymmetry cases, we evaluate whether line broadening occurred. This is assessed by verifying if the consecutive two points of the fitted Gaussian width (FWHM) exceeds the FWHM of the original H$\alpha$ profile (typically, 4.2$\sim$5.0 {\AA} for EK Dra and 4.7$\sim$7 {\AA} for V889 Her, depending on dates because of the weather=water vapor absorption and activity level) beyond the error range. 



\begin{figure*}
\epsscale{0.5}
\plotone{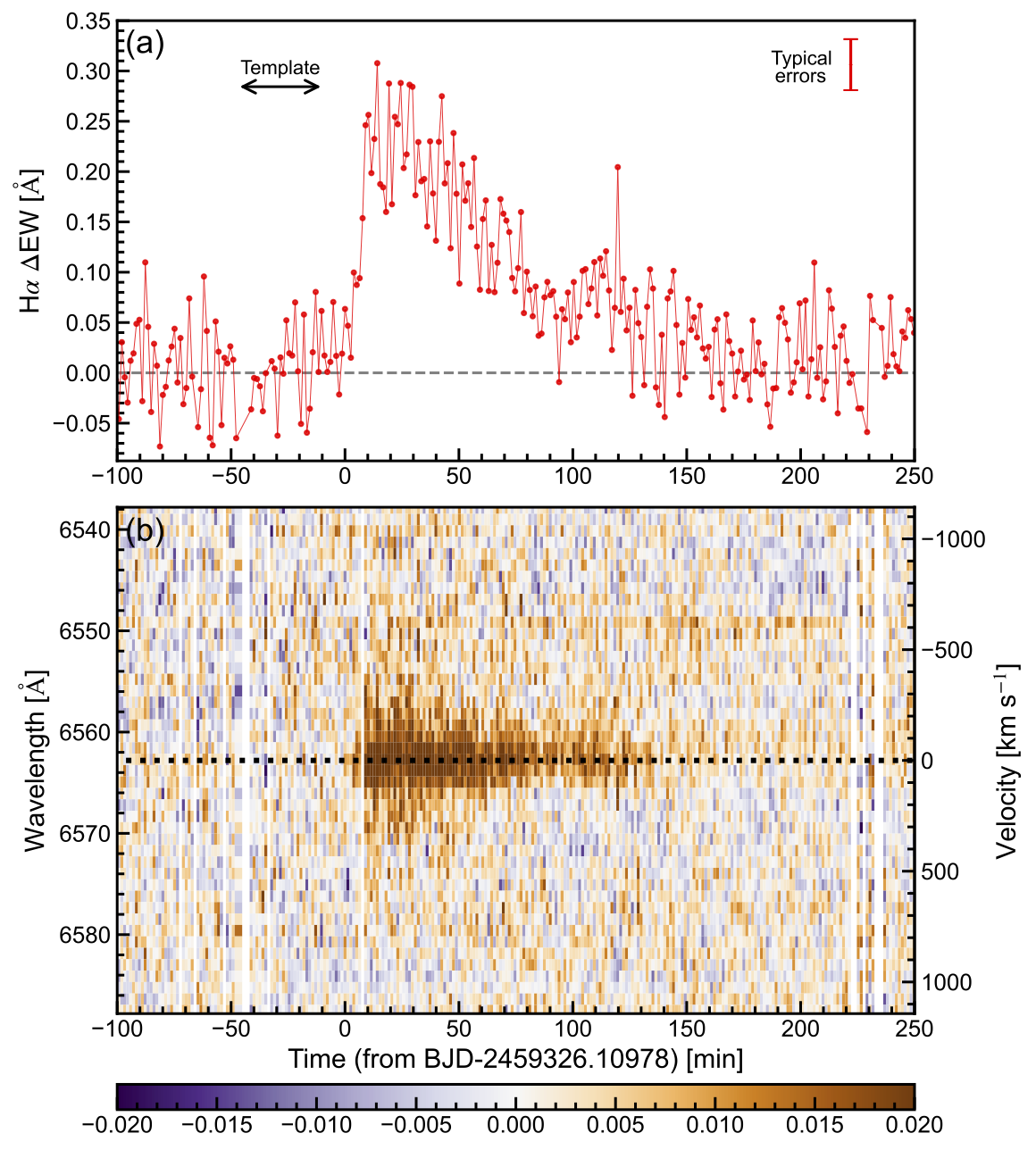}
\epsscale{0.29}
\plotone{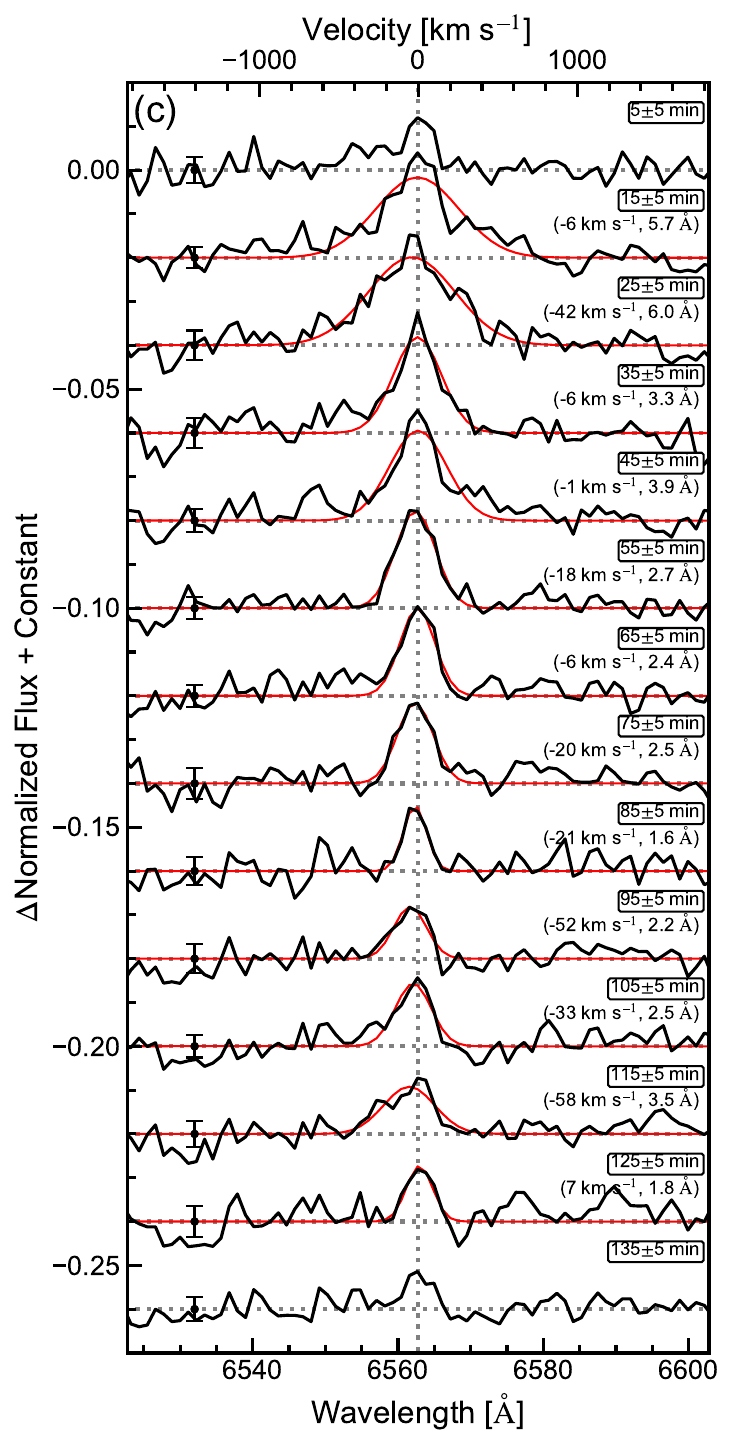}
\caption{H$\alpha$ spectra of the flare event EK3 on 21 April 2021.
(a) Light curves from TESS (black) and H$\alpha$ equivalent width (red). Error bars are shown in the top right. The gray dashed line indicates the background level. The black arrow marks the time used to create the pre-flare template median spectrum.
(b) Dynamic spectrum of the H$\alpha$ line after subtracting the pre-flare spectrum. The dotted line shows the line center.
(c) Time evolution of the pre-flare subtracted H$\alpha$ spectra, shown from top to bottom. 
Each spectrum is time-averaged to improve S/N. 
\Add{Black error bars with points} show uncertainties. 
Red lines are fitted spectra with the Gaussian function for data with sufficient S/N. Each fitting parameter of Doppler shift and line width is indicated in the upper right of each spectrum. The above explanations are applied to Figures \ref{fig:ek3}--\ref{fig:v3}.
}
\label{fig:ek3}
\end{figure*}

\begin{figure*}
\epsscale{0.5}
\plotone{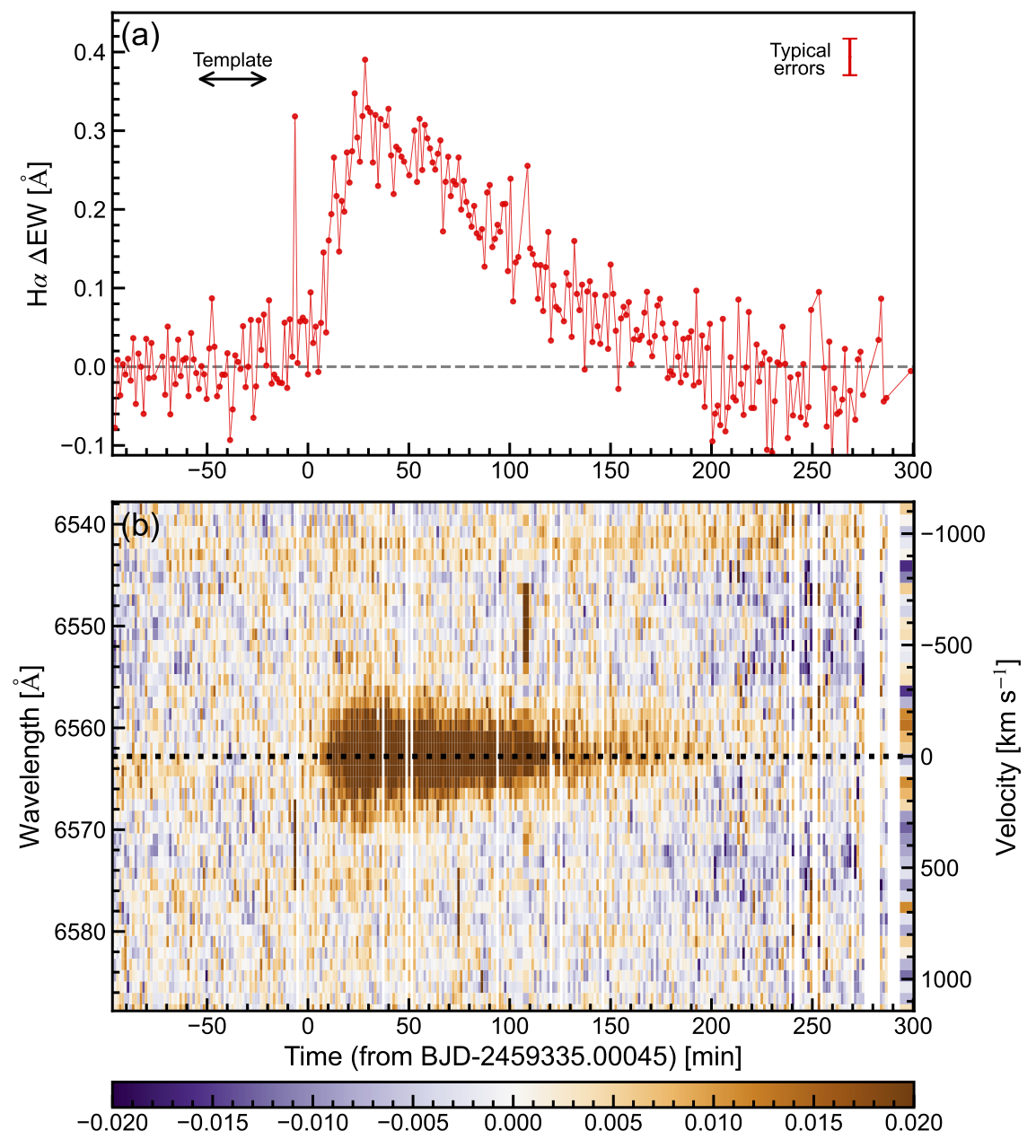}
\epsscale{0.29}
\plotone{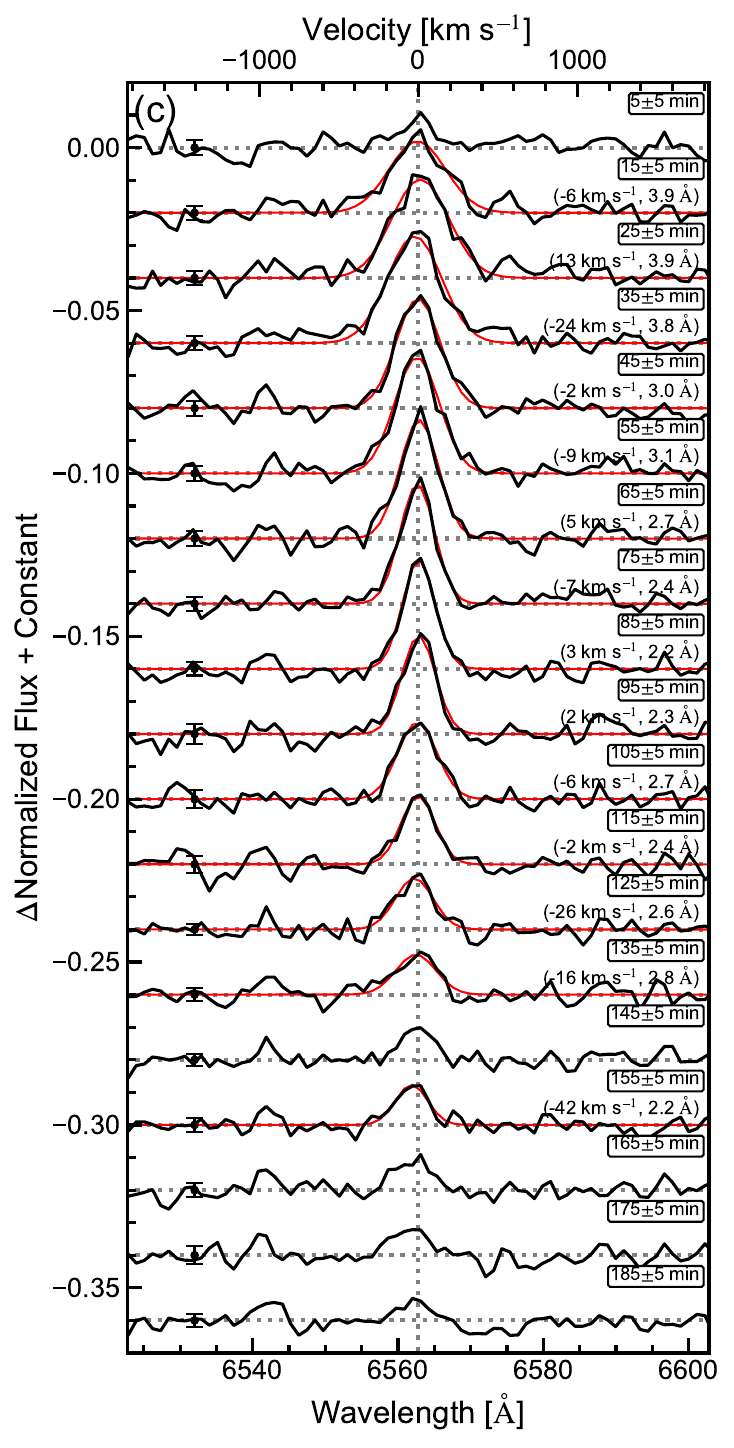}
\caption{Same as Figure \ref{fig:ek3}, but for the flare event EK4 on 30 April 2021.}
\label{fig:ek4}
\end{figure*}

\begin{figure*}
\epsscale{0.5}
\plotone{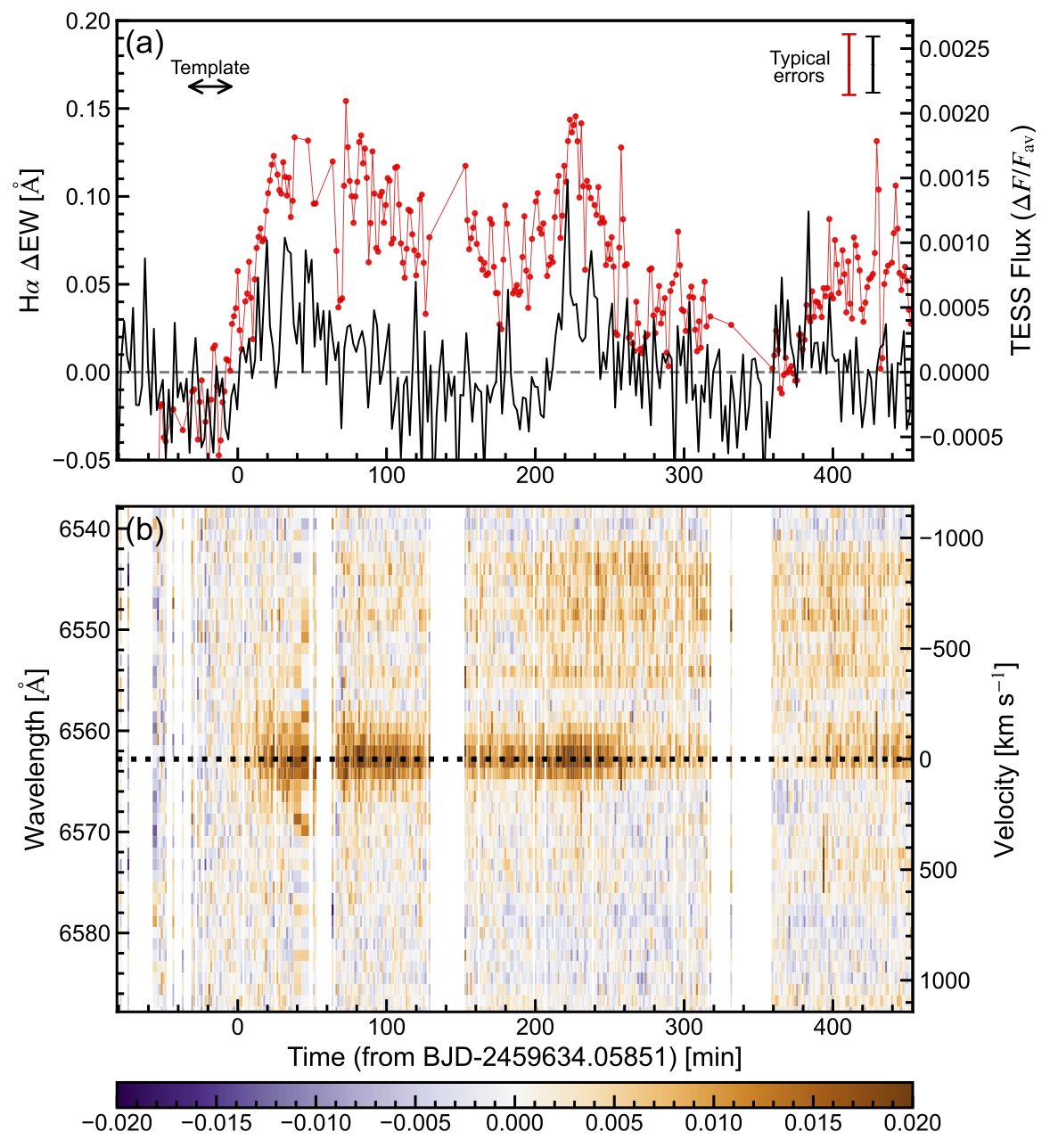}
\epsscale{0.29}
\plotone{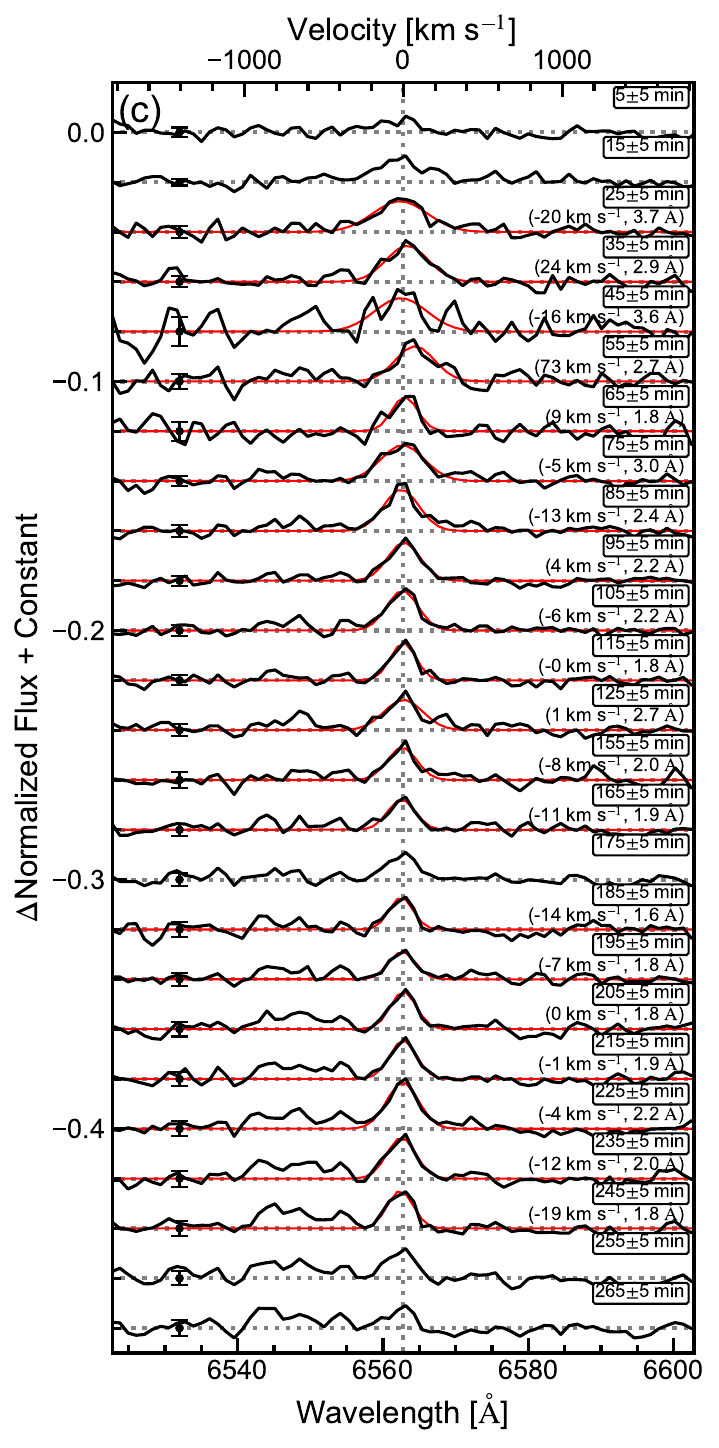}
\caption{Same as Figure \ref{fig:ek3}, but for the flare event EK5 on 23 February 2022. The TESS light curve shows three potential white-light flares around $t \sim 10$, $\sim 210$, and $\sim 380$ minutes, although their significance is low compared to the noise level. The H$\alpha$ data appear to exhibit corresponding peaks; however, the first and second flares do not return to the quiescent level and are instead connected. The final H$\alpha$ peak at $t \sim 400$ minutes falls below the flare threshold ($\Delta$EW $> 0.1$). Therefore, although multiple H$\alpha$ peaks are present, they are treated as a single, complex flare in this study.}
\label{fig:ek5}
\end{figure*}

\begin{figure*}
\epsscale{0.5}
\plotone{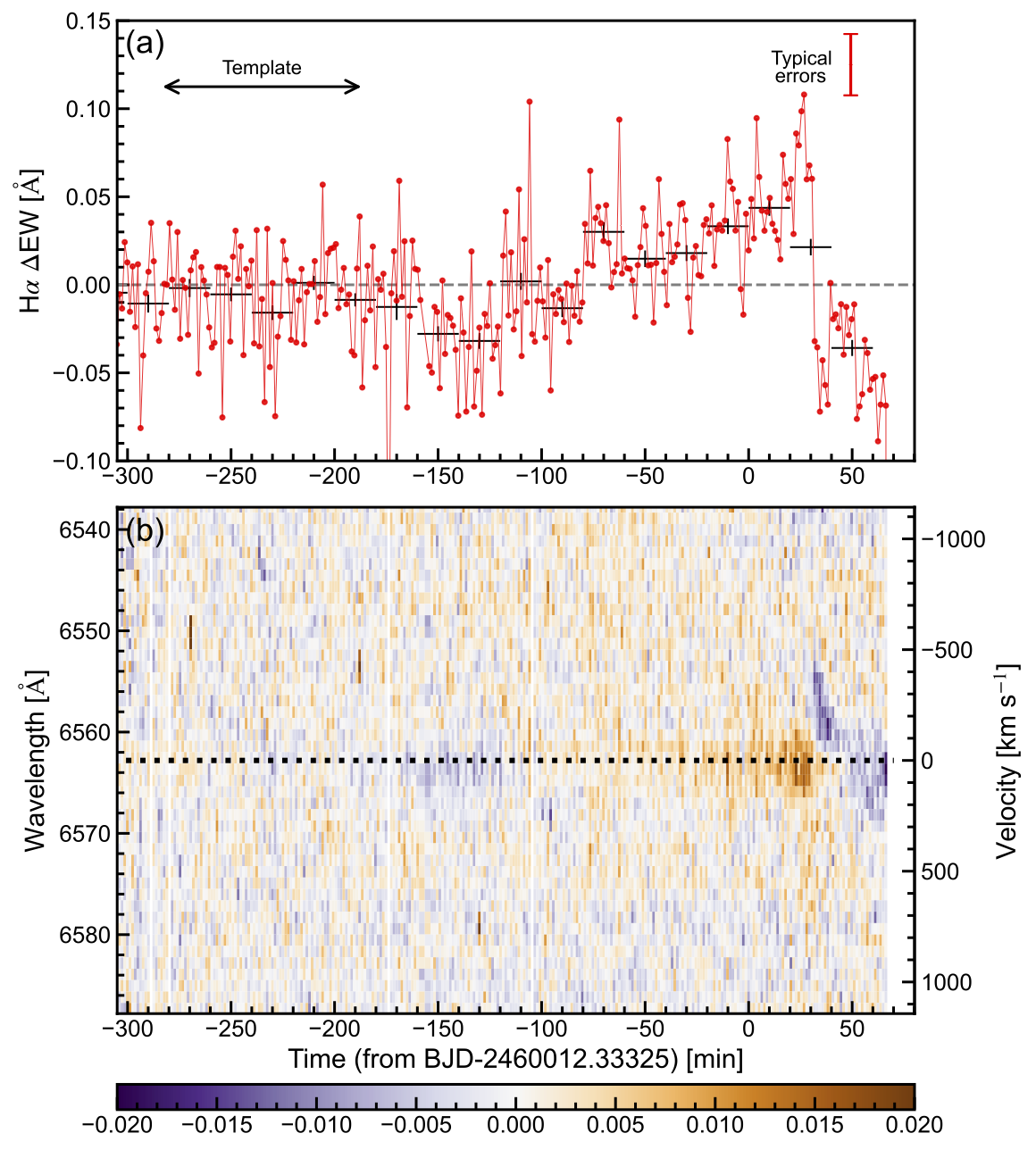}
\epsscale{0.29}
\plotone{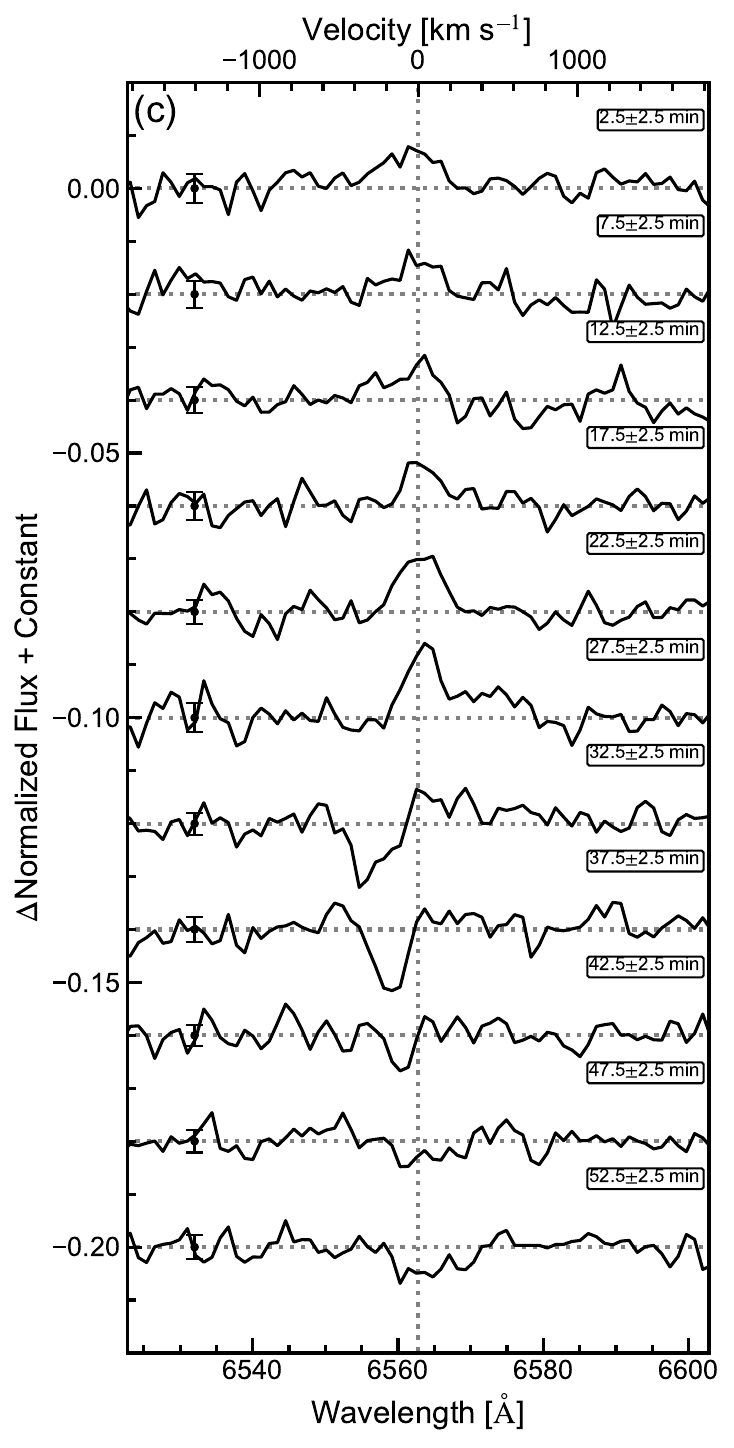}
\caption{Same as Figure \ref{fig:ek3}, but for the flare event EK9 on 8 March 2023.}
\label{fig:ek9}
\end{figure*}

\begin{figure*}
\epsscale{0.5}
\plotone{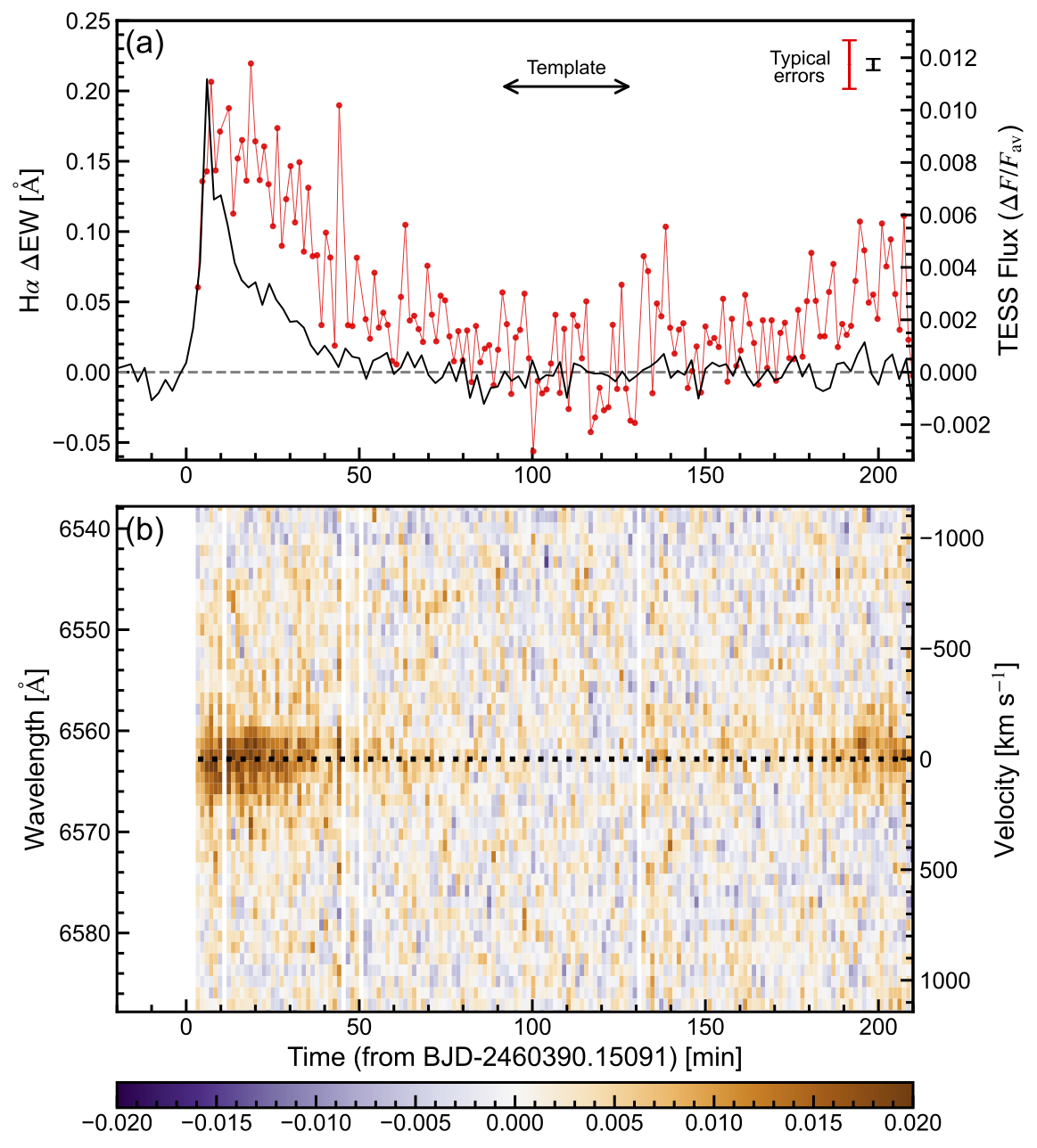}
\epsscale{0.29}
\plotone{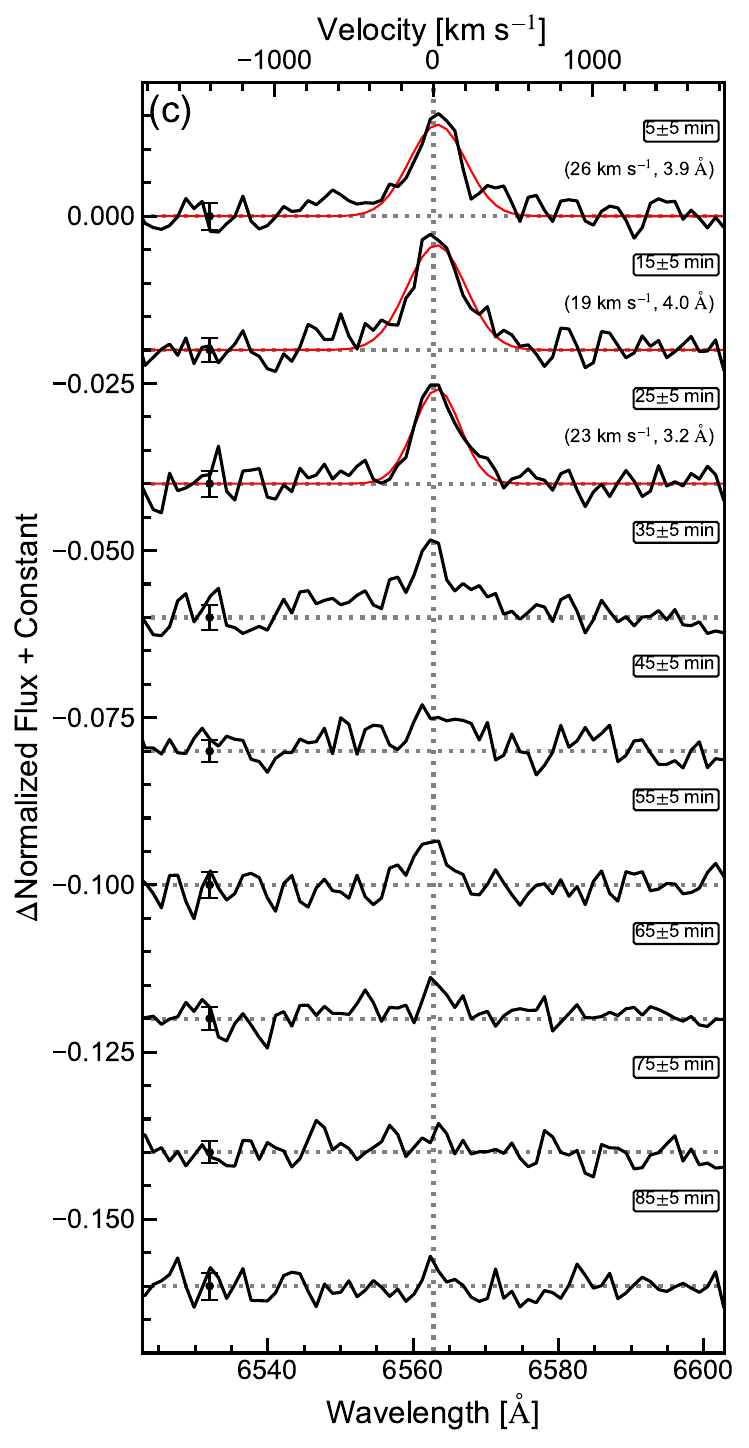}
\caption{Same as Figure \ref{fig:ek3}, but for the flare event EK10 on 20 March 2024.}
\label{fig:ek10}
\end{figure*}

\begin{figure*}
\epsscale{0.5}
\plotone{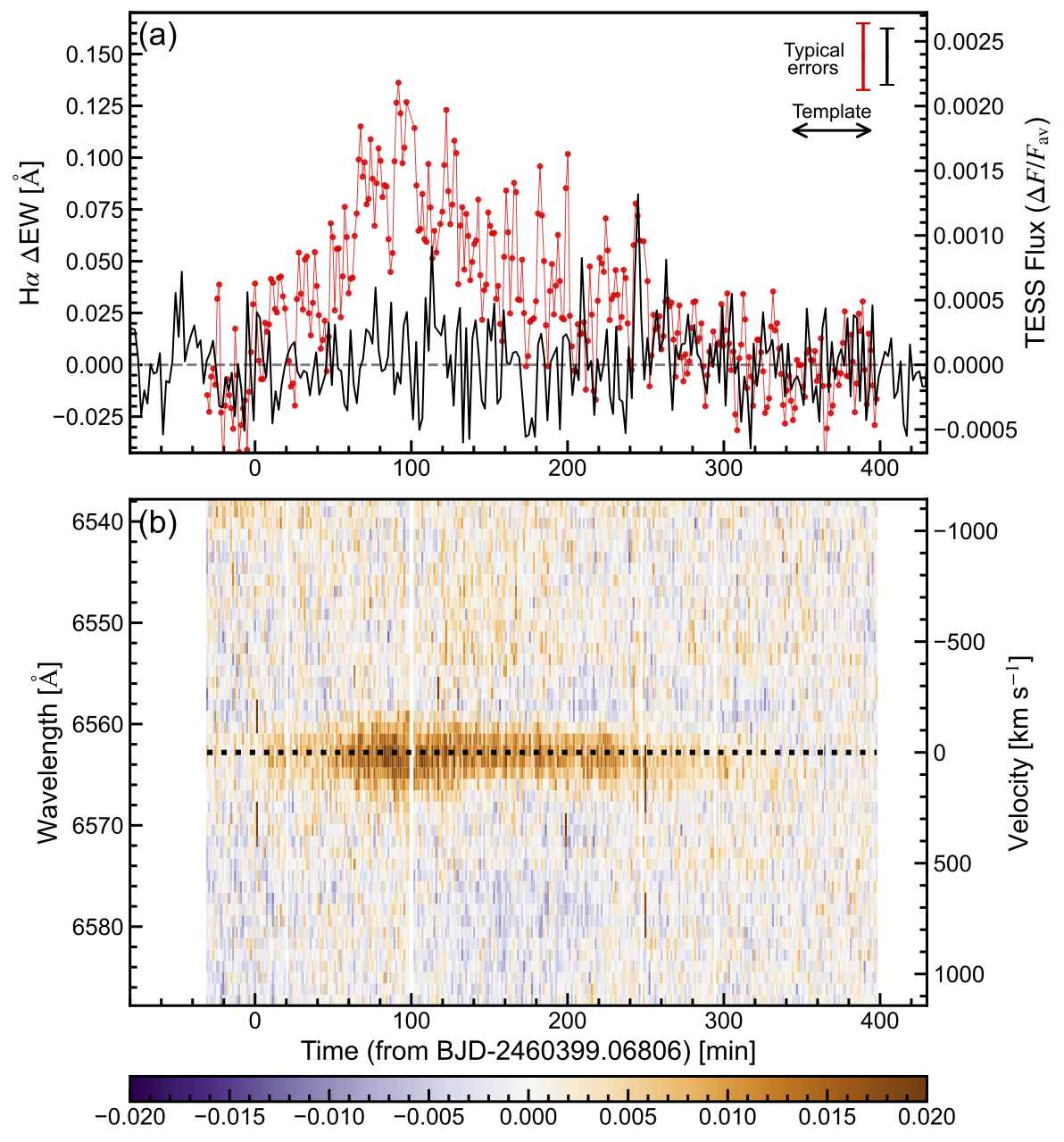}
\epsscale{0.29}
\plotone{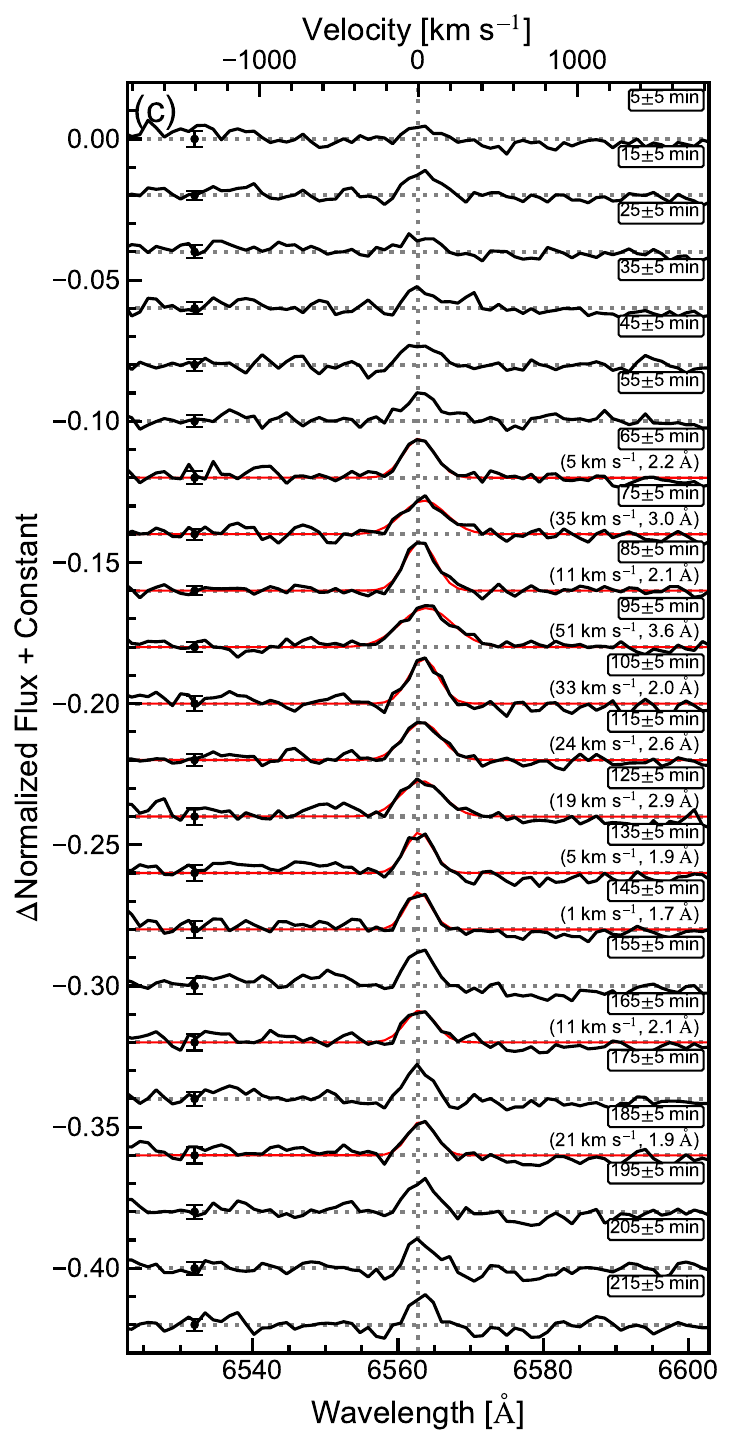}
\caption{Same as Figure \ref{fig:ek3}, but for the flare event EK11 on 29 March 2024.}
\label{fig:ek11}
\end{figure*}

\begin{figure*}
\epsscale{0.5}
\plotone{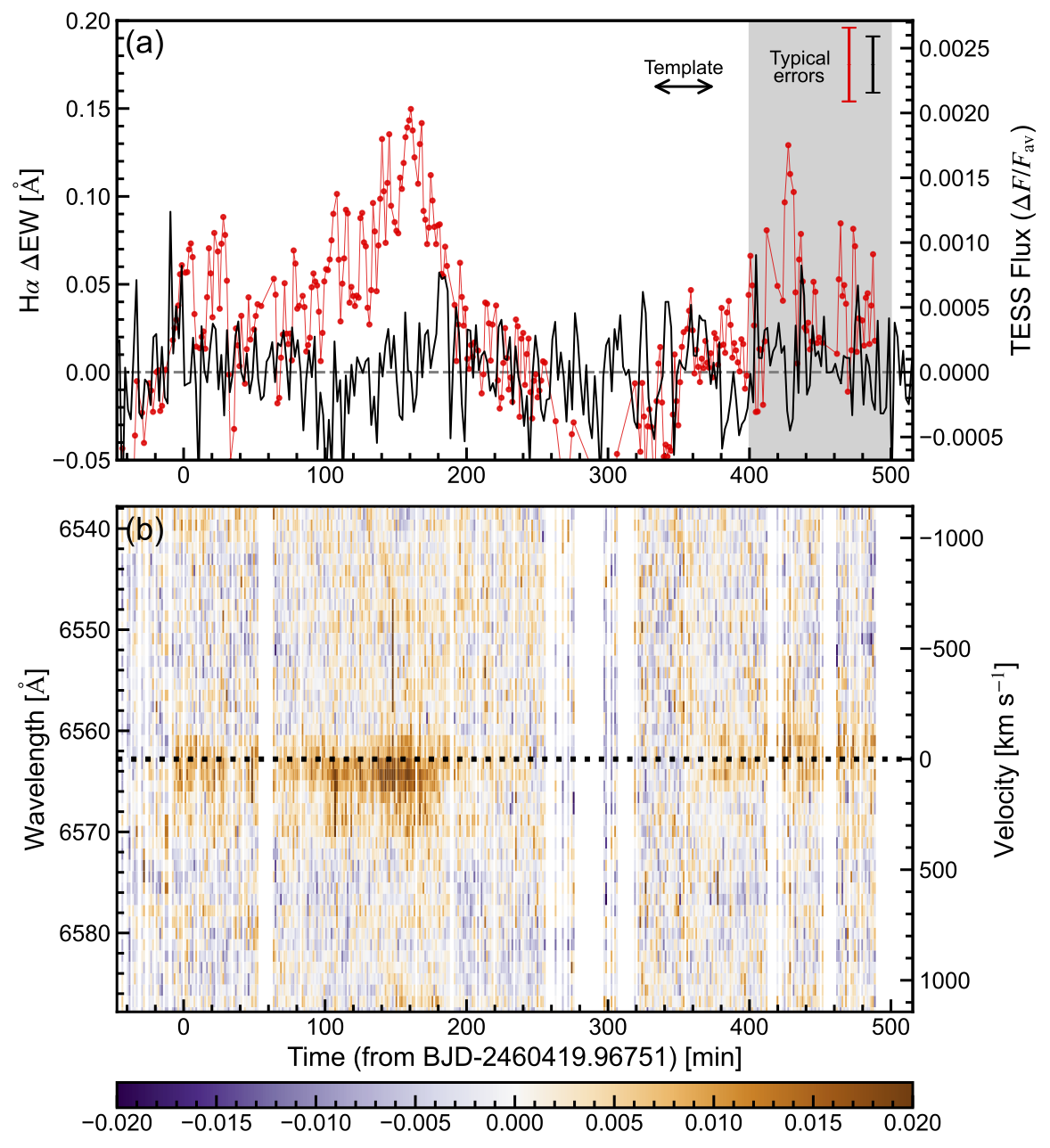}
\epsscale{0.29}
\plotone{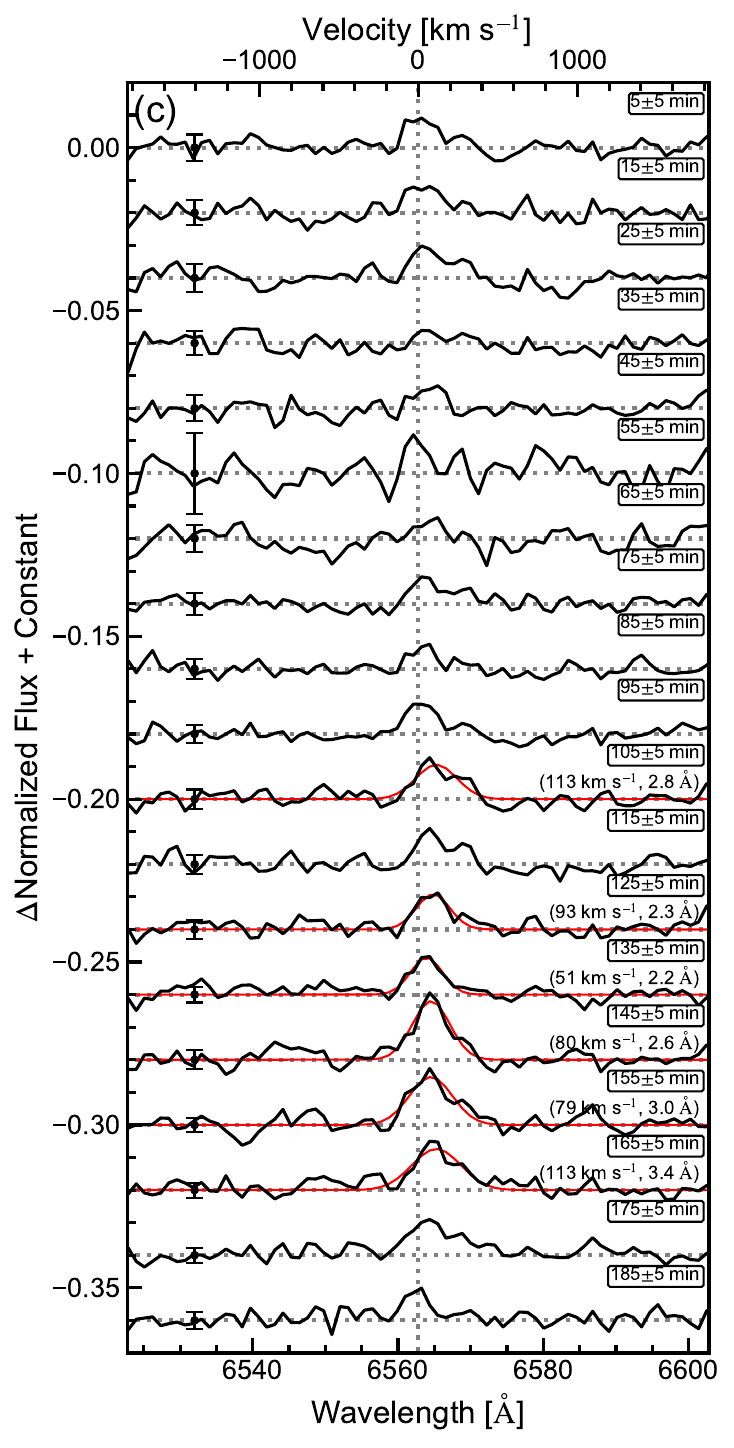}
\caption{Same as Figure \ref{fig:ek3}, but for the flare event EK12 on 19 April 2024. The initial bump observed around 0–30 minutes falls below the detection threshold ($\Delta$EW $>$ 0.1) and is therefore not considered significant. Consequently, the subsequent bump at around 100-200 minutes is treated as the primary superflare. Note that after approximately $t = 260$ minutes, the weather conditions began to be worse, and around $t = 420$ minutes, a slight bump is visible. However, this coincides with another period of poor data quality, and thus we conservatively excluded it as a flare candidate. These exclusions, however, do not significantly affect the estimated flare/CME occurrence rate.}
\label{fig:ek12}
\end{figure*}

\begin{figure*}
\epsscale{0.5}
\plotone{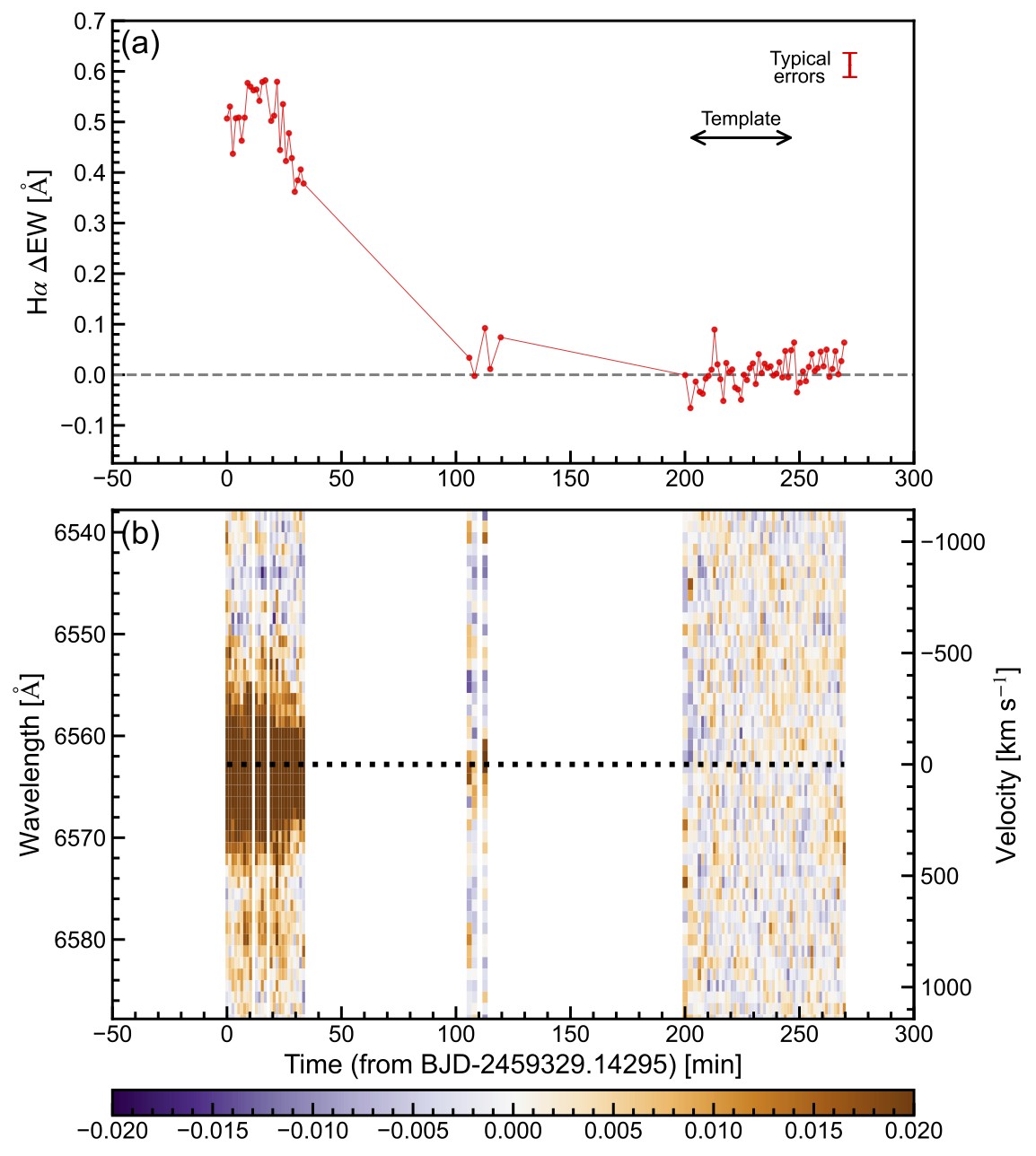}
\epsscale{0.29}
\plotone{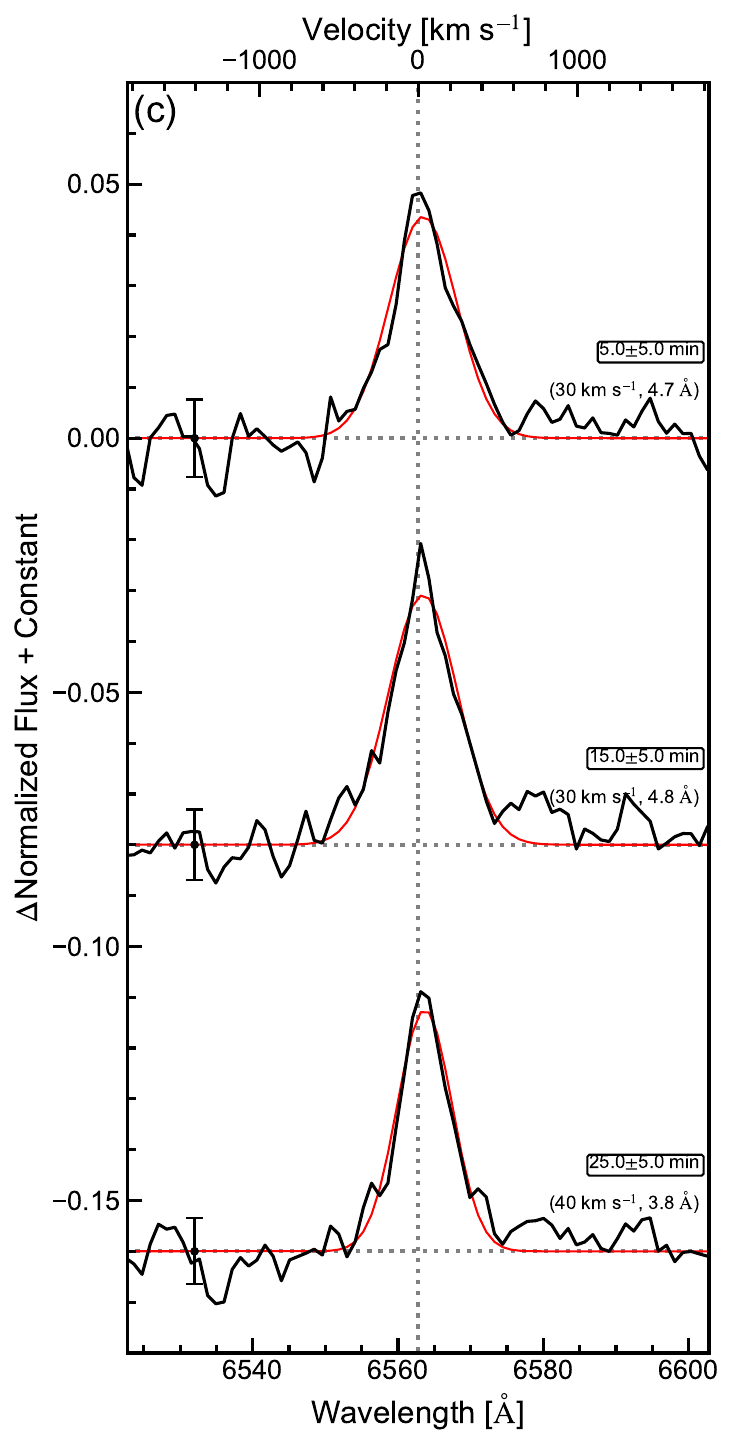}
\caption{Same as Figure \ref{fig:ek3}, but for the flare event V1 on 24 April 2021.}
\label{fig:v1}
\end{figure*}

\begin{figure*}
\epsscale{0.5}
\plotone{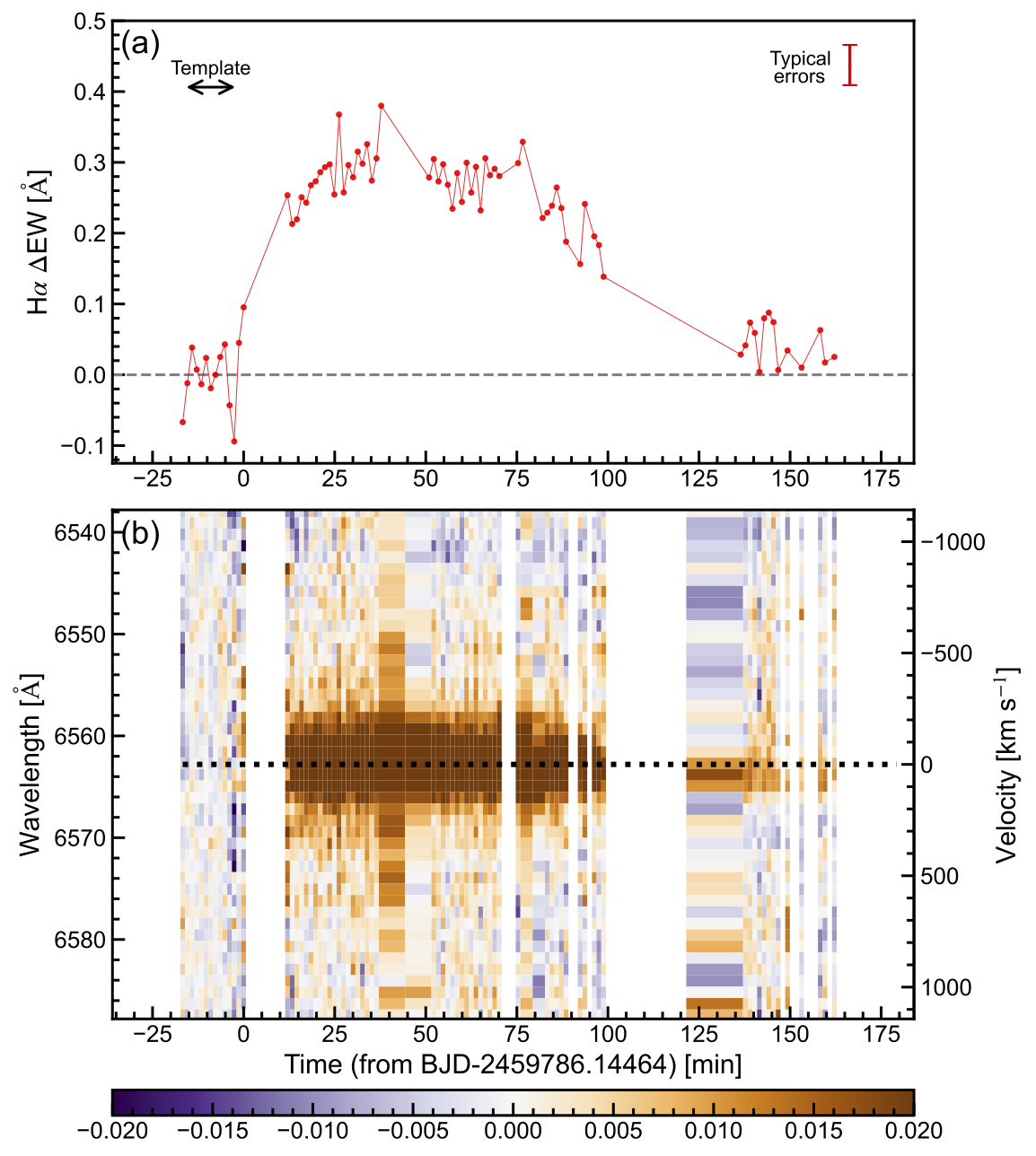}
\epsscale{0.29}
\plotone{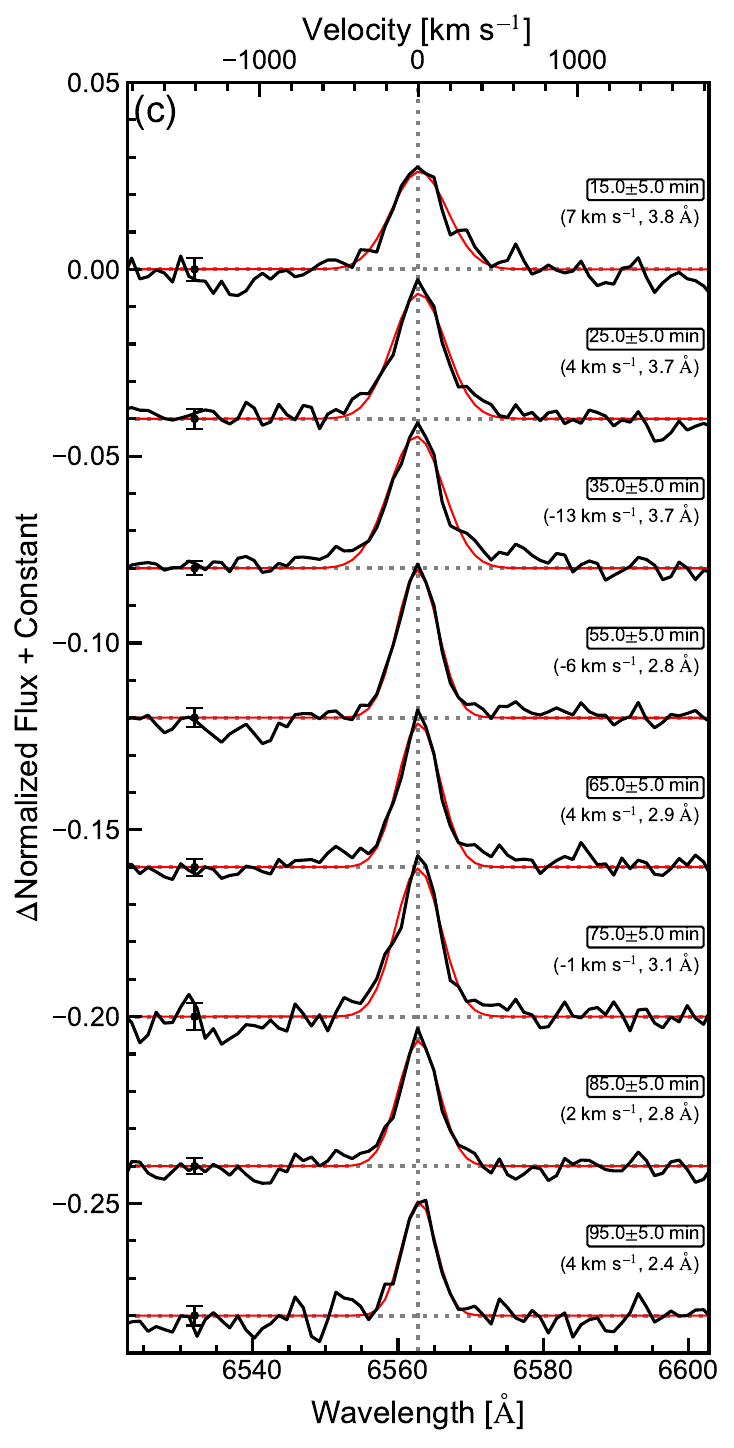}
\caption{Same as Figure \ref{fig:ek3}, but for the flare event V2 on 25 July 2022.}
\label{fig:v2}
\end{figure*}

\begin{figure*}
\epsscale{0.5}
\plotone{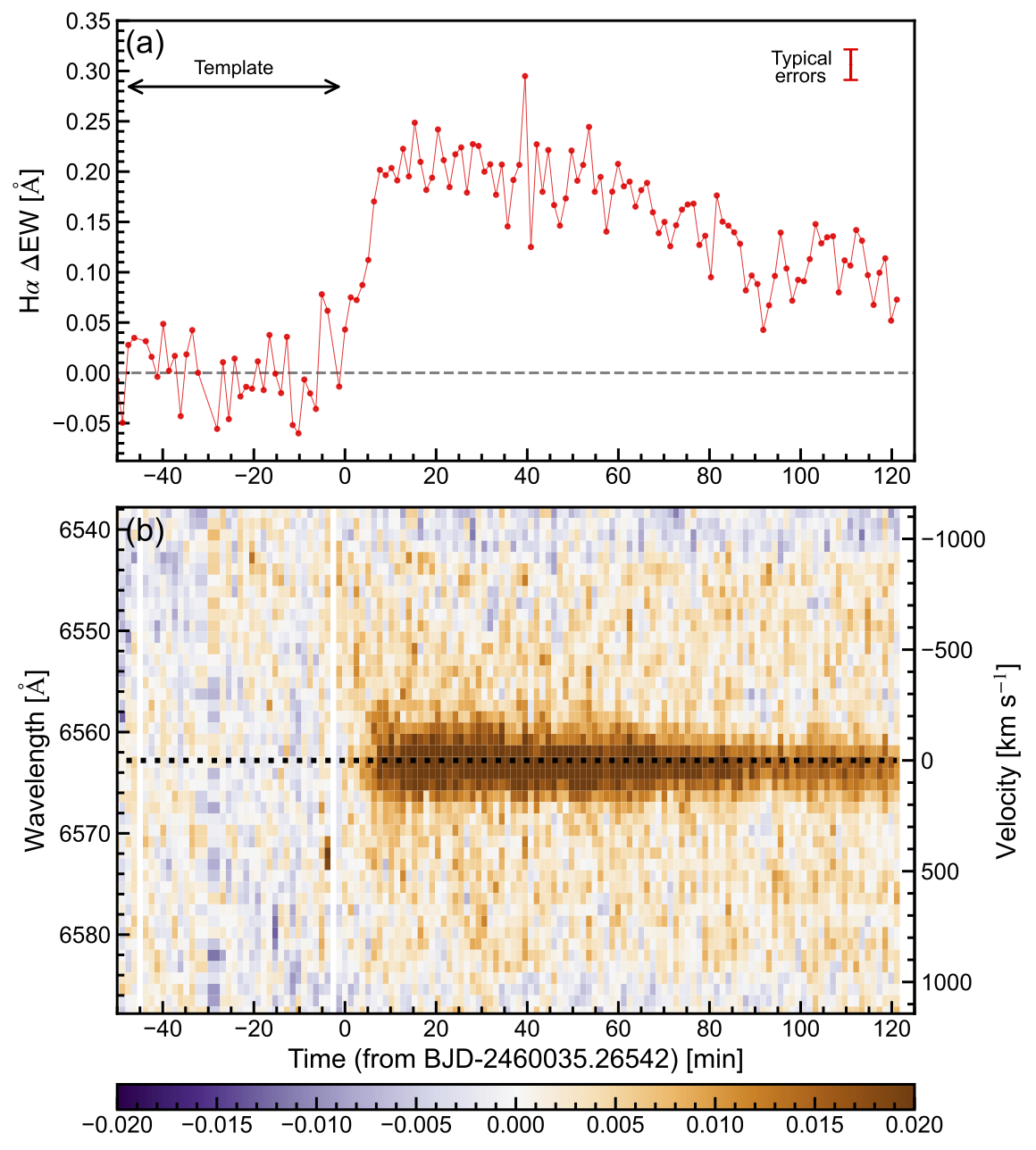}
\epsscale{0.29}
\plotone{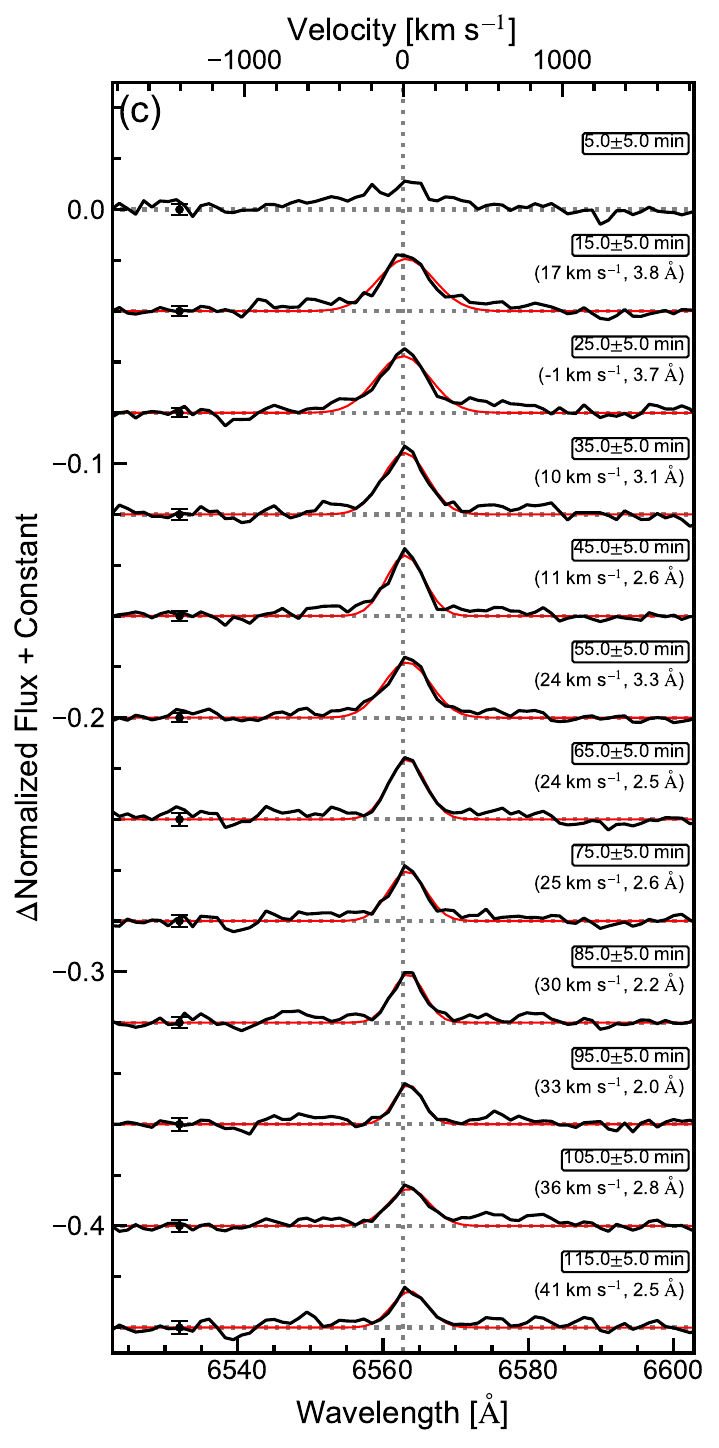}
\caption{Same as Figure \ref{fig:ek3}, but for the flare event V3 on 31 March 2023.}
\label{fig:v3}
\end{figure*}

\begin{figure*}
\epsscale{1.0}
\plotone{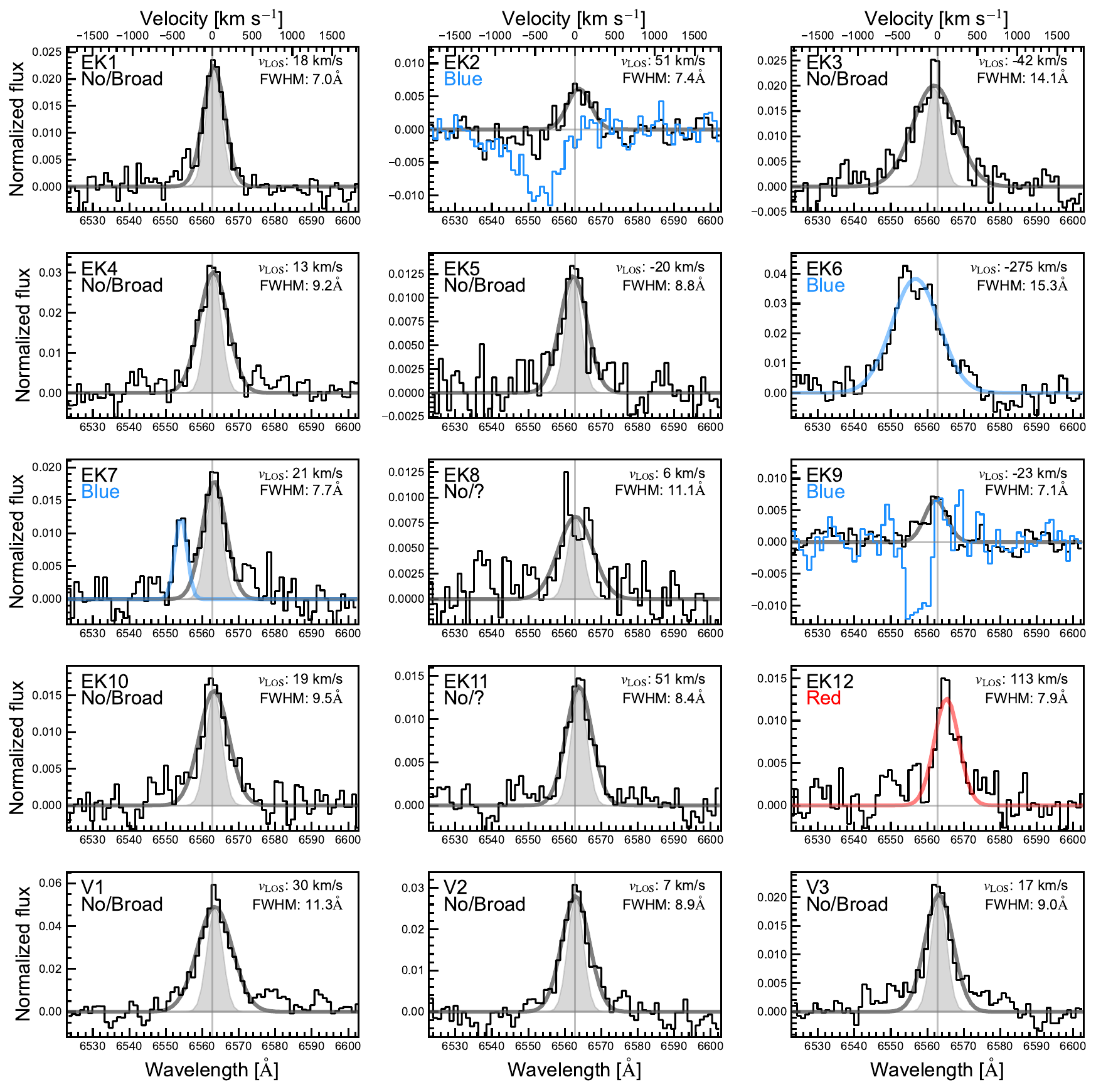}
\caption{The flare catalogue. Each panel displays a typical pre-flare-subtracted H$\alpha$ line spectrum for the superflare events EK1-12 and V1-3, as labeled in each panel. 
The black solid lines correspond to the observed spectrum, highlighting the most prominent time for each flare. The gray thick lines represent the Gaussian fits to the observed spectra. 
For events exhibiting significant blue/redshifts at velocities exceeding 50 km s$^{-1}$, the fitted Gaussians are shown in blue (EK6) or red (EK12), respectively. 
EK7 is fitted with two Gaussian components as described by \cite{2024ApJ...961...23N}. 
For EK2 and EK9, where post-flare blueshifted absorption components appear, these spectra are plotted in blue. The velocity and Gaussian standard deviation from fitting the Gaussian to the flare H$\alpha$ central wavelength radiation are indicated in the top right of each panel. Additionally, as an indicator of broadening, a Gaussian with a standard deviation of 2.5 {\AA} (FWHM of 5.9 {\AA}), which is comparable with a typical H$\alpha$ line width with our spectrograph (see Section \ref{sec:3-2-3}), is plotted in shaded gray.}
\label{fig:catalog}
\end{figure*}

\section{Results}\label{sec:4}

\subsection{Overview of Detected Flares}\label{sec:4-1}

As shown in Table \ref{tab:flare-basic}, we identified 15 H$\alpha$ flares from young solar-type stars: 12 from EK Dra and 3 from V889 Her.
The larger number of detected flares on EK Dra does not imply that EK Dra is more flare-active, as its net exposure time (19.09 days) was significantly longer than that of V889 Her (3.08 days).
The resulting H$\alpha$ flare frequency was 0.62$_{\pm 0.19}$ events per day for EK Dra, which is actually lower than the 0.97$_{\pm 0.65}$ events per day for V889 Her.
The H$\alpha$ flare frequency is consistent with the white-light flare frequency from TESS photometry, averaged over all TESS's sectors: 0.50$_{\pm 0.04}$ events per day for EK Dra and 1.30$_{\pm 0.13}$ events per day for V889 Her.

The detected flares on EK Dra and V889 Her are labeled as EK``XX" and V``XX", where XX indicates the order of discovery\footnote{Note that in our previous study \citep{2024ApJ...961...23N}, we used different labels (E``XX"), and the numbering for the same events differs from that in this study.}.
Figures \ref{fig:ek3}--\ref{fig:v3} show the light curves, dynamic spectra, and pre-flare subtracted spectra for all flares newly reported in this study.
Among the 15 events, 5 on EK Dra were previously reported in \cite{2022NatAs...6..241N,2022ApJ...926L...5N,2024ApJ...961...23N}.
For easier comparison between events, Figure \ref{fig:catalog} presents a catalog of representative spectral line profiles.
The representative profile refers to the period showing clear line asymmetry; if no asymmetry is present, it corresponds to the time near the flare peak when line broadening is most prominent.

The detected H$\alpha$ flare energies range from $1.1 \times 10^{31}$ erg to $4.9 \times 10^{32}$ erg, with FWHM durations spanning from 7.8 minutes to $\ge$242 minutes.
All flare parameters are summarized in Table \ref{tab:flare-basic}. 
Figure \ref{fig:energy-duration} shows that flare duration and energy are positively correlated.
All of the associated white-light flares exceed $\sim10^{33}$ erg.
Even for flares without detected white-light emission or without simultaneous TESS data, their bolometric energies are estimated to be $\gtrsim10^{33}$ erg, considering that H$\alpha$ typically contributes only $\lesssim$1\% of the bolometric X-ray and/or white-light flare energy \citep{2024ApJ...961...23N}.
Therefore, all of the flares are likely classified as superflares ($\gtrsim10^{33}$ erg).


\subsection{Line Profile Asymmetry and Broadening}\label{sec:4-2}

\begin{figure*}
\gridline{
\fig{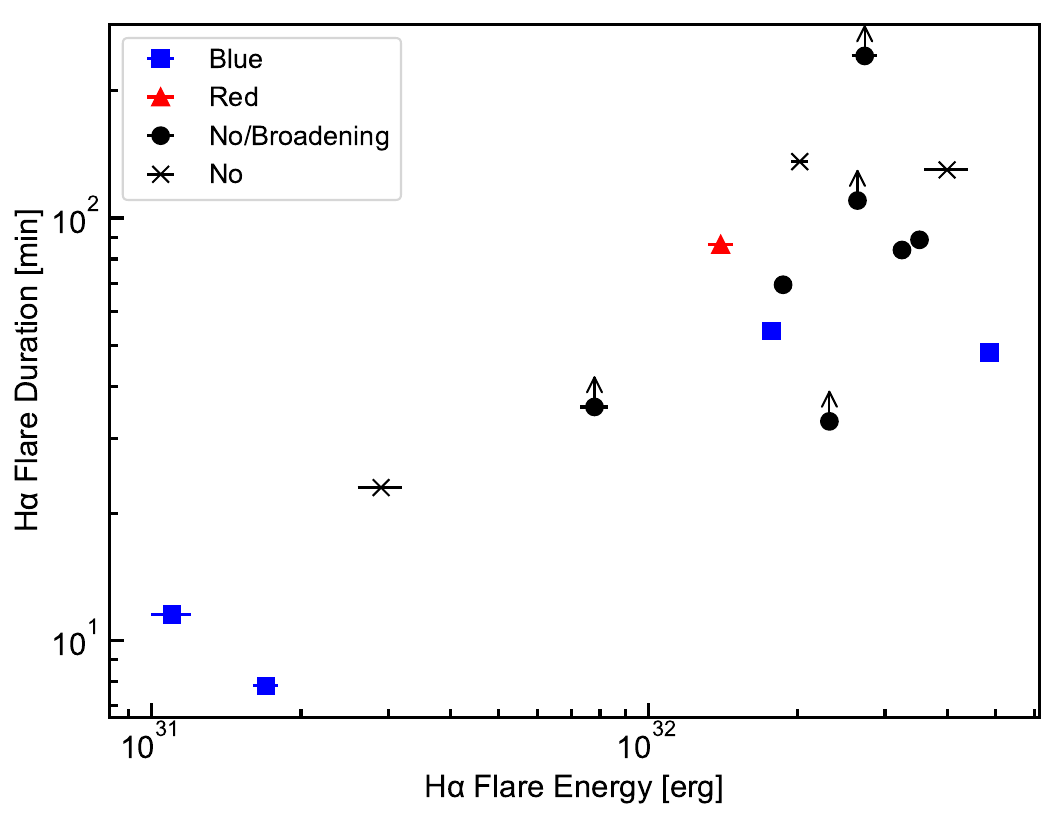}{0.45\textwidth}{\vspace{0mm} (a)}
\fig{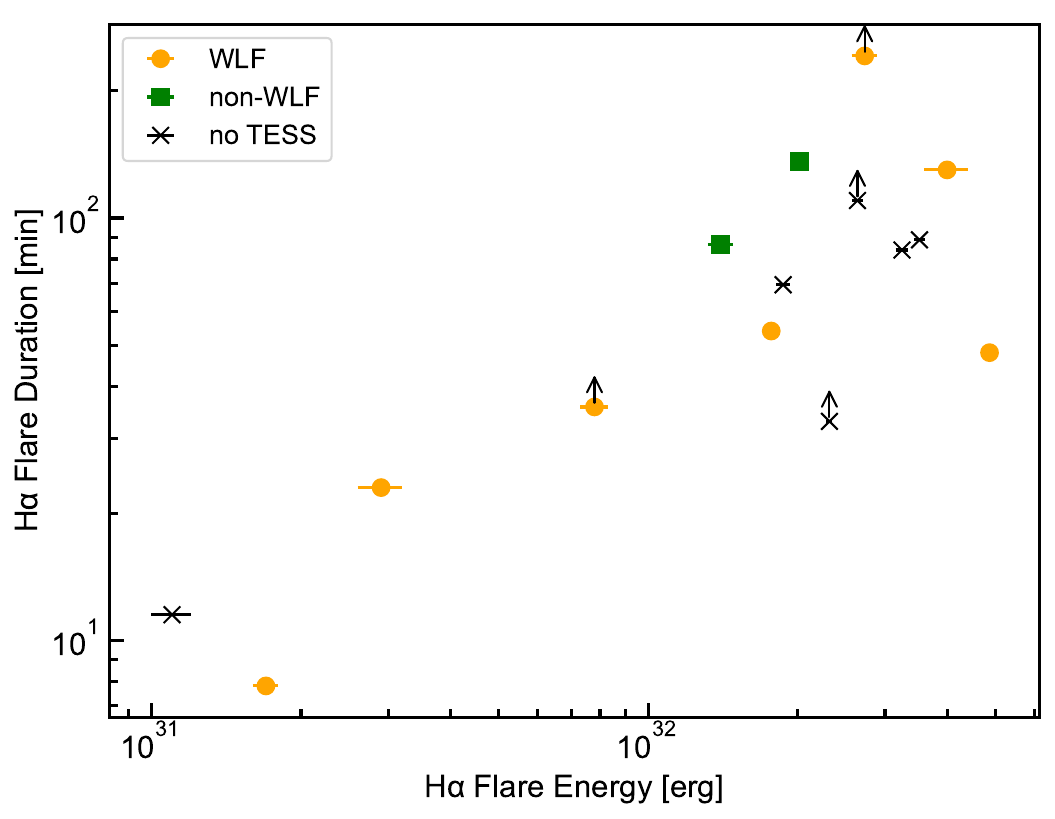}{0.45\textwidth}{\vspace{0mm} (b)}
}
\caption{Relationship between H$\alpha$ flare energy and FWHM duration. (a) Red triangles indicate events with red asymmetry. Blue squares show events with blue absorption or emission (filament/prominence eruptions). Black circles are events with line broadening but no asymmetry. Black crosses show no asymmetry or broadening. (b) Orange circles are WLF events. Green \Add{squares} are non-WLF events. \Add{Black crosses are events withougt simultaneous TESS observations.}}
\label{fig:energy-duration}
\end{figure*}

We identified five H$\alpha$ flares with asymmetries: two show blueshifted absorption profiles (EK2 and EK9), two show blueshifted emission profiles (EK6 and EK7), and one shows a redshifted emission profile (EK12).
All of these events were detected on EK Dra; none of the three flares on V889 Her exhibited asymmetries, despite their relatively large energies ($E_{\rm H\alpha} > 10^{32}$ erg).
Among the four blueshifted events, three (EK2, EK6, and EK7) were previously reported in \cite{2024ApJ...961...23N}.
In this study, we newly report one blueshifted absorption event (EK9) and one redshifted emission event (EK12).
The interesting event EK9 is further discussed in Section \ref{sec:dis:ke9}. 
As shown in Figure \ref{fig:energy-duration}(a), blueshifted flares do not show any distinct energy-duration distribution. 
However, it may be noteworthy that both blueshifted absorption profiles appear in the least energetic and shortest events.
Further statistical aspects including any observational biases will be discussed in Section \ref{sec:dis:frequency}.

We also found that most (eight) of the ten non-asymmetry events, except for EK8 and EK11, exhibit significant line broadening.
The FWHM line width range from 7.0 {\AA} to 14.1 {\AA}, as summarized in Table \ref{tab:linewidth}.
Due to the limited number of non-asymmetry flares, there is no clear separation between broadening and non-broadening events in the flare energy-duration diagram (Figure \ref{fig:energy-duration}).
However, Figure \ref{fig:linewidth}(a) shows a weak positive correlation between flare peak luminosity and line width.
Figure \ref{fig:linewidth}(b) presents the time evolution of line width and luminosity for EK3, EK4, and V3, whose flare evolution is well characterized.
In all three events, line width and luminosity are positively correlated throughout the flare evolution, and the data align along similar power-law trends.
The power-law indices are less than unity, ranging from 0.30$\pm$0.11 to 0.76$\pm$0.10.
These are further discussed in Section \ref{sec:dis:rad}

\begin{deluxetable*}{ccccc}
\label{tab:linewidth}
\tablecaption{Properties of the H$\alpha$ flare broadening.}
\tablewidth{0pt}
\tablehead{
\colhead{ID} & \colhead{Peak Time} & \colhead{FWHM} & \colhead{$L_{\rm H\alpha}$} & \colhead{$L_{\rm WLF}$}  \\
\colhead{} & [min] & \colhead{[{\AA}]} & \colhead{[$10^{28}$ erg s$^{-1}$]}  & \colhead{[$10^{30}$ erg s$^{-1}$]}
}  
\startdata
EK1 & 25±5 & 7.0±1.2 & 3.7±0.9 & 3.6±0.2 \\
EK3 & 25±5 & 14.1±2.4 & 6.6±1.5 & (--) \\
EK4 & 25±5 & 9.2±1.3 & 6.6±1.2 & (--) \\
EK5 & 25±5 & 8.8±2.5 & 2.5±1.0 & no(0.34±0.15) \\
EK10 & 15±5 & 9.5±1.8 & 3.5±0.9 & 6.0±0.1 \\
EK11 & 95±5 & no & 2.7±0.8 & no \\ 
EK12 & 165±5 & 7.9±2.1 & 2.3±0.8 & no \\ 
V1 & 15±5 & 11.3±2.2 & \Add{13.8±3.5} & (--) \\
V2 & 15±5 & 8.9±1.6 & \Add{6.2±1.5} & (--) \\
V3 & 15±5 & 9.0±1.3 & \Add{4.6±0.9} & (--) \\
\enddata
\tablecomments{``Peak Time" means the time when the line width is the largest. Not necessarily flare peak.
``EK8" has too weak spectrum which did not satisfied any criteria for fitting (i.e., $\Delta$EW $>$ 0.05 {\AA} and ``amp" is 0.01.)
}
\end{deluxetable*}

\subsection{Association of White-light Flares}\label{sec:4-3}

Of the 15 events, 9 flares had simultaneous TESS observations (see Table \ref{tab:flare-basic}), all of which were from EK Dra.
Among the 9 H$\alpha$ flares, 7 were associated with white-light flares (WLFs), while 2 (EK11 and EK12) were non-white-light flares (non-WLFs).
As shown in Figure \ref{fig:energy-duration}(b), although the number of events is limited, non-white-light flares (non-WLFs) may tend to have longer durations for a given energy.
In terms of line broadening (Table \ref{tab:linewidth}, Figure \ref{fig:linewidth}(a)), non-WLFs show relatively narrow H$\alpha$ profiles, with FWHM values of none or 7.9$\pm$2.1 {\AA}, compared to $\gtrsim$9 {\AA} for most other flares (except EK1).
Although the sample size is small, they are intriguing and may offer insight into the radiation mechanisms of stellar superflares.






\section{Discussion}\label{sec:5}

\subsection{Summary of New Discoveries}

In the following, we summarize the new findings of this study and outline the direction of the discussion in the subsequent subsections.

\begin{itemize}
    \item We successfully obtained flaring H$\alpha$ spectra from V889 Her for the first time (Section \ref{sec:4-1}), which is an important addition, as all previous studies of solar-type stars relied solely on EK Dra ($\sim$50-125 Myr age) \citep{2022NatAs...6..241N,2022ApJ...926L...5N,2024ApJ...961...23N,2024MNRAS.532.1486L} and we expanded our sample to the younger ZAMS star ($\sim$30 Myr age). None of the three flares on V889 Her showed any H$\alpha$ line asymmetries (\Add{although the number of flares is not enough to statistically discuss the diversity of asymmetries on V889 Her}), but they did show significant line broadening (Section \ref{sec:4-2} and \ref{sec:4-3}).
    
    \item We increased the sample size of flaring H$\alpha$ spectra from young solar-type stars up to unprecedented 15 events (cf. five events in \citealt{2024ApJ...961...23N}), enabling us to discuss statistical aspects such as relationship among flare radiation properties (Section \ref{sec:dis:rad}), CME frequency (Section \ref{sec:dis:frequency}) and CME-mass loss rate (Section \ref{sec:dis:massloss}).
    
    \item The second evidence of ``filament" eruption was detected as blueshift absorption with an intriguing time evolution (EK9, Section \ref{sec:dis:ke9}). 
    
    \item Clear detections of H$\alpha$ line broadening and red asymmetry, along with their time evolution, were made in this study (Section \ref{sec:4-2}). While some indications of these features were previously reported by \cite{2022ApJ...926L...5N,2024MNRAS.532.1486L}, the detection of high-velocity components and clear time evolution in particular is noteworthy.
    Regarding line broadening, we have gathered sufficient data to characterize its statistical properties, which is important for understanding the flare radiation mechanism (Sections \ref{sec:dis:rad} and \ref{sec:dis:ek12}).
    
    \item Simultaneous TESS observations reveal that some H$\alpha$ flares (EK11 and EK12) are non-white-light flares (non-WLFs).
    This phenomenon has been reported in solar flares \citep{2017ApJ...850..204W} and M-dwarf flares \citep{2020PASJ...72...68N,2020PASJ..tmp..253M,Notsu2023,2025ApJ...979...93K}, but this is the first indication for a G-dwarf.
    These findings offer new insights into the possible commonality of radiation mechanisms in solar and stellar flares (Sections \ref{sec:dis:rad} and \ref{sec:dis:ek12}). 
\end{itemize}

Building on these new findings, we discuss implications for the flare radiation mechanism in Section \ref{sec:dis:rad}, focusing on line width, red asymmetries, and white-light emission.
In Section \ref{sec:dis:ek12}, we discuss the origin of the redshift observed in the only flare that shows significant redshifts.
In Section \ref{sec:dis:ke9}, we analyze and interpret the newly observed filament eruption in detail.
Then, in Sections \ref{sec:dis:mass} and \ref{sec:dis:frequency}, we discuss the statistical properties of filament/prominence eruptions on young solar-type stars, focusing on their mass, velocity, and kinetic energy in Section \ref{sec:dis:mass}, and their occurrence frequency in Section \ref{sec:dis:frequency}.
Finally, in Section \ref{sec:dis:massloss}, we provide the first observational estimate of the CME mass-loss rate and compare it with wind mass-loss rates.

\subsection{Insight into Radiation Mechanism for Superflares on the young solar-type stars}\label{sec:dis:rad}

\begin{figure*}
\gridline{
\fig{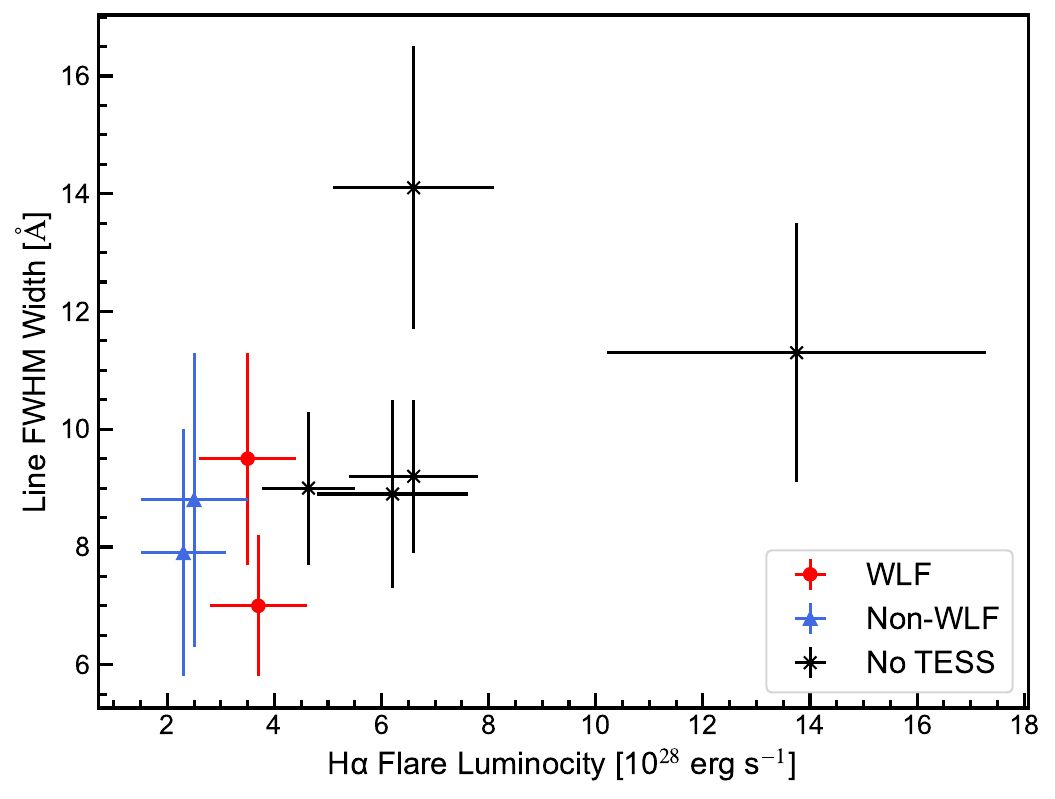}{0.45\textwidth}{\vspace{0mm} (a) }
\fig{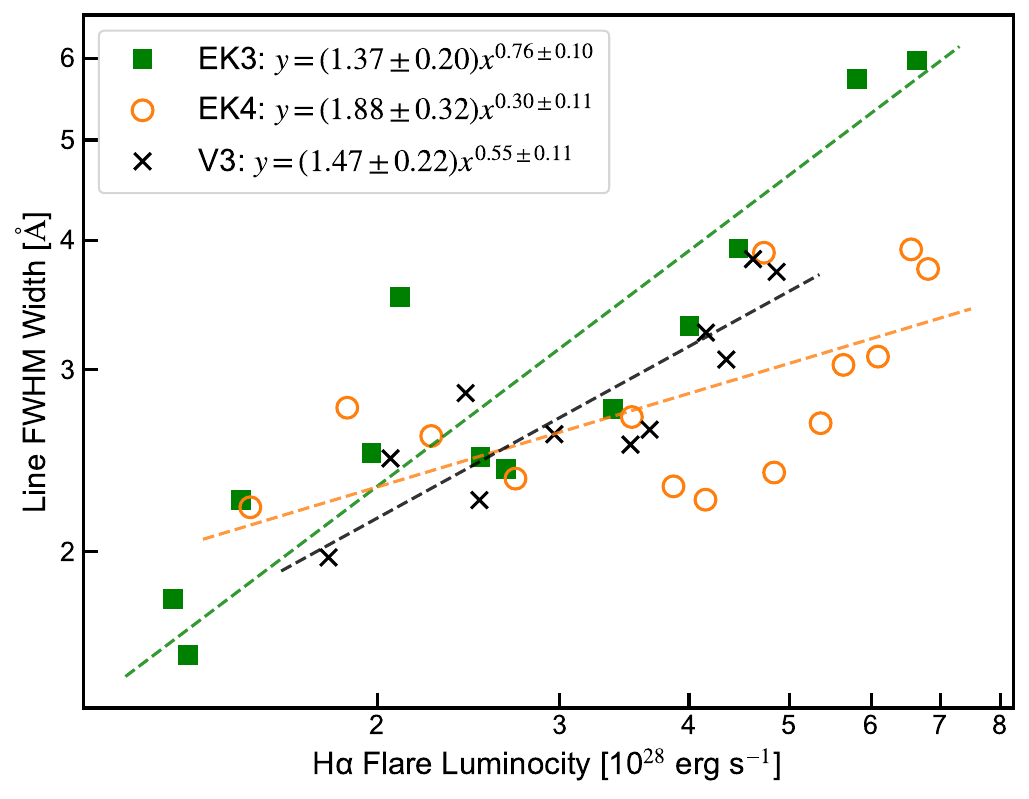}{0.45\textwidth}{\vspace{0mm} (b) }
}
\gridline{
\fig{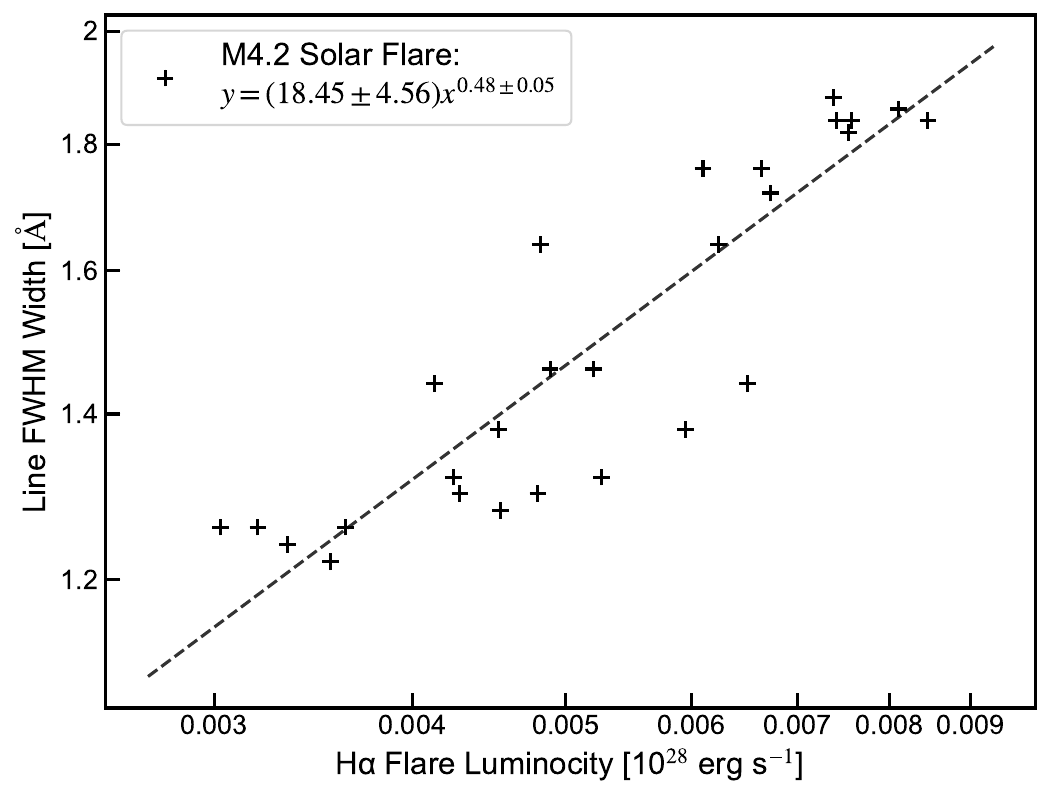}{0.45\textwidth}{\vspace{0mm} (c) }
\fig{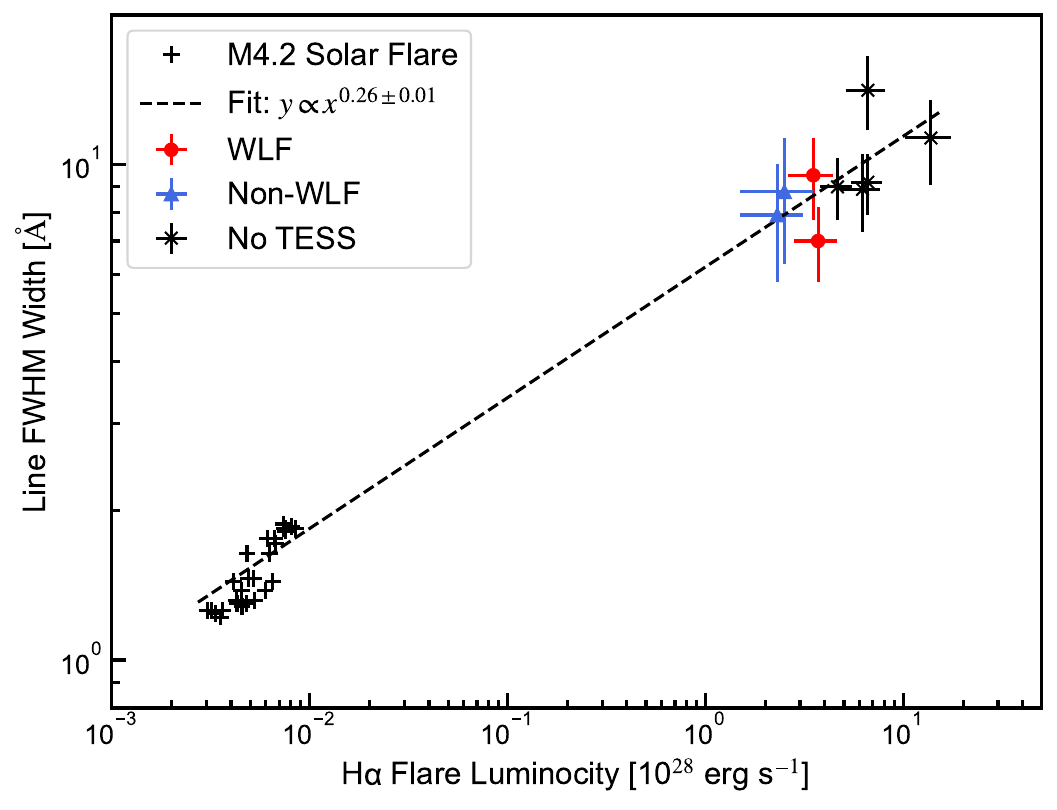}{0.45\textwidth}{\vspace{0mm} (d) }
}
\caption{H$\alpha$ line width vs. luminosity for solar and stellar flares.
(a) Flares on young solar-type stars, marked by flare type: white-light flares (WLF) \Add{with red circles}, non-white-light flares (non-WLF) \Add{with blue triangles}, and flares without TESS data \Add{with black crosses}. 
Values are from the peak time of line broadening.
(b) Time evolution for EK3 (\Add{green squares}), EK4 (\Add{orange circles}), and V3 (\Add{black crosses}). Each colored line shows a power-law fit for the event.
(c) Time evolution of an M4.2-class solar flare from the Sun-as-a-star analysis by \cite{2022ApJ...933..209N}. \Add{The dashed line shows a power-law fit for the event.}
(d) Comparison of solar and stellar data. \Add{The symbols are the same as the panel (a) and (c).}  All data are fitted with a single power-law relation \Add{as indicated with the dashed line}. 
}
\label{fig:linewidth}
\end{figure*}

We found that line broadening is widely seen in young solar-type star's flares, but redshift is absent in most cases (one exception of EK12 as discussed in Section \ref{sec:dis:ek12}).
\cite{2022ApJ...926L...5N} and \cite{2024ApJ...961...23N} report two non-eruptive flares and showed no significant red asymmetry (EK1 and EK8).
\cite{2024MNRAS.532.1486L} also reported the broad spectral features up to FWHM width of 6.8--10.4 {\AA} from four EK Dra flares, and all four events have slight redshifts with the spectral resolution $R$ of 2700, almost similar with our spectrograph ($R$=2000). Note that we have relatively conservative threshold ($\gtrsim$75 km s$^{-1}$) for determining the asymmetry in the spectrum as in Section \ref{sec:3-2-3} while \cite{2024MNRAS.532.1486L} did not use any quantitative threshold. 
By increasing the sample, we are now obtaining further statistical view for line broadening up to 9 events with the largest FWHM width of 14.1$\pm$2.4 {\AA}. 
As a reference, an M-dwarf superflare observed with the same spectrograph showed FWHM broadening up to $\sim$10 {\AA} \citep{2020PASJ...72...68N}, highlighting the particularly broad cases in EK3 and V1.

In general, H$\alpha$ line broadening during solar and stellar flares is thought to result from a combination of Stark broadening and opacity broadening (\citealt{2017ApJ...837..125K,2020PASJ...72...68N}; see also Section 10.2 of \citealt{2024LRSP...21....1K} for review). Stark broadening is sensitive to the chromospheric electron density, which in turn depends on the flux and hardness of the electron beam. Opacity broadening is influenced by the height at which chromospheric flare heating occurs, with deeper heating by energetic electrons typically producing apparently broader line profiles. Therefore, broader H$\alpha$ lines indicate either a higher electron-beam flux or a harder energy spectrum and/of heating electrons.

Building on this understanding, the positive correlation between flare luminosity and line width shown in Figure \ref{fig:linewidth}(a) and (b) suggests that larger flares tend to produce stronger fluxes or harder spectra of flare heating per unit surface area.
Figure \ref{fig:linewidth}(b) also shows that, within each flare event, both the luminosity and the line width increase together. This indicates that the heating parameters per unit area change significantly during the course of a flare, rather than staying constant.
These findings are important for interpreting flares on solar-type stars. For example, most flares from solar-type stars have been studied using photometry from TESS or Kepler, often under the assumption of a blackbody spectrum with a constant temperature of 10,000 K.
However, our results suggest that the chromospheric flare heating significantly vary with both the flare energy and its temporal evolution.
Although this variability is known for solar and M-dwarf flares \citep[see][for review]{2024LRSP...21....1K}, it has not been fully addressed for G-dwarf flare studies.
We suggest that future studies should aim to model the spectra of G-dwarf flares using these observational insights, to improve the accuracy of energy estimates derived from photometric data.

To place the stellar flares in a solar context, we compared our results with Sun-as-a-star observations of an M4.2 class solar flare analyzed by \citet{2022ApJ...933..209N}.
The FWHM width is re-estimated from the data in the literature.
As shown in Figure \ref{fig:linewidth}(c), this solar flare exhibited FWHM line broadening of only 1 to 2 {\AA}.
This clearly highlights that stellar flares on average involve much higher chromospheric heating rate, not just a simple spatial scale up of solar flares with the same heating rate.
This stronger heating rate can be understood by the stronger magnetic field strength of such active stars because the reconnection heating rate can be described as a function of magnetic field strength $\sim \frac{B^2}{4 \pi} V_{\rm A} \propto B^3$, where $B$ is the coronal magnetic field strength and $V_{\rm A}$ is the Alfv\'{e}n velocity \citep{2011LRSP....8....6S}.
The power law index between line width and luminosity for this solar flare is 0.48 with an uncertainty of 0.05, which is consistent with the stellar values ranging from 0.30 to 0.76 (Figure \ref{fig:linewidth}(b)). 
This suggests that the temporal evolution of chromospheric heating is similar between solar and stellar flares, even though the overall energy scale differs.
However, the coefficient of the power-law fit for the solar flare is approximately 18, which is about ten times larger than those for the stellar flares (ranging from \Add{1.4} to 1.9). 
This implies that, although stellar flares are more energetic in total, the average surface heating may be lower than we expect from its luminosity scale.
\citet{2022ApJ...933..209N} also pointed out that Sun-as-a-star spectra tend to underestimate line broadening compared to the brightest flare cores. 
This suggests that stellar superflares can exhibit even stronger broadening in their flaring cores than what is currently observed, indicating more extreme localized heating.
Finally, Figure \ref{fig:linewidth}(d) shows the combined solar and stellar data fitted with a single power law of index 0.25. Although there is currently no theoretical explanation for this value, future studies may help clarify the physical connection between solar and stellar flare heating.

If line broadening reflects strong flare heating, it is reasonable to expect a correlation between the H$\alpha$ line width and both white-light emission and red asymmetry, as already known for M-dwarf flares \citep{2020PASJ...72...68N,2023ApJ...945...61N}. However, our study found no clear correlation between line broadening and white-light emission, likely due to the limited number of flares with simultaneous TESS observations. A larger sample will be necessary to investigate this trend more robustly.
Also, most flares in our sample also did not show extended redshifted components, with the exception of EK12. 
Given that strong heating are inferred for the observed superflares in the above, one might expect more prominent downward plasma motions, with redshift velocities exceeding those typically seen in solar flares (several tens to about 100 km s$^{-1}$; \citealt{2022ApJ...933..209N}). Such motions should be detectable with moderate spectral resolution (R$\sim$2000).
One possibility is observational bias: the redshifted velocity may fall below our detection threshold, or the red asymmetry, typically observed as H$\alpha$ line wing enhancements due to chromospheric condensation in solar flares \citep{2022ApJ...933..209N}, may be buried in the noise. 
In M-dwarf flares, where H$\alpha$ enhancement can often $\gtrsim$100\% relative to the continuum, such features are detectable \citep[e.g.,][]{Notsu2023,2025ApJ...979...93K}. 
However, in G-dwarf flares, where the enhancement is only $\sim$1\%, detecting such asymmetries is much more challenging.


\subsection{New Case Study (EK12): The Only Flare Showing Persistent Redshifts}\label{sec:dis:ek12}

In the event EK12, we detected one flare that exhibited redshifted emission spectra without any prior signature of blueshifts. As shown in Figure \ref{fig:ek12}, EK12 displays two primary peaks, occurring at approximately $t = 0$ minutes and $t = 100$ minutes. The first peak shows a slight redshift, while the second peak exhibits a prominent redshift with a velocity of approximately 100 km s$^{-1}$. The redshift appears to persist throughout the duration of the second peak, but no clear time evolution in the redshift velocity or correlation with luminosity was observed.

In Sun-as-a-star flare observations, redshift evolution due to chromospheric condensation typically transitions from fast to slow phases \citep{2022ApJ...933..209N,2012PASJ...64...20A}. 
However, such a temporal pattern is not seen in EK12, making chromospheric condensation an unlikely explanation. An alternative possibility is a backward-directed or failed prominence eruption. 
However, such events are expected to show velocity evolution, whereas in EK12 the redshift velocity remains nearly constant for more than one hour, rendering this scenario improbable as well.

A more plausible explanation in terms of time evolution can be a post-flare loop emission. 
\citet{2024ApJ...974L..13O} reported a solar flare with both primary and secondary peaks, where the first peak originates from flare ribbon emission and the second from post-flare loops. 
In solar flares, post-flare loop downflows can manifest as redshifted absorption, but in the case of a superflare, the higher coronal density in flaring loops may result in redshifted emission instead.
The observed redshifted velocity in the stellar flare $\sim$100 km s$^{-1}$ is slightly higher than that reported for Sun-as-a-star observations of X-class solar flares $\sim$50 km s$^{-1}$ \citep{2024ApJ...974L..13O}, likely because superflares can exhibit larger downflows due to their greater spatial scales.
Therefore, EK12 may represents observational evidence of post-flare loop emission on a young solar-type star. 
This can be related to the redshifts seen in far-ultraviolet lines during the decay phase of a superflare on EK Dra \citep{2015AJ....150....7A}.

\subsection{New Case Study (EK9): Gigantic Filament Eruption and Indication of Pre-Eruption Activity}\label{sec:dis:ke9}

\begin{figure*}
\gridline{
\fig{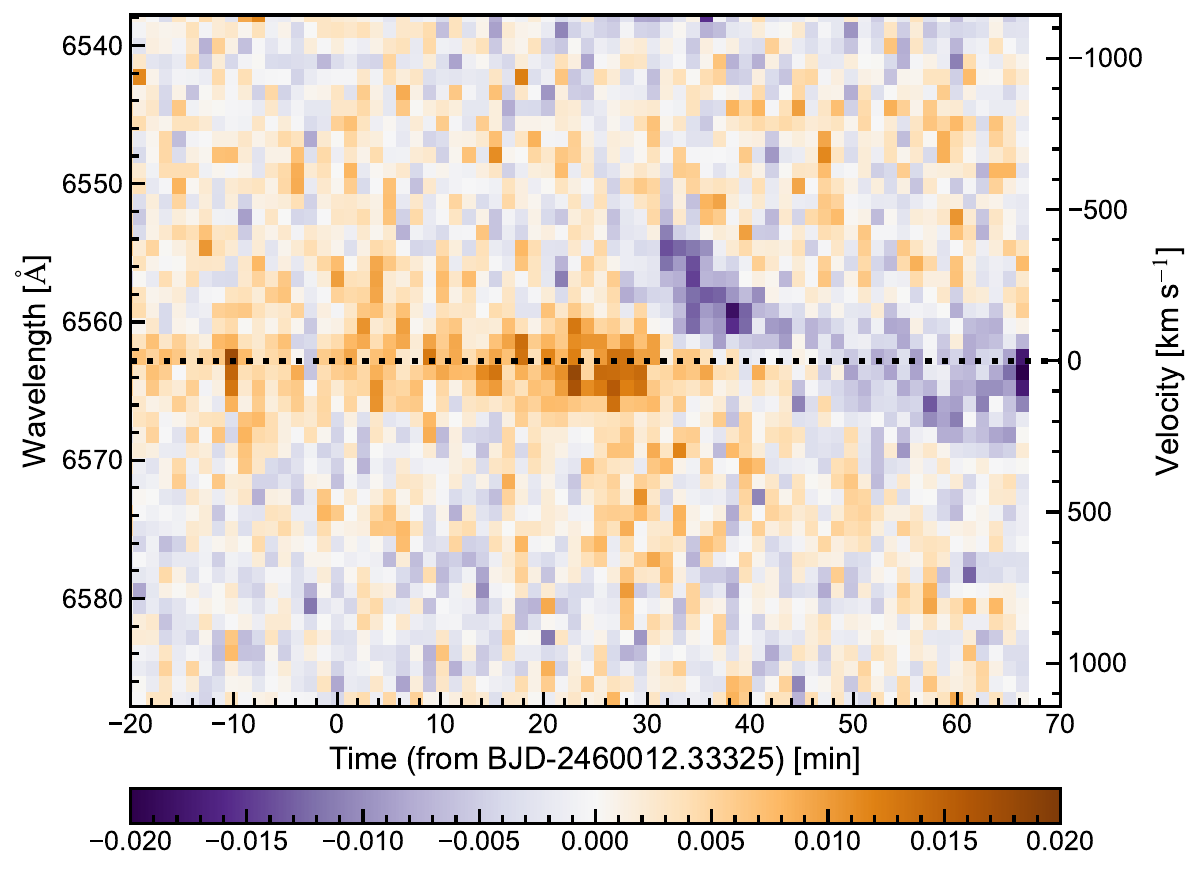}{0.5\textwidth}{\vspace{0mm} (a) }
\fig{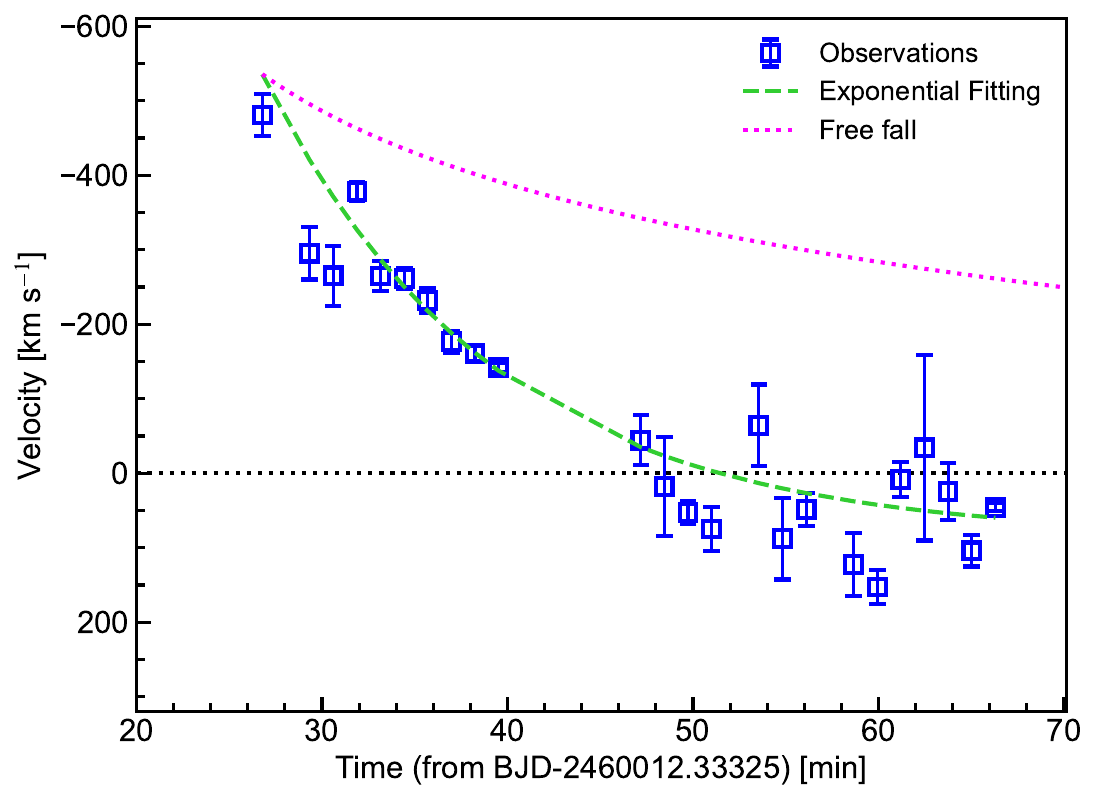}{0.5\textwidth}{\vspace{0mm} (b) }
}
\caption{Time evolution of the filament eruption from flare event EK9.
(a) Zoom-in view of the dynamic H$\alpha$ spectrum. 
\Add{The dotted line is the H$\alpha$ line center wavelength.}
(b) Absorption velocity over time \Add{(blue squares)}. The green line shows an exponential fit and the magenta line shows the free-fall velocity.
}
\label{fig:vel.prom.EK9}
\end{figure*}

We discovered a new case of blueshifted H$\alpha$ absorption profile in the event EK9, undoubtable evidence of stellar filament eruption (see also \citealt{2022NatAs...6..241N}).
The first event (EK2) was reported and analyzed by \cite{2022NatAs...6..241N}, and here, we did the same analysis for the second event.
The H$\alpha$ radiation energy is $1.1\times10^{31}$ erg. Considering the H$\alpha$ and bolometric white-light energy relation estabilished in \cite{2024ApJ...961...23N}, the bolometric white-light energy is estimated to be $\sim10^{33}$ erg, which is comparable with the first event (EK2, $2\times 10^{33}$ erg).
The maximum blueshifted velocity is 480 km s$^{-1}$, which is also similar with the first event (EK2) of 510 km s$^{-1}$. 
The equivalent width ($\Delta$EW) of the absorption is $-$0.067$\pm$0.015 {\AA}, which is not so much different from, but a factor of 3 smaller than, the first event (EK2) of $-$0.16 {\AA}.
These may indicate that the second event (EK9) is comparable in order of magnitude to, but smaller by a few factor than, the first event (EK2) in terms of flare energy, velocity, and area/mass (see Table \ref{tab:summary-of-prominence}).

On the other hand, the velocity evolution of this event shows unexpected behavior that differs from previous studies.
Figure \ref{fig:vel.prom.EK9} presents the time evolution of the blueshifted velocity.
We found that the initial deceleration was 0.83±0.12 km s$^{-2}$, which is significantly larger than both the stellar surface gravity (0.30±0.05 km s$^{-2}$) and the deceleration observed in other events (0.12–0.34 km s$^{-2}$).
In previous studies, the same analysis was applied to EK2 and EK6, and the velocity evolution could be explained by gravitational deceleration alone \citep{2022NatAs...6..241N,2024ApJ...961...23N}, and this is also supported by numerical models \citep{2024ApJ...963...50I,2024ApJ...961...23N}.
However, the time evolution of the new event EK9 suggests that gravity alone cannot explain the observed behavior, and additional interpretations are required.
One possible explanation is that the fast-moving components rapidly disappeared due to heating or expansion, allowing slower plasma to dominate in the later phase, giving the apparent high deceleration.
Such behavior is indeed observed in solar filament eruptions/surges  \citep{2022NatAs...6..241N}.
In this scenario, the open question is why this effect appears in EK9 but not in earlier events like EK2.
An alternative explanation is that the eruption was suppressed by strong overlying magnetic fields \citep{2018ApJ...862...93A}.
In this case, EK2 might have occurred in an open field region, while EK9 may have taken place within a closed magnetic loop.
At present, the exact cause of this unusual behavior remains unclear. Further solar observations and detailed numerical modeling are needed to better interpret this interesting case.

It is also noteworthy that potential pre-flare filament activity, without significant Doppler shifts, is observed from approximately $-170$ minutes to $-120$ minutes before the flare onset, followed by a possible flux increase leading into the impulsive phase, as shown in Figure \ref{fig:ek9}. The binned light curve (in black) in Figure \ref{fig:ek9}(a) exhibits a clear decrease and subsequent increase in flux, exceeding the error bars during the pre-flare period, in contrast to the long-term stable background level.
With 20-min integrated spectra, the decrease in the H$\alpha$ line center, normalized by the continuum level, was approximately $-0.005$, which exceeds three times the standard deviation of the spectra ($\sim$0.003–0.004).
This may provide insight into the activation phase of eruptive phenomena on stars other than the Sun. Further comparison with theoretical models and solar observations is necessary to validate this interpretation.

From the measured $\Delta$EW of 0.067±0.015 {\AA} and a typical velocity of 247±71 km s$^{-1}$, we calculated the mass and kinetic energy using the same methodology as \cite{2022NatAs...6..241N}, based on a cloud model and assumed plasma parameters.
A detailed explanation is provided in Appendix \ref{app:calcmass}.
The resulting mass is $3.4^{+13.8}_{-2.8} \times 10^{17}$ g, and the kinetic energy is $3.4^{+4.2}_{-0.9} \times 10^{32}$ erg.
These values are about a factor of three smaller than those of the first filament eruption case (EK2), but remain broadly consistent with the solar/stellar paradigm (see Section \ref{sec:dis:mass}).

\begin{deluxetable*}{lcccccccccc}
\label{tab:summary-of-prominence}
\tablecaption{Properties of the prominence and filament eruptions on EK Dra.}
\tablewidth{0pt}
\tablehead{
\colhead{ID} & \colhead{H$\alpha$ Asymmetry} & $V_{\rm max}$ & $V_{\rm typical}$ & $V_{\rm disp,typ}$ & --d$V$/d$t$ & $L_{\rm H\alpha, blue}$ & $M_{\rm p}$ & $E_{\rm kin}$ &  Ref.  \\
\colhead{} & \colhead{} & \colhead{[km s$^{-1}$]}  & \colhead{[km s$^{-1}$]} &  \colhead{[km s$^{-1}$]} & \colhead{[km s$^{-2}$]} & \colhead{[$10^{28}$ erg s$^{-1}$]} &  \colhead{[10$^{18}$ g]}  &  \colhead{[10$^{32}$ erg]} &   
}
\startdata
EK2 & Blue absorption & 510 & 260 & 220 & 0.34$_{\pm 0.04}$ & -- & 1.1$^{+4.2}_{-0.9}$ & 3.5$^{+14.0}_{-3.0}$ & (1) \\
EK6 & Blue emission & 330-690 & 300 & 300 & 0.34$_{\pm 0.15}$ & 16.6$_{\pm 1.7}$ & $130^{+290}_{-90}$ & $580^{+1280}_{-400}$ & (2)  \\
 & (two comp.)$^\S$ & (490-690) & (390) & (230) & (0.23$_{\pm 0.14}$) & (9.8$_{\pm 1.3}$) & (--) & (--) &  \\
EK7 & Blue emission & 430 & 380  & 55 & 0.12$_{\pm 0.20}$ & 0.47$_{\pm 0.16}$   & $1.7^{+3.7}_{-1.1}$ & $12^{+27}_{-8}$ & (2) \\
EK9 & Blue absorption & 480 & 250 & 110 & 0.83$_{\pm 0.12}$ & -- & $0.34_{-0.28}^{+1.38}$ & $1.0^{+4.2}_{-0.9}$ & (3) \\
\enddata
\tablecomments{
$V_{\rm max}$ is the maximum blueshifted velocity ($max(-V)$).
$V_{\rm typical}$ is the typical blueshifted velocity around when the EW takes a maximum value.
$V_{\rm disp,typ}$ is the typical velocity dispersion ($\sigma$) in the fitted Gaussian function.
--d$V_{\rm blue}$/d$t$ is the deceleration of the blueshifted component. EK Dra's surface gravity is 0.30$_{\pm 0.05}$ km s$^{-2}$.
$L_{\rm H\alpha, blue}$ is the luminosity of the blueshifted component.
$M_{\rm p}$ is the mass of the erupted prominences/filaments.
$E_{\rm kin}$ is the kinetic energy of the prominence/filament eruptions.
$^\S$The values in parentheses ``()"  are the values of the two component fitting of the flare E1.
References: (1) \cite{2022NatAs...6..241N}, (2) \cite{2024ApJ...961...23N}, (3) This study. 
}
\end{deluxetable*}


\subsection{Statistical View of Velocity, Mass, and Kinetic Energy of Stellar Filament/Prominence Eruptions}\label{sec:dis:mass}

\begin{figure}
\epsscale{1.0}
\plotone{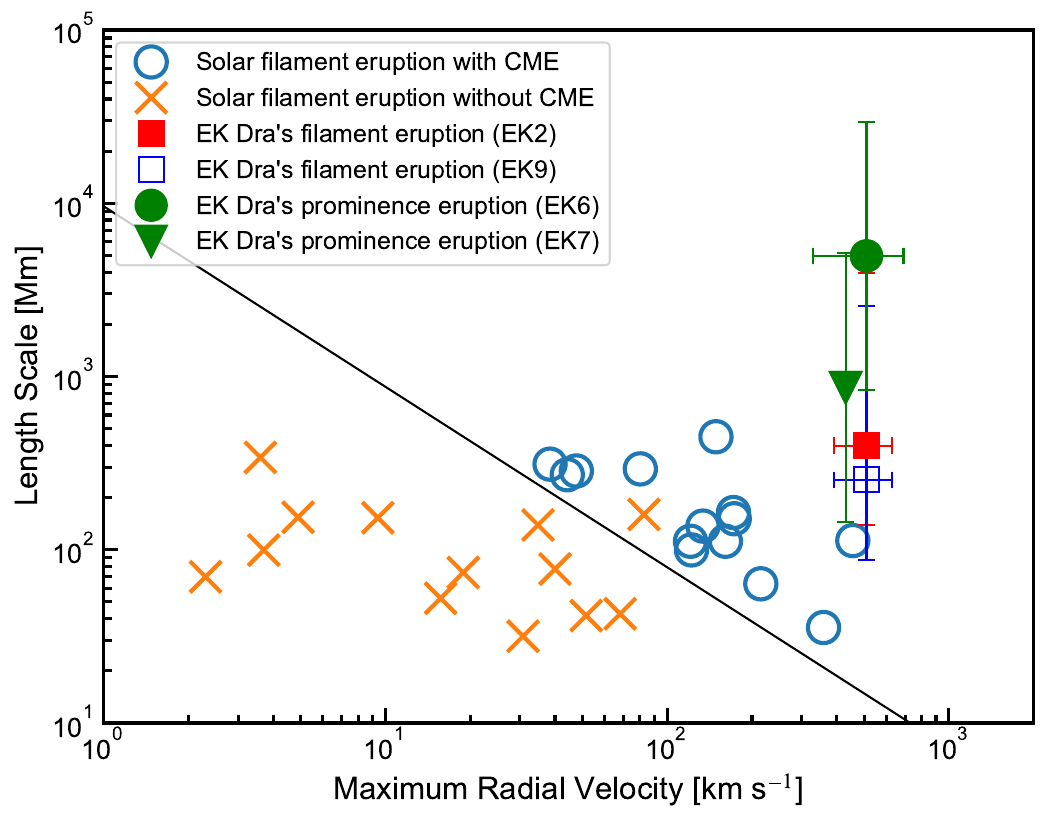}
\caption{
Comparison of velocity and length scale for filament/prominence eruptions on the Sun and EK Dra.
Blue circles and orange crosses show solar filament eruptions with and without associated CMEs, respectively \citep{2021EP&S...73...58S}.
\Add{The dark green symbols (circle and downward-pointing triangles) and red/blue squares} show stellar prominence and filament eruptions on EK Dra, \Add{respectively}.
Stellar velocities are maximum line-of-sight values, while solar velocities are maximum radial values.
The solid black line indicates an empirical solar threshold that roughly separates CME-associated and non-CME filament eruptions \citep{2021EP&S...73...58S}.
}
\label{fig:stat-length}
\end{figure}

\begin{figure}
\epsscale{1.0}
\plotone{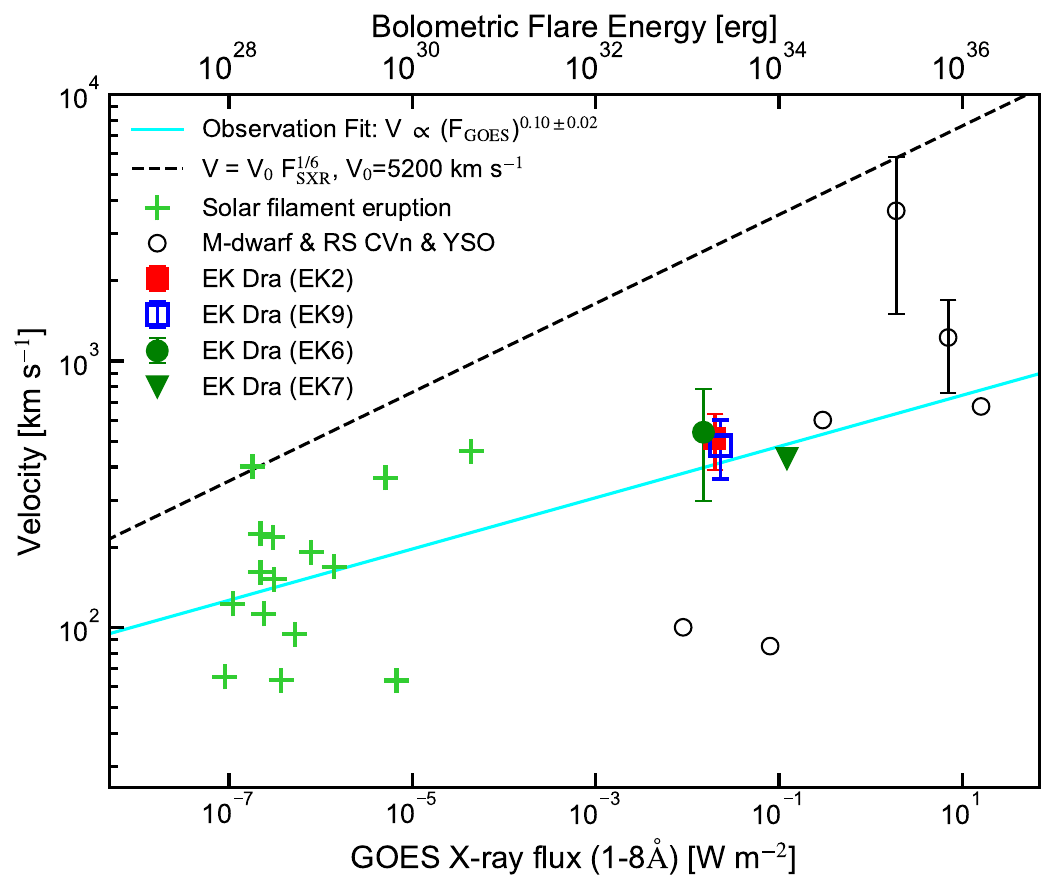}
\caption{
Filament/prominence eruption velocity vs. flare energy, shown as GOES flux (bottom axis) or bolometric white-light flare (WLF) energy (top axis).
\Add{The dark green symbols (circle and downward-pointing triangles) and red/blue squares} show stellar prominence and filament eruptions on EK Dra, \Add{respectively}.
Solar filament eruption data \Add{(lime green pluses)} are from \cite{2021EP&S...73...58S}, and data for M-dwarfs, RS CVn, and young stellar objects \Add{(black circles)} are from the list in \cite{2024ApJ...961...23N}.
The cyan solid line shows a power-law fit to all solar and stellar data. The black dashed line represents a power-law with index 1/6, tracing the upper limit of the data \citep{2016ApJ...833L...8T}.
}
\label{fig:stat-velosity}
\end{figure}

\begin{figure*}
\gridline{
\fig{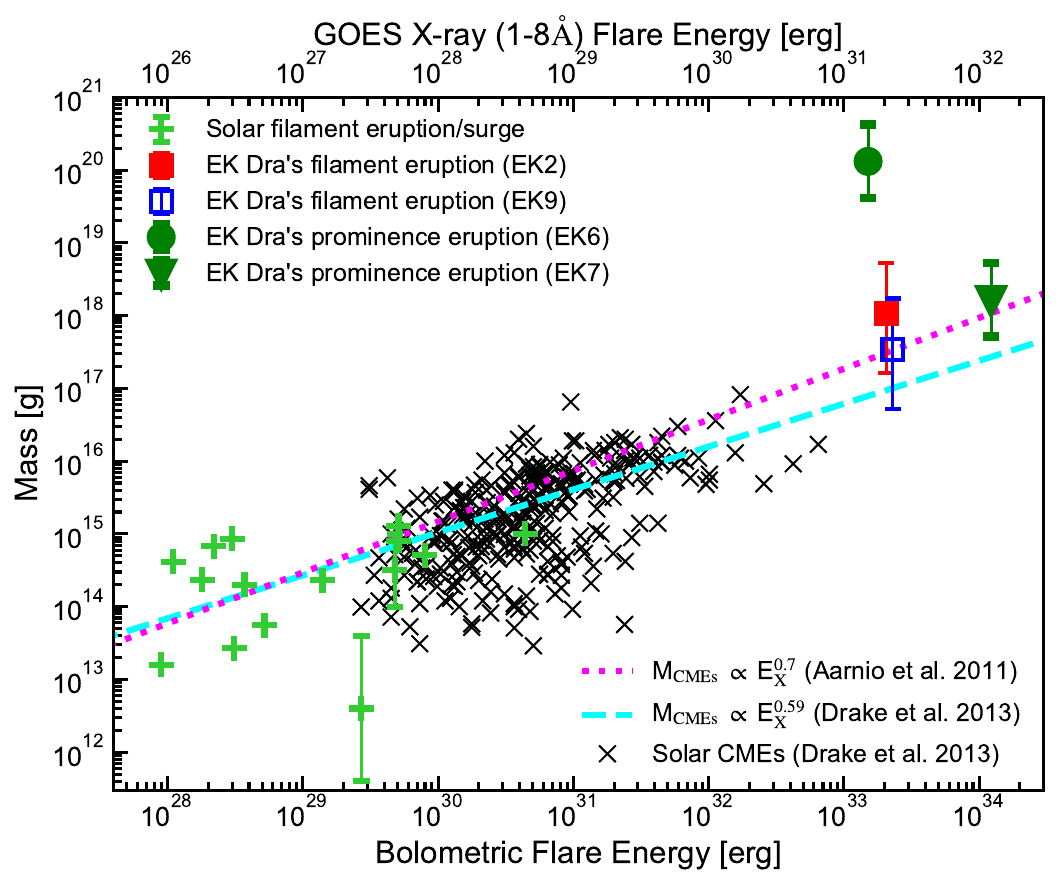}{0.5\textwidth}{\vspace{0mm} (a) }
\fig{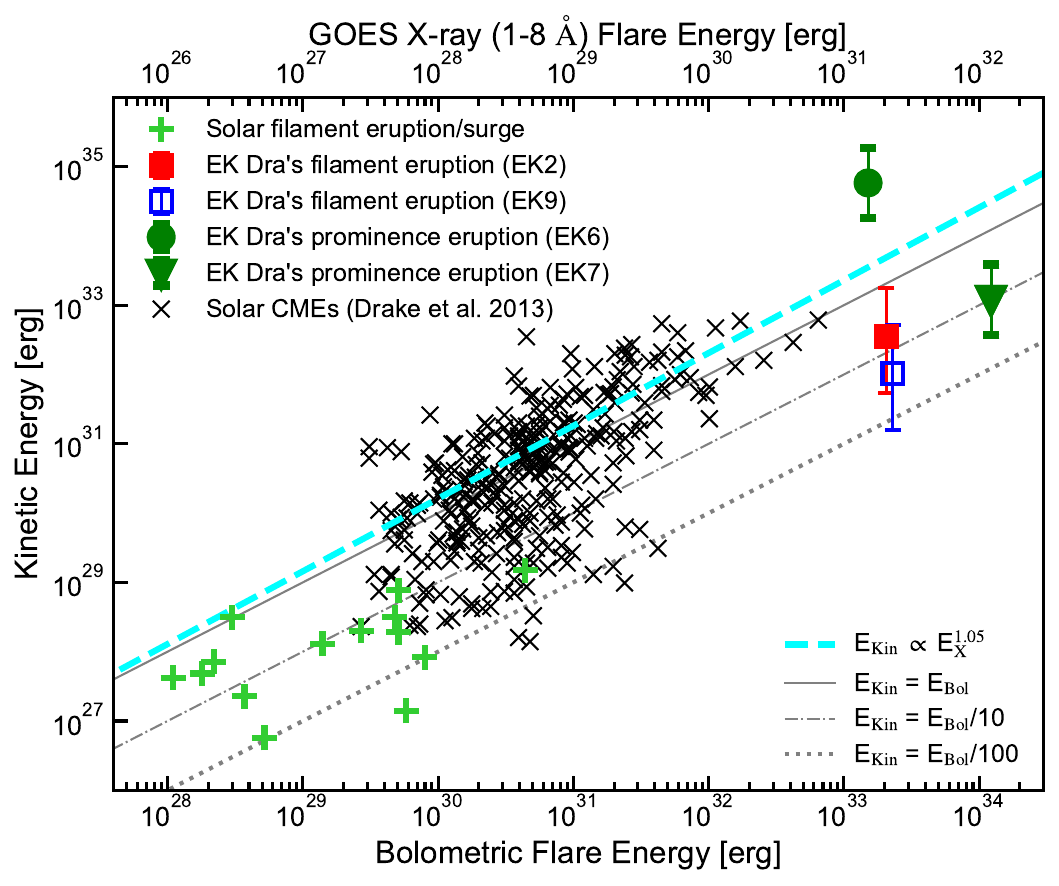}{0.5\textwidth}{\vspace{0mm} (b) }
}
\caption{
Mass and kinetic energy of solar and stellar filament/prominence eruptions and CMEs as a function of bolometric WLF energy or GOES X-ray energy.
\Add{The dark green symbols (circle and downward-pointing triangles) and red/blue squares} show stellar prominence and filament eruptions on EK Dra, \Add{respectively}.
Black crosses represent solar CME data \citep[][]{2009IAUS..257..233Y,2013ApJ...764..170D}, and \Add{lime green} pluses show solar filament/prominence eruptions \citep[see,][]{2022NatAs...6..241N,2023ApJ...943..143K}.
\Add{The large error bars on the two lime green data points in panel (a) are taken from the original sources by \citet{1987JApA....8..295J} and \citet{2015ApJ...804..147C}, whereas error estimates for the other data points were not provided in the publications \citep{2004ApJ...610.1136L,1999spro.proc..367O,2022NatAs...6..241N,2023ApJ...943..143K}; therefore, these two error bars do not necessarily represent the typical uncertainties. Note also that the data in panels (a) and (b) are not in a one-to-one correspondence.}
The cyan dashed and magenta dotted lines are observational fits for solar CMEs:
$M_{\rm CME} \propto E_{\rm bol}^{0.59}$ and $E_{\rm kin} \propto E_{\rm bol}^{1.05}$ (cyan; \citealt{2013ApJ...764..170D}),
and $M_{\rm CME} \propto E_{\rm bol}^{0.7}$ (magenta; \citealt{2012ApJ...760....9A}).
Gray lines in the right panel indicate $E_{\rm kin} = E_{\rm bol}$, $E_{\rm bol}/10$, and $E_{\rm bol}/100$.
GOES X-ray energy/flux is converted to bolometric energy using $E_{\rm bol} = 100,E_{\rm GOES}$ \citep{2012ApJ...759...71E} and
$E_{\rm bol},[\mathrm{erg}] = 10^{35},F_{\rm GOES},[\mathrm{W,m^{-2}}]$ \citep{2013PASJ...65...49S}.
Bolometric energy is assumed to be approximately equal to bolometric WLF energy \citep{2010NatPh...6..690K,2011A&A...530A..84K,2015ApJ...809...79O}.
}
\label{fig:stat-mass}
\end{figure*}

Here, we review the statistical results for the length scale, velocity, mass, and kinetic energy of stellar filament/prominence eruptions. 
As described in \Add{Section~\ref{sec:dis:ke9}}, we added one new event in this paper, bringing the total to four events including those reported in previous studies. 
The physical parameters of the newly identified event are broadly consistent with those of earlier events.
In this regard, the statistical properties of these physical parameters remain largely unchanged from those reported by \citet{2024ApJ...961...23N}.
Here, we briefly summarize the conclusion in \citet{2024ApJ...961...23N}.

The parameters of all events are listed in Table \ref{tab:summary-of-prominence}.
Overall, the velocities, ranging from 330 to 690 km s$^{-1}$, are higher than those of typical solar filament/prominence eruptions associated with CMEs, which are usually a few hundred km s$^{-1}$ \citep{2003ApJ...586..562G,2021EP&S...73...58S}.
Note that solar velocities are often measured as the top speed of the erupting structure, while the stellar values presented here are mean line-of-sight velocities and should be considered lower limits. 
Therefore, the actual difference between solar and stellar cases is likely much greater than the above comparison suggests.
The velocity dispersion, interpreted as filament expansion \citep{2024ApJ...961...23N}, implies top speeds of 660 to 1080 km s$^{-1}$ for these events. This exceeds the escape velocity of about 670 km s$^{-1}$, suggesting that at least part of the filament/prominence material has escaped as a CME.
Furthermore, \cite{2003ApJ...586..562G} showed that solar CME velocities are typically 2--4 times higher than those of associated filament/prominence eruptions due to self-similar expansion.
Given this, all of the eruptions detected here are likely associated with successful CMEs that exceeded the escape velocity.

Figure \ref{fig:stat-length} shows a comparison between the velocity and length scale of filament/prominence eruptions on the Sun and EK Dra. \citet{2021EP&S...73...58S} proposed a threshold separating solar filament eruptions with and without CMEs (a black line in Figure \ref{fig:stat-length}). 
This solar standard has been applied to solar-type stars by \citet{2022NatAs...6..241N} and \citet{2024ApJ...961...23N}, suggesting that stellar filament/prominence eruptions, including EK9, are likely associated with stellar CMEs.

Figure \ref{fig:stat-velosity} shows filament/prominence eruption velocities as a function of flare energy for both solar and stellar flares, not limited to solar-type stars. A weak positive correlation is seen between the two parameters, although the correlation is not strong. This trend is also observed in solar CME-flare studies. \citet{2016ApJ...833L...8T} proposed that the positive correlation observed only for the upper limit of CME velocity at a given flare energy can be explained by physical constraints.
Building on this idea, \citet{2024ApJ...961...23N} extended the theoretical power-law relation, $V_{\rm upper} \propto E^{1/6}$, to stellar filament/prominence eruptions and found that the upper envelope of the velocity distribution can be described by this scaling.
It is worth noting that the weak correlation may be partly due to line-of-sight uncertainty in stellar observations. Even so, this framework helps place stellar CME and eruption velocities within the broader context of the solar paradigm.

Figure \ref{fig:stat-mass} shows the mass and kinetic energy of filament/prominence eruptions as a function of flare energy. As suggested by several studies \citep[e.g.,][]{2019ApJ...877..105M}, the masses of stellar filaments are generally consistent with extrapolations from solar scaling relations, with the exception of the EK6 event. This consistency may indicate a common physical mechanism underlying solar and stellar CMEs/eruptions, as theoretically proposed by \citet{2016ApJ...833L...8T} and \citet{2023ApJ...943..143K}.
Given the large scatter of typically two to three orders of magnitude seen in solar data, a similar level of variation in stellar data, including the EK2 event, is plausible.
In contrast, the kinetic energies of the stellar eruptions appear somewhat lower than expected from solar extrapolations, generally following a relation of $E_{\rm Kin} \sim E_{\rm Bol}/10$, again with the exception of EK2. One possible explanation is the magnetic suppression of CMEs in active stars \citep[][]{2018ApJ...862...93A}. 
Another explanation involves the difference between filament/prominence eruptions and overall CMEs \citep{2024ApJ...961...23N}. 
As mentioned above, on the Sun, the velocities of filament/prominence eruptions are much smaller than CME velocities \citep{2003ApJ...586..562G}. 
Therefore, the apparently small velocities of stellar filaments may result from such a comparison mismatch. 
Indeed, solar filament eruptions and surges fall below the CME-based scaling relation, supporting this second scenario.

\subsection{Frequency of Super-CMEs from young solar-type stars}\label{sec:dis:frequency}

\begin{figure}
\epsscale{1.0}
\plotone{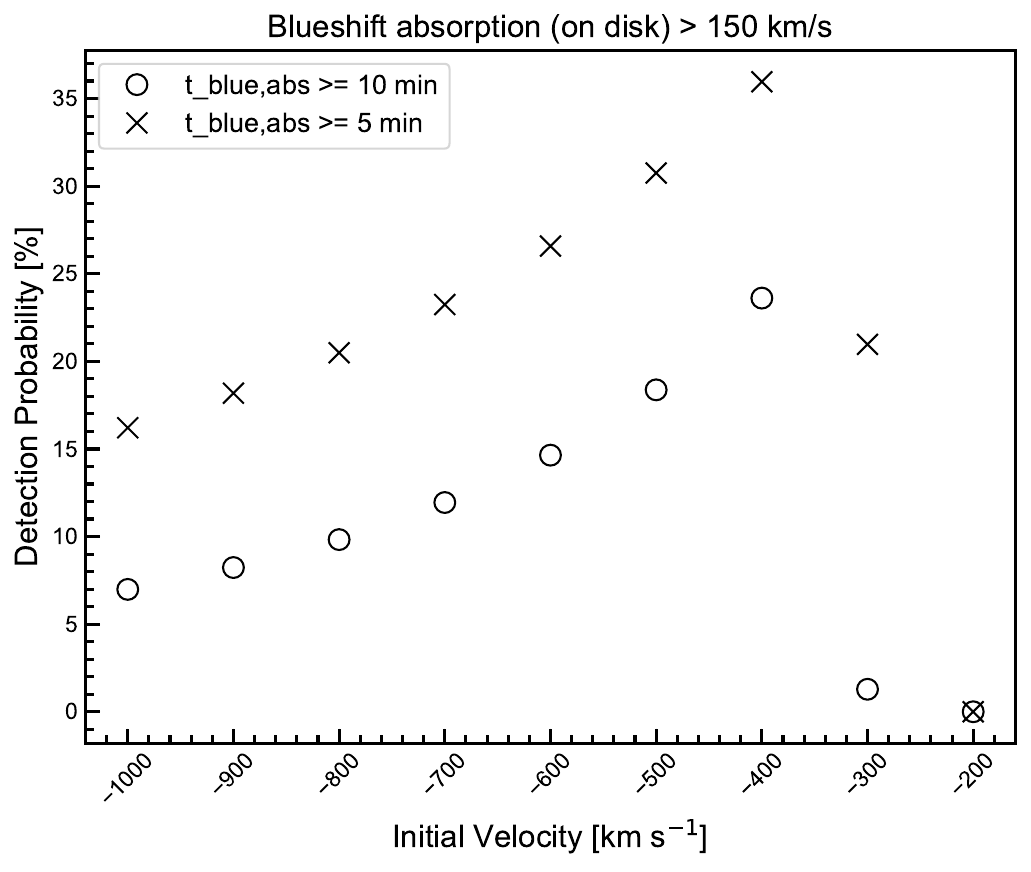}
\caption{
Detection probability of blueshifted absorption as a function of initial velocity of erupted filament. \Add{The assumptions we made are described in Section \ref{sec:dis:frequency}. The crosses and circles indicate the probabilities that the events can be observed in absorption for more than 5 and 10 minutes, respectively.}
}
\label{fig:detectionprob}
\end{figure}

Table \ref{tab:summary-of-prominence} summarizes the filament/prominence eruptions detected to date on young solar-type stars.
As described in Section \ref{sec:4-1}, we identified four fast and massive filament/prominence eruptions among 15 superflares with energies $\gtrsim 10^{33}$ erg.
Based on this, the lower limit of the observable filament/prominence eruption association rate with superflares is approximately 27\%.
According to the Wilson score interval (\texttt{python} package \texttt{statsmodels.stats.proportion.proportion\_confint}), the error range can be estimated to be (11--52)\%.
Previously, the estimated association rate was 60\% with only five samples \citep{2024ApJ...961...23N}.
In this study, by increasing the sample size to 15, we could derive more reliable value, which is a major contribution of our work\footnote{Here, we did not include the four events reported in \citet{2024MNRAS.532.1486L}, as our aim is to ensure statistical consistency based on homogeneous data and their low time cadence may be insufficient to capture rapid, short-duration blueshifted events.}.
Although no eruptions were detected among the three flares observed on V889 Her, this is not necessarily surprising given the 27$_{-16}^{+25}$\% association rate and does not imply that V889 Her is a non/less-eruptive star.
Regarding the CME, solar–stellar comparison and numerical models suggest that such massive filament/prominence eruptions are probably associated with super-CMEs \citep{2022NatAs...6..241N,2024ApJ...961...23N,2024ApJ...963...50I}, therefore,
this 27$_{-16}^{+25}$\% can also be interpreted as a super-CME occurrence rate associated with superflares ($\gtrsim 10^{33}$ erg).
Note that this represents just a lower limit because eruptions directed perpendicular to the line of sight or those with low velocities may not be detectable\footnote{Indeed, \cite{namekata2025natas} discovered much slower, smaller H$\alpha$ eruptions from a 10$^{32}$-erg flare on EK Dra with high-spectral-resolution spectroscopy with R$\sim$30,000. It is likely that such small-scale and low-velocity eruptions are indeed occurring; however, whether they lead to CMEs or influence the CME occurrence rate remains a subject for further investigation.}.
Considering these limitations, the actual association rate is likely higher.
Also, the definition of flares introduced in Section \ref{sec:3-2} may also affect the estimation of the rate.


\Add{To evaluate the likelihood of detecting blueshifted absorption following stellar eruptions, we computed the trajectories of freely falling plasma parcels under the influence of stellar gravity.
The probability was calculated under the assumptions that (1) flares/eruptions occur uniformly across the visible hemisphere, (2) eruptions are launched perpendicular to the stellar surface, and (3) the eruptions originate at the height of 0.01$R_{\rm star}$ from the stellar surface.
Assuming initial radial ejection velocities in the range of $-1000$ to $-200$ km s$^{-1}$, we numerically integrated the one-dimensional equations of motion using the Euler method with a time step of 0.01 s.
The stellar parameters were set to $M_{\rm star} = 0.95 M_\odot$ and $R_{\rm star} = 0.94 R_\odot$.
For each initial velocity and angle from the disk center, we calculated the line-of-sight (LoS) velocity as a function of time.
We then determined the time period during which the eruption remains projected against the stellar disk (i.e., visible in absorption) and the LoS velocity exceeds $-150$ km s$^{-1}$ (approximately the velocity resolution of the spectrograph).
Based on this, we estimated the probability that blueshifted absorption can be detected for more than 5 and 10 minutes after the eruption onset.
Figure \ref{fig:detectionprob} shows the resulting detection probabilities as a function of the initial velocity.
Broadly speaking, the probability of detecting blueshifted absorption is 10–30\%, which is consistent with our observational result of 13$_{-10}^{+25}$\% (2 detections out of 15 events).
These results indicate that the low detection rate of blueshifted events does not necessarily imply a low occurrence rate of eruptions, but can instead be reasonably explained by projection effects.
This suggests that eruptions likely accompany a larger fraction of flare events than are actually observed in blueshifted absorption.}

On the Sun, nearly 100\% of flares above the X class ($\sim10^{31}$ erg) are associated with CMEs, although the association rate decreases for M-, C-, and B-class flares \citep{2009IAUS..257..233Y}. 
In active stars such as EK Dra and V889 Her, the threshold flare energy above which CME association becomes nearly 100\% is likely higher than the solar value ($\sim10^{31}$ erg), owing to stronger magnetic suppression. \citet{2018ApJ...862...93A} numerically modeled stellar CMEs and demonstrated that large-scale magnetic field configurations can establish a suppression threshold around $\sim3\times10^{32}$ erg in CME kinetic energy. 
Similarly, \citet{2024MNRAS.533.1156S} modeled CMEs on the young K-dwarf AB Dor, finding a threshold in the range of $10^{31}$–$10^{33}$ erg. 
The CME association rate inferred for superflares ($\gtrsim10^{33}$ erg) on young solar-type stars in this study is consistent with these numerical predictions.

We also mention that the association rate is slightly higher than that observed in M-dwarf flares, which is $\sim$15\% \citep[10 eruption candidates out of 68 flares;][]{Notsu2023,2025ApJ...979...93K}.
This difference may result from difference in magnetic topology, or from differences in the visibility of prominences/filaments between G- and M-dwarfs, as discussed by \cite{2024ApJ...961...23N,2025ApJ...979...93K}.
Future modeling studies on the radiation processes of prominence/filaments in G/M dwarfs will also be important for further discussion \citep[cf.][]{2022MNRAS.513.6058L}.

The lower limit of the occurrence frequency of super-CMEs or filament/prominence eruptions can be directly estimated from observations.
Averaging over EK Dra and V889 Her, we estimate a super-CME frequency of 0.18 events per day, based on 4 events observed over a combined exposure time of 22.17 days.
However, the rotation period of EK Dra (2.77 days) is significantly longer than that of V889 Her (1.34 days), and as a result the flare frequency differs by a factor of 2–3 between the two stars (see Table \ref{tab:freq} and Appendix \ref{app:tess})\footnote{Kepler statistical data suggest that there may be no significant difference in flare frequency among stars with rotation periods of 1-3 days \citep{2020arXiv201102117O}, although the saturation regime at periods shorter than a few days still shows considerable scatter. Here, we base our discussion solely on statistical data from TESS observations and limit our discussion to the cases of EK Dra and V889 Her.}.
Therefore, it is more appropriate to estimate the frequencies separately for each star.
As a result, the super-CME frequency for EK Dra is 0.21$_{\pm0.12}$ events per day, while the upper limit for V889 Her is $<$0.32$^{+0.46}_{-0.32}$ events per day.
Although the sample size remains limited, we can calibrate flare rates using long-term TESS monitoring.
The white-light flare frequency ($E_{\rm WLF} \ge 10^{33}$ erg) is 0.50$_{\pm0.04}$ and 1.30$_{\pm0.13}$ events per day for EK Dra and V889 Her, respectively.
Applying the 27\% association rate, we estimate super-CME frequencies of 0.17$_{\pm0.01}$ and 0.43$_{\pm0.04}$ events per day for EK Dra and V889 Her, respectively.
These indirect estimates are broadly consistent with the direct observational values above.

Here, we directly estimated the super-CME frequency on young solar-type stars aged 30–100 Myr, providing a valuable reference for studies of the young Sun.
This information can help assess the cumulative impact of the young Sun’s CMEs on the chemistry and escape of the early Solar System planets and young exoplanets \citep[e.g.,][]{2016NatGe...9..452A,2018Icar..315..146J,2022MNRAS.509.5858H,2023Life...13.1103K}.

\begin{deluxetable*}{cccccccccc}
\tablecaption{Frequencies of flares and eruptions.}
\tablewidth{0pt}
\tablehead{
\colhead{} & unit & \colhead{\textsf{EK Dra}}  & \colhead{\textsf{V889 Her}}  & \colhead{Note} 
}
\startdata
Net H$\alpha$ Exposure Time & [d] & 19.09 & 3.08 & (1) \\
\hline
H$\alpha$ flare ($\ge 10^{31}$ erg) & [d$^{-1}$] & 0.62$_{\pm 0.19}$ & 0.97$_{\pm 0.65}$ & (1) \\
White-Light Flares ($>10^{33}$ erg) & [d$^{-1}$] & 0.50$_{\pm 0.04}$ & 1.30$_{\pm 0.13}$ & (2) \\
(XUV Flares; $3\times10^{33-34}$ erg) & [d$^{-1}$] & (4.24) & (--) & (3) \\
\hline
Eruptions\_from\_H$\alpha$\_Blueshift & [d$^{-1}$] & 0.21$_{\pm 0.12}$ & $<$0.32$^{+0.46}_{-0.32}$ & (4) \\
Eruptions\_from\_WLFs & [d$^{-1}$] & 0.17$_{\pm 0.01}$ & 0.43$_{\pm 0.04}$ & (5) \\
\hline
$\dot{M}$\_from\_Super-CMEs & [$M_{\rm \odot}$ yr$^{-1}$] & $4\times$($10^{-13}$-$10^{-12}$) & -- & (6) \\
$\dot{M}$ from \citealt{2015ApJ...809...79O} & [$M_{\rm \odot}$ yr$^{-1}$] & $9\times$($10^{-13}$-$10^{-12}$) & -- & (3) \\
\enddata
\tablecomments{
References: (1) The net exposure time (not the number of observing nights) by the Seimei telescope  include the data with an error in the H$\alpha$ EW less than 0.2 {\AA}. This is almost equivalent to the data that has been plotted. 
(2) From two minutes cadence data (except for Sector 50).
(3) From \cite{2015ApJ...809...79O}.
(4) The number of H$\alpha$ blueshift events divided by the net H$\alpha$ observation time. For V889 Her, we calculated the upper limit.
(5) TESS flare frequency times CME association rates. For V889 Her, we calculated the upper limit.
(6) Total prominence masses divided by the net H$\alpha$ observation time.
The error bar of each frequency is derived by $\sqrt{N+1}/T_{\rm obs}$. For the mass loss rate, the error range reflects only the uncertainty in the estimated eruptive masses which are the dominant sources of errors.
}
\label{tab:freq}
\end{deluxetable*}


\subsection{CME-Driven Mass-loss Rate of young solar-type stars}\label{sec:dis:massloss}

Here we estimate the lower limit of super-CME mass-loss rate based on observations of stellar filament/prominence eruptions.
For this purpose, we assumed that super-CMEs carry masses comparable with filament/prominence eruptions.
Solar observations show that the CME mass-flare energy relationship is similar to the filament mass-flare energy relationship \citep{2023ApJ...943..143K,2024ApJ...961...23N}, supporting this assumption.
By combining the estimated plasma mass (Section~\ref{sec:dis:mass}) and occurrence frequency (Section~\ref{sec:dis:frequency}), we derive a lower-limit estimate of the super-CME mass-loss rate for EK Dra as $4\times(10^{-13}$–$10^{-12})$ $M_{\odot}$ yr$^{-1}$.
Note that the quoted error range reflects only the uncertainty in the estimated eruptive masses which are the dominant sources of errors.
This provides the first direct constraint on CME-driven mass loss for young solar-type stars.
It is important to note that this value represents a lower limit as discussed in Section \ref{sec:dis:frequency}. 
Also, it is based on a relatively short total exposure time of $\sim$20 days, and thus likely reflects a short-term average. 
To estimate the general mass-loss rate, further long-term averaging may be required. 

Previously, \cite{2015ApJ...809...79O} estimated the CME mass-loss rate of EK Dra as $9\times(10^{-13}$–$10^{-12})$ $M_{\odot}$ yr$^{-1}$ by combining the X-ray flare frequency of 4.2 events per day for the energy range of $3\times10^{33}$--$3\times10^{34}$ erg \citep{1999ApJ...513L..53A} and solar CME mass-flare energy relationship \citep{2012ApJ...760....9A,2013ApJ...764..170D}.
Our direct estimate is consistent with the indirect estimate by \citet{2015ApJ...809...79O}. 
However, we need to be careful when interpreting this consistency. 
First, \citet{2022ApJ...926L...5N} reported that the flare occurrence frequency of EK~Dra observed with TESS optical photometry in 2019–2020 was an order of magnitude lower than that observed in X-rays in 1995 \citep{1999ApJ...513L..53A}, suggesting a possible long-term change in stellar activity (cf. a spot cycle period of EK Dra to $\sim$9 yrs with a continuing long-term fading trend; \citealt{2018AandA...620A.162J}).
If we estimate the mass loss rate using the TESS (or H$\alpha$) flare frequency following the method of \citet{2015ApJ...809...79O}, the resulting value is an order of magnitude smaller.
In contrast, the average filament or prominence mass in our sample is an order of magnitude larger than predicted by the solar CME mass–flare energy relationship.
As a result, the increase in mass compensates for the decrease in frequency, yielding a mass loss rate consistent with that of \citet{2015ApJ...809...79O}.
This highlights the potential importance of accounting for stellar activity cycle variations and refining the CME mass-flare energy relationship over a broader dynamic range, in order to accurately characterize CME-driven mass loss.

Stellar (ambient) wind mass loss has often been compared with CME-driven mass loss in terms of their respective contributions to stellar mass and angular momentum evolution, as well as their relative impact on planetary atmospheres and magnetospheres. Observationally, stellar wind mass-loss rates have been estimated primarily via astrospheric Ly$\alpha$ absorption \citep[e.g.,][]{2005ApJ...628L.143W}, and more recently through observations of slingshot prominences \citep[e.g.,][]{2020MNRAS.491.4076J}.
\citet{2021LRSP...18....3V} derived an empirical scaling relation for wind mass-loss rate as a function of stellar coronal X-ray flux, $\dot{M} \propto F_{\rm X}^{0.66}$. Based on this relation, the estimated mass-loss rate for EK Dra is $\sim 10^{-12}$ $M_{\odot}$ yr$^{-1}$. Numerical simulations by \citet{2011ApJ...741...54C} and \citet{2023ApJ...957...71S} also suggest wind mass-loss rates on the order of $10^{-12}$ $M_{\odot}$ yr$^{-1}$ at the age of EK Dra ($\sim$100 Myr or $P_{\rm rot}\sim$3 days). 
Although numerical models still have many uncertainties in the input parameters and further development is needed, these simulations broadly agree with observational estimates.
With these background in mind, we found that our estimate of the lower limit of CME mass-loss rate is comparable to the ambient wind mass-loss rate. 
As discussed above, our estimates represents only a lower limit, and the actual CME mass loss can be significantly higher than wind mass loss. 
\citet{2015ApJ...809...79O} showed that the CME mass-loss rate could increase by up to an order of magnitude when unobservable lower-energy flares ($<10^{33}$ erg) are included, and empirical models provided by \citet{2017ApJ...840..114C} also suggested much higher CME mass-loss rates.
The high CME mass-loss rates estimated in these studies, being supported by our direct observations, highlight the potentially dominant role of CME-driven mass loss over steady stellar winds in the evolution of young solar-type stars and their impact on surrounding exoplanets at an age of around 100 Myr.


\section{Summary and Conclusion}

We conducted a five-year (132-night) dedicated observational campaign targeting two nearby young solar-type stars, EK Dra (age of 50–125 Myr) and V889 Her (age of 30 Myr), using the 3.8m Seimei Telescope and TESS. These young solar-type stars serve as excellent proxies for the infant Sun near the zero-age main sequence, a critical epoch when Earth’s atmosphere transitioned from a primary to a secondary atmosphere.
The ultimate goal of this project is to characterize the radiation and plasma environment around the young Sun. In this context, the specific aim of our campaign is to observationally constrain various aspects of superflares, particularly flare radiation/heating mechanisms and the occurrence of CMEs/eruptions, primarily through high time-cadence H$\alpha$ spectroscopy.
As initial reports, we previously presented five superflares on young solar-type stars \citep{2022NatAs...6..241N,2022ApJ...926L...5N,2024ApJ...961...23N}. In this paper, we report a total of 15 superflares, enabling a statistical investigation.
The main results are summarized as follows.

\begin{enumerate}
    \item \textit{Flare Numbers} (Section \ref{sec:4-1}): We report 12 superflares from EK Dra and 3 from V889 Her. We successfully obtained flaring H$\alpha$ spectra from V889 Her for the first time, which is an important addition, as all previous studies of solar-type stars relied solely on EK Dra ($\sim$50-125 Myr age) and we expanded our sample to the younger ZAMS star ($\sim$30 Myr age). \Add{None of the flares on V889 Her showed any H$\alpha$ line asymmetry.} Two show blueshifted absorption profiles (EK2 and EK9), two show blueshifted emission profiles (EK6 and EK7), and one shows a redshifted emission profile (EK12). Nine flares show significant line broadening.
    
    \item \textit{Interpretation of Line Broadening} (Section \ref{sec:dis:rad}): We found a positive correlation between the peak H$\alpha$ line width and flare luminosity, indicating that larger flares and/or more impulsive phases exhibit stronger Stark and/or opacity broadening. A comparison with Sun-as-a-star observations reveals a similar power-law index of the relation between flare luminosity and line width, suggesting a shared picture of flare evolution. 
    
    \item \textit{Interpretation of Redshift} (Section \ref{sec:dis:ek12}): In the event EK12, we detected one flare that exhibited long-lasting redshifted emission spectra with nearly constant velocity of $\sim100$ km s$^{-1}$. We proposed a dense post-flare loop emission as one of the plausible explanation.
    
    \item \textit{Interpretation of Blueshift} (Section \ref{sec:dis:ke9}): The H$\alpha$ blueshifted absorptions and emissions are unambiguous evidence of stellar filament and prominence eruptions, respectively \citep{2022NatAs...6..241N,2024ApJ...961...23N}. The newly discovered filament eruption (EK9) exhibits unexpectedly strong deceleration, exceeding the stellar surface gravity, and also shows possible signs of pre-flare filament activation.
    
    \item \textit{Mass, Velocity, and Kinetic Energy of CMEs/Eruptions} (Section \ref{sec:dis:mass}): We newly added one filament eruption (EK9) to the sample; however, the discussions related to mass, velocity, and kinetic energy remain unchanged from those in \citet{2024ApJ...961...23N}.
    (i) The relation between length scale and velocity supports the occurrence of CMEs associated with these filament/prominence eruptions.
    (ii) The upper limit of eruption velocity as a function of flare energy can be explained by the theoretical model proposed by \citet{2016ApJ...833L...8T}.
    (iii) The eruption masses are consistent with solar-based scaling relations, whereas the kinetic energies appear to be smaller than expected from solar extrapolations.
    
    \item \textit{Frequency of CMEs/Eruptions} (Section \ref{sec:dis:frequency}): We derived a lower limit for the super-CME/eruption association rate with superflares as 27$_{-16}^{+25}$\% based on the currently available dataset, combining observations of EK Dra and V889 Her. The lower limit of the super-CME occurrence frequency is estimated to be 0.21$_{\pm 0.12}$ events per day for EK Dra and less than 0.32$^{+0.46}_{-0.32}$ events per day for V889 Her. It should be noted that these estimates correspond to CMEs associated with flares of energy $\gtrsim 10^{33}$ erg.
    
    \item \textit{CME-Driven Mass-Loss Rate} (Section \ref{sec:dis:massloss}): We derived the first direct estimate of the CME-driven mass-loss rate for EK Dra as $4 \times (10^{-13}$–$10^{-12})$ $M_{\odot}$ yr$^{-1}$. This estimate is consistent with the previous empirical value reported by \citet{2015ApJ...809...79O}. Moreover, the derived rate is of the same order of magnitude as the wind-driven mass-loss rates estimated observationally \citep{2021LRSP...18....3V} and numerically \citep{2011ApJ...741...54C,2023ApJ...957...71S} for stars with similar age and rotation rate to EK Dra. 
\end{enumerate}

These findings provide valuable observational constraints on the mass and angular momentum evolution of the infant Sun and young solar-type stars, particularly during evolutionary stages that remain poorly understood \citep[e.g.,][]{2011ApJ...741...54C,2017ApJ...840..114C,2021LRSP...18....3V,2021ApJ...907...89H,2023ApJ...957...71S}. 
In addition, the obtained frequency information is valuable for evaluating the impact of CMEs from the young Sun on young rocky planets, including the early Earth, Mars, and Venus \citep[e.g.,][]{2016NatGe...9..452A,2023Life...13.1103K}.

Lastly, we summarize the future direction of this study.
First, to better understand the evolutionary history of the Sun, it is essential to extend this study to a broader range of stellar ages and rotation rates, beyond the current focus on stars aged 30–125 Myr with rotation periods of 1.34–2.77 days.
Although constraining older stars is challenging due to the significant decline in flare occurrence rates, extending this work to much younger, pre-main-sequence stars is a promising future direction. 
Second, high-resolution spectroscopic observations represent an important next step. The current spectral resolution ($R \sim 2000$) may mask slow plasma motions, limiting our ability to detect detailed kinematic signatures.
To address this, we are developing a higher-resolution spectrograph ($R \sim 17{,}000$), the Medium Dispersion Spectrograph for Stellar Activity Research (MIDSSAR) on the 3.8m Seimei Telescope at Okayama Observatory, which covers the H$\alpha$ and Ca II H\&K lines. 
Finally, we plan to extend this kind of CME study to the X-ray and ultraviolet wavelengths using observatories such as XMM-Newton, XRISM, and the Hubble Space Telescope, as well as future UV missions including ESCAPE \citep{2019SPIE11118E..08F} and LAPYUTA \citep{2024SPIE13093E..0IT}. Expanding the sample across multiple wavelengths and obtaining coordinated multi-wavelength datasets will help to reveal a more comprehensive picture of stellar astrospheres.

\section*{Acknowledgment}

\Add{We thank the referee for their careful review, which has helped improve the quality of our manuscript.}
This work was supported by JSPS (Japan Society for the Promotion of Science) KAKENHI Grant Numbers 21J00316 (K.N.), 25K01041 (K.N., H.M.), 24K17082 (K.I.), 24K00685 (H.M.), 21H01131 (H.M., K.I., and K.S.), 24H00248 (K.N., K.I., H.M., S.H., D.N.), and 24K00680 (K.N., H.M., S.H., D.N., K.S.). 
This work was supported by the Operation Management Laboratory (OML) of the National Institutes of Natural Sciences (NINS), Japan (K.N., Y.N. and H.M.). 
Y.N. acknowledge the support from NASA ADAP award Number 80NSSC21K0632,  NSF/AGS Award Number 1916509, NASA TESS Cycle 6 Program 80NSSC24K0493, and NASA TESS Cycle 7 Program 80NSSC25K7906
The spectroscopic data used in this paper were obtained through the program 20A-N-CN08, 21A-N-CN02, 21B-N-CN09, 22A-N-CN06, 22B-N-CN11, 23A-N-CN07 and 24A-N-CT08 (PI: K.N.) with the 3.8m Seimei telescope, which is located at Okayama Observatory of Kyoto University.
We thank Dr. Takeo Minezaki for kindly arranging an exchange of observation time with the Seimei telescope.
This paper includes data collected with the TESS mission, obtained from the MAST data archive at the Space Telescope Science Institute (STScI). Funding for the TESS mission is provided by the NASA Explorer Program. STScI is operated by the Association of Universities for Research in Astronomy, Inc., under NASA contract NAS 5-26555. 
Some of the data presented in this paper were obtained from the Mikulski Archive for Space Telescopes (MAST) at the Space Telescope Science Institute. The specific observations analyzed can be accessed via \dataset[10.17909/yf9r-br37]{https://doi.org/10.17909/yf9r-br37}.
The authors acknowledge ideas from the participants in the workshop ``Blazing Paths to Observing Stellar and Exoplanet Particle Environments" organized by the W.M. Keck Institute for Space Studies.
The authors also would like to acknowledge the the relevant discussions in the International Space Science Institute (ISSI)
Workshop ``Stellar Magnetism and its Impact on (Exo)Planets (\url{https://workshops.issibern.ch/stellar-magnetism/})".
We thank Dr. Munehito Shoda and Prof. Vladimir Airapetian for discussion on the ambient stellar wind mass loss.

\facilities{3.8 m Seimei telescope, TESS}

\software{\texttt{astropy} \citep{2018AJ....156..123A} , \texttt{IRAF} \citep{Tody1986}, \texttt{PyRAF} \citep{2012ascl.soft07011S}}







\appendix

\section{Observational Log}\label{app:obslog}

Table \ref{tab:obslog:tess} summarizes the TESS observational log, and Tables \ref{tab:obslog:1} and \ref{tab:obslog:2} summarize the observational log of the Seimei Telescope.

\section{Removal of Trend of Water Vapor Lines}\label{app:watervapor}

Regarding the event of V889 Her on 25 July 2022, due to the high humidity of the Japanese summer, the prominent water vapor absorption lines was observed as the elevation decreased (i.e., air mass increased). 
The background trend was completely the same as that observed on 26 July 2022, indicating that the variation comes not from intrinsic stellar variability (e.g., rotational modulation) but from telluric lines.
This significantly polluted the spectral region near the H$\alpha$ line. 
We corrected the intensity changes of the water vapor absorption caused by the increased air mass using the absorption line intensities from the same time on the following day (26 July 2022 ). 
As a result, the light curve and dynamic spectrum were greatly improved. 
However, due to the correction of the water vapor absorption lines, we should be careful regarding the asymmetry in the latter part of the flare (when the air mass became larger).

\section{TESS flare frequency}\label{app:tess}

\begin{figure*}
\gridline{
\fig{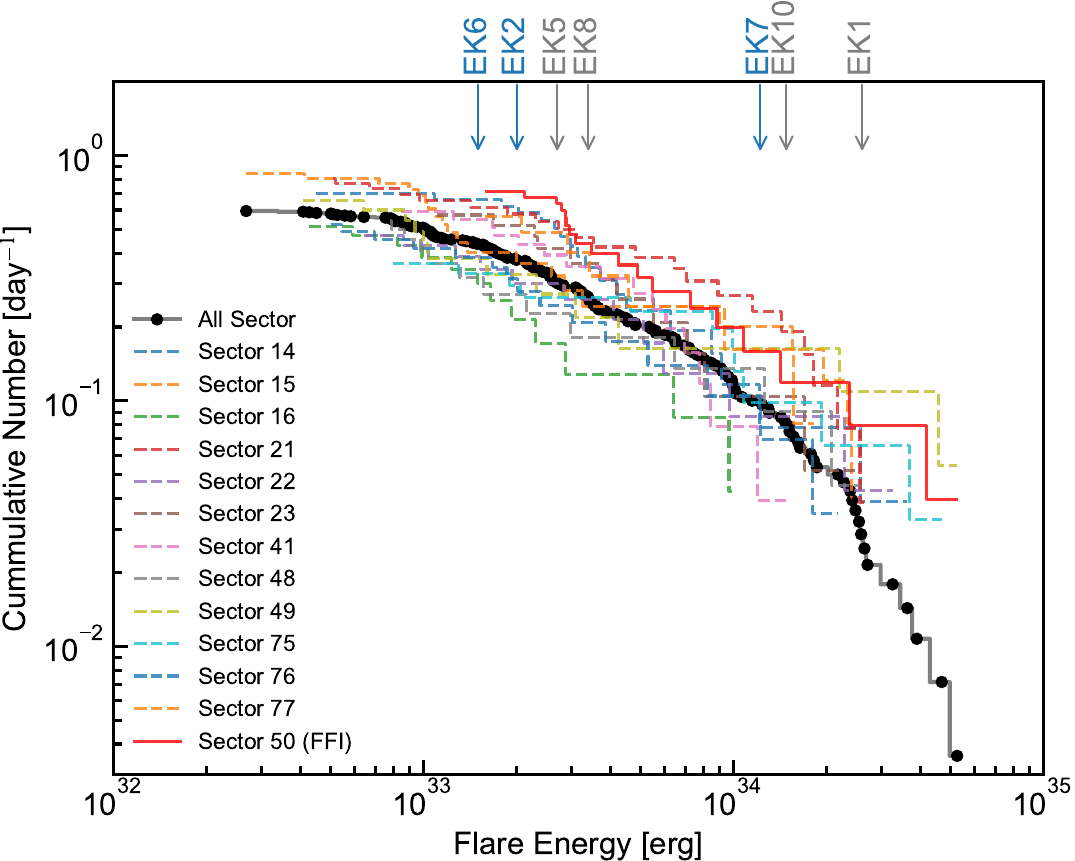 }{0.5\textwidth}{\vspace{0mm} (a) EK Dra}
\fig{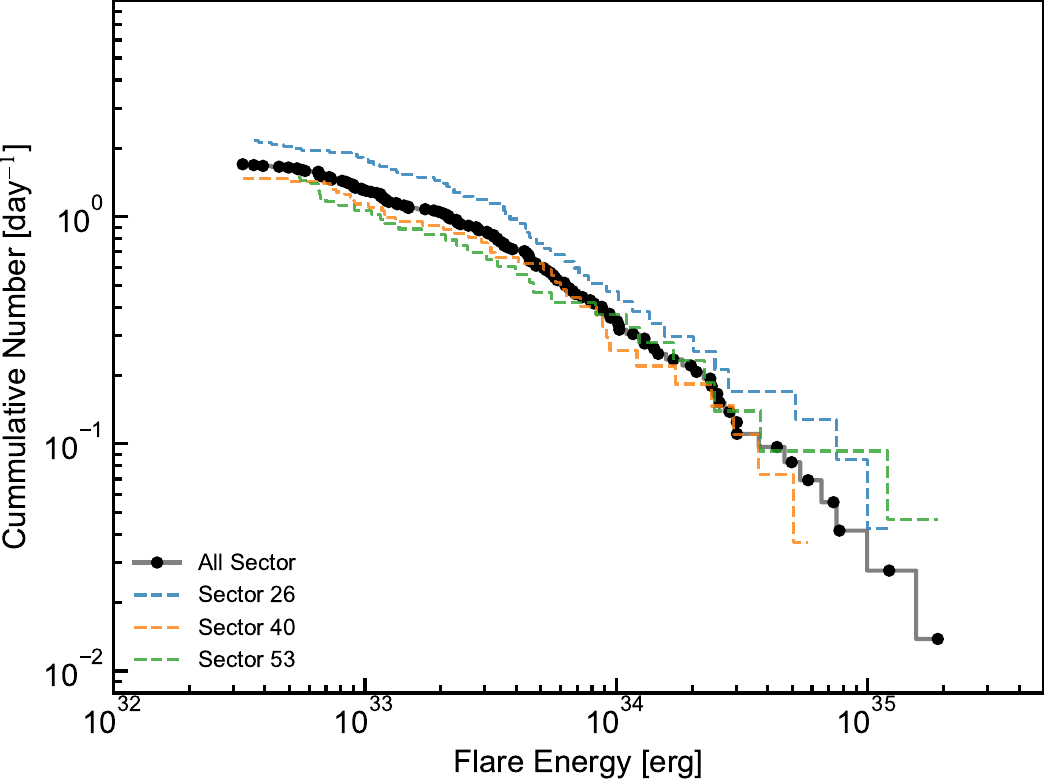 }{0.5\textwidth}{\vspace{0mm} (b) V889 Her}
}
\caption{Cumulative flare frequency as a function of flare energy for EK Dra and V889 Her. Data from all TESS sectors are shown in black, while individual sectors are shown with colored lines.}
\label{fig:ffd}
\end{figure*}

Figure \ref{fig:ffd} shows the flare frequency distributions for EK Dra and V889 Her. 
The flare energies detected simultaneously with H$\alpha$ are indicated with arrows, demonstrating that our sample covers well the energy range of flares detectable by TESS.
It also shows that V889 Her produces more frequent and energetic superflares than EK Dra.
Table \ref{tab:obslog:tess} lists the number of detected flares and the total flare energy radiated per total exposure time.

\section{Calculation of the mass, kinetic energy, and length scale of filament eruption.}\label{app:calcmass}

In our study, we estimated the mass and kinetic energy using the methodology by \cite{2022NatAs...6..241N}.
The mass was determined by calculating the average equivalent width (abs($\Delta$EW)=$0.07  \pm 0.02$ {\AA}) at the time of maximum EW ($t = 31.9 \sim 38.2$ min) and the corresponding velocity at that time ($V_{\rm typical} = 247\pm71$ km s$^{-1}$). 
The error range of the equivalent width and velocity was determined based on the scattering observed in the data during this period. Furthermore, the EW utilized here was derived from the results of Gaussian fitting.
The error range of masses and length scales primary come from the assumed conservative ranges of plasma parameters (see \citealt{2022NatAs...6..241N}).
The kinetic energy was subsequently derived from the mass and its velocity. It is important to note that the velocity does not necessarily peak when the EW is at its maximum and vice versa. The reason why we need velocity for estimating the mass is that the change in the background intensity due to velocity changes affects the degree of absorption\footnote{This does not imply that we are considering the effects of ``Doppler dimming". Eventually, the influence of Doppler dimming might need to be addressed, but it is not considered here for simplicity.}.


\clearpage

\begin{deluxetable*}{clcccccccc}
\tablecaption{Observational Log of TESS and Their Flare Parameters for EK Dra and V889 Her.}
\tablewidth{0pt}
\tablehead{
\colhead{Sector} & \colhead{Date$^{\dagger}$} & \colhead{$T_{\rm cadence}$} & \colhead{Data-Type} & \colhead{$T_{\rm obs}$} & \colhead{$N_{\rm Flare}$} & \colhead{$E_{\rm Flare,tot}$$^{\S}$} \\
\colhead{} & \colhead{} & \colhead{[sec]} & \colhead{} & \colhead{[d]} & \colhead{} & \colhead{[$10^{33}$ erg d$^{-1}$]} 
}
\startdata
\multicolumn{2}{l}{\textsf{EK Dra (TIC 159613900)}} & \\
\hline
14 & 2019 Jul 18--Aug 14 & 120 & PDC-SAP &  25.7 & 18 & 3.02  \\
15 & 2019 Aug 15--Sep 10 & 120 & PDC-SAP &  24.8 & 21 & 4.10  \\
16 & 2019 Sep 12--Oct 06 & 120 & PDC-SAP &  23.4 & 12 & 0.83  \\
21 & 2020 Jan 21--Feb 18 & 120 & PDC-SAP &  26.0 & 21 & 5.65  \\
22 & 2020 Feb 19--Mar 17 & 120 & PDC-SAP &  23.2 & 11 & 2.47  \\
23 & 2020 Mar 19--Apr 15 & 120 & PDC-SAP &  19.2 & 11 & 2.50  \\
41 & 2021 Jul 23--Aug 20 & 20,120$^{\ddag}$ & PDC-SAP &  25.4 & 15 & 2.23  \\
48 & 2022 Jan 28--Feb 26 & 20,120$^{\ddag}$ & PDC-SAP &  22.1 & 12 & 2.31  \\
49 & 2022 Feb 26--Mar 26 & 20,120$^{\ddag}$ & PDC-SAP &  18.3 & 12 & 5.00  \\
50 & 2022 Mar 26--Apr 22 & 600 & FFI-RAW/PCA &  25.1 & 18 & 5.60  \\
75 & 2024 Jan 30--Feb 26 & 20,120$^{\ddag}$ & SAP$^{\ast}$ &  30.4 & 11 & 4.04  \\
76 & 2024 Feb 26--Mar 26 & 20,120$^{\ddag}$ & SAP$^{\ast}$ &  28.6 & 15 & 1.83  \\
77 & 2024 Mar 26--Apr 23 & 20,120$^{\ddag}$ & SAP$^{\ast}$ &  12.4 & 7 & 2.97  \\
\hline
\hline
\multicolumn{2}{l}{\textsf{V889 Her (TIC 471000657)}} & \\
\hline
26 & 2020 Jun 09--Jul 04 & 120 & PDC-SAP & 23.5 & 51 & 19.58 \\
40 & 2020 Jun 09--Jul 04 & 120 & PDC-SAP & 27.2 & 40 & 9.84  \\
53 & 2020 Jun 09--Jul 04 & 120 & PDC-SAP & 21.5 & 32 & 16.30  \\
\enddata
\tablecomments{$^{\dagger}$Obtained from \url{https://heasarc.gsfc.nasa.gov/docs/tess/sector.html}. $^{\ddag}$We used 120 sec cadence mode for holding the consistency with the observations in 2019 and 2020. $^{\S}$ The total radiation energy divided by the total observation time $T_{\rm obs}$ of each sector for the energy of $>$ $5\times 10^{33}$ erg. 
$^{\ast}$For the data in 2024 (sector 75-77), we used SAP data because of the significant observational gaps in PDC-SAP data. 
}
\label{tab:obslog:tess}
\end{deluxetable*}

\begin{deluxetable*}{lccc|lccc|lccc}
\tablecaption{Observational Log of the Seimei Telescope/KOOLS-IFU for EK Dra.}
\tablewidth{0pt}
\tablehead{
\colhead{Date} & \colhead{$N_{\rm exp}$} & \colhead{$T_{\rm exp}$} & \colhead{$T_{\rm ro}$} & \colhead{Date} & \colhead{$N_{\rm exp}$} & \colhead{$T_{\rm exp}$} & \colhead{$T_{\rm ro}$} & \colhead{Date} & \colhead{$N_{\rm exp}$} & \colhead{$T_{\rm exp}$} & \colhead{$T_{\rm ro}$} \\   
\colhead{} & \colhead{} & \colhead{[sec]} & \colhead{[sec]} & \colhead{} & \colhead{} & \colhead{[sec]} & \colhead{[sec]} & \colhead{} & \colhead{} & \colhead{[sec]}  & \colhead{[sec]}
}
\startdata
2020 Feb 14 & 249 & 30 & 21.1 & 2021 Apr 24 & 152 & 60 & 17.3 & 2023 Mar 09 & 242 & 60 & 16.5 \\
2020 Feb 16 & 53 & 30 & 21.1 & 2021 Apr 25 & 194 & 60 & 17.3 & 2023 Mar 10 & 303 & 60 & 16.5 \\
2020 Feb 24 & 128 & 30 & 21.1 & 2021 Apr 26 & 55 & 120 & 17.3 & 2023 Mar 27 & 129 & 60,120 & 16.5 \\
2020 Feb 26 & 470 & 30 & 21.1 & 2021 Apr 30 & 294 & 60 & 17.2 & 2023 Mar 28 & 299 & 60 & 16.5 \\
2020 Feb 27 & 267 & 30 & 21.1 & 2021 May 25 & 42 & 60 & 17.2 & 2023 Mar 29 & 231 & 60 & 16.5 \\
2020 Mar 01 & 181 & 30 & 17.1 & 2022 Jan 29 & 222 & 60 & 16.5 & 2023 Mar 30 & 315 & 60 & 16.5 \\
2020 Mar 02 & 169 & 30 & 17.1 & 2022 Jan 30 & 273 & 60 & 16.5 & 2023 Mar 31 & 204 & 60 & 16.5 \\
2020 Mar 06 & 543 & 30 & 17.3 & 2022 Feb 22 & 396 & 60 & 16.6 & 2023 Apr 01 & 312 & 60 & 16.5 \\
2020 Mar 08 & 398 & 30 & 21.3 & 2022 Feb 23 & 392 & 60 & 16.5 & 2023 Apr 02 & 286 & 60 & 16.5 \\
2020 Mar 10 & 279 & 30 & 21.3 & 2022 Mar 14 & 61 & 60 & 16.5 & 2023 May 01 & 148 & 60 & 16.5 \\
2020 Mar 11 & 508 & 30 & 21.3 & 2022 Mar 16 & 260 & 60 & 16.5 & 2023 May 02 & 158 & 60 & 16.5 \\
2020 Mar 12 & 304 & 30 & 17.3 & 2022 Mar 18 & 100 & 60 & 16.5 & 2023 May 03 & 3 & 60 & 16.5 \\
2020 Mar 14 & 636 & 30 & 21.3 & 2022 Mar 20 & 35 & 60,120 & 16.5 & 2023 May 05 & 97 & 60 & 16.5 \\
2020 Mar 16 & 387 & 30 & 17.3 & 2022 Mar 22 & 440 & 60,120 & 16.5 & 2024 Feb 23 & 160 & 60 & 16.5 \\
2020 Apr 04 & 91 & 30 & 17.3 & 2022 Mar 23 & 250 & 60,120 & 16.5 & 2024 Feb 25 & 311 & 60 & 16.5 \\
2020 Apr 05 & 723 & 20,30 & 17.3 & 2022 Mar 24 & 411 & 60,120 & 16.5 & 2024 Feb 26 & 149 & 60 & 16.5 \\
2020 Apr 06 & 258 & 30 & 17.3 & 2022 Apr 10 & 288 & 60,120 & 16.5 & 2024 Mar 18 & 107 & 60 & 16.5 \\
2020 Apr 07 & 464 & 30 & 17.3 & 2022 Apr 11 & 104 & 60,120 & 16.5 & 2024 Mar 19 & 118 & 60 & 16.5 \\
2020 Apr 08 & 431 & 30 & 17.3 & 2022 Apr 12 & 181 & 60,120 & 16.5 & 2024 Mar 20 & 190 & 60 & 16.5 \\
2020 Apr 09 & 526 & 30 & 17.3 & 2022 Apr 15 & 100 & 60,120 & 16.5 & 2024 Mar 21 & 190 & 60 & 16.5 \\
2020 Apr 10 & 165 & 30 & 17.3 & 2022 Apr 16 & 440 & 60,120 & 16.5 & 2024 Mar 29 & 335 & 60 & 16.5 \\
2020 Apr 13 & 424 & 30 & 17.3 & 2022 Apr 17 & 353 & 60,120 & 16.5 & 2024 Mar 30 & 350 & 60 & 16.5 \\
2021 Mar 30 & 123 & 60 & 17.2 & 2022 Apr 18 & 174 & 60,120 & 16.5 & 2024 Mar 31 & 155 & 60 & 16.5 \\
2021 Mar 31 & 320 & 60 & 17.2 & 2022 Apr 19 & 366 & 60,120 & 16.5 & 2024 Apr 01 & 268 & 60 & 16.5 \\
2021 Apr 01 & 398 & 60 & 17.3 & 2022 Apr 20 & 299 & 60,120,160 & 16.5 & 2024 Apr 04 & 140 & 60 & 16.5 \\
2021 Apr 02 & 94 & 60 & 17.3 & 2023 Mar 02 & 301 & 60 & 16.5 & 2024 Apr 05 & 147 & 60 & 16.5 \\
2021 Apr 03 & 200 & 60 & 17.3 & 2023 Mar 03 & 436 & 60 & 16.5 & 2024 Apr 06 & 28 & 60 & 16.5 \\
2021 Apr 19 & 205 & 60 & 17.3 & 2023 Mar 04 & 261 & 60,120 & 16.5 & 2024 Apr 07 & 48 & 60 & 16.5 \\
2021 Apr 20 & 146 & 60 & 17.3 & 2023 Mar 05 & 440 & 60 & 16.5 & 2024 Apr 17 & 67 & 60 & 16.5 \\
2021 Apr 21 & 410 & 60 & 17.3 & 2023 Mar 06 & 336 & 60 & 16.5 & 2024 Apr 18 & 131 & 60 & 16.5 \\
2021 Apr 22 & 206 & 60 & 17.3 & 2023 Mar 07 & 330 & 30,60 & 16.5 & 2024 Apr 19 & 438 & 60 & 16.5 \\
2021 Apr 23 & 76 & 60 & 17.3 & 2023 Mar 08 & 373 & 60 & 16.5 & 2024 Apr 25 & 261 & 60 & 16.5\\ 
\enddata
\tablecomments{$N_{\rm exp}$ is the number of the exposure. $T_{\rm exp}$ is the exposure time. $T_{\rm ro}$ is the typical readout time.}
\label{tab:obslog:1}
\end{deluxetable*}

\begin{deluxetable*}{lccc|lccc|lccc}
\tablecaption{Observational Log of the Seimei Telescope/KOOLS-IFU for V889 Her.}
\tablewidth{0pt}
\tablehead{
\colhead{Date} & \colhead{$N_{\rm exp}$} & \colhead{$T_{\rm exp}$} & \colhead{$T_{\rm ro}$} & \colhead{Date} & \colhead{$N_{\rm exp}$} & \colhead{$T_{\rm exp}$} & \colhead{$T_{\rm ro}$} & \colhead{Date} & \colhead{$N_{\rm exp}$} & \colhead{$T_{\rm exp}$} & \colhead{$T_{\rm ro}$} \\  
\colhead{} & \colhead{} & \colhead{[sec]} & \colhead{[sec]} & \colhead{} & \colhead{} & \colhead{[sec]} & \colhead{[sec]} & \colhead{} & \colhead{} & \colhead{[sec]}  & \colhead{[sec]}
}
\startdata
2021 Apr 01 & 24 & 60 & 17.2 & 2021 Aug 25 & 235 & 60 & 16.5 & 2023 Mar 27 & 155 & 60 & 16.5 \\
2021 Apr 02 & 24 & 60 & 17.2 & 2021 Sep 15 & 199 & 60 & 16.4 & 2023 Mar 28 & 184 & 60 & 16.5 \\
2021 Apr 20 & 203 & 60 & 17.2 & 2022 Jul 25 & 112 & 60 & 17.3 & 2023 Mar 29 & 199 & 60 & 16.5 \\
2021 Apr 22 & 166 & 60 & 17.2 & 2022 Jul 26 & 139 & 60 & 16.5 & 2023 Mar 30 & 70 & 60 & 16.5 \\
2021 Apr 24 & 116 & 60,120 & 17.3 & 2022 Aug 08 & 27 & 60 & 16.5 & 2023 Mar 31 & 146 & 60 & 16.5 \\
2021 Apr 25 & 201 & 60 & 17.2 & 2022 Aug 09 & 108 & 60 & 12.6 & 2023 Apr 01 & 130 & 60 & 16.5 \\
2021 Apr 26 & 202 & 60,120 & 17.2 & 2022 Aug 11 & 179 & 60,120 & 16.5 & 2023 Apr 02 & 144 & 60 & 16.5 \\
2021 May 01 & 196 & 60 & 17.2 & 2022 Aug 12 & 155 & 60 & 16.5 & 2023 May 01 & 264 & 60 & 16.5 \\
2021 May 25 & 62 & 60 & 17.2 & 2022 Aug 14 & 274 & 60 & 16.5 & 2023 May 02 & 267 & 60 & 16.5 \\
2021 Aug 21 & 37 & 60 & 16.4 & 2022 Aug 15 & 177 & 60,120 & 16.5 & 2023 May 03 & 71 & 60 & 16.5 \\
2021 Aug 23 & 69 & 60 & 16.5 & 2022 Aug 18 & 175 & 60 & 16.5 & 2023 May 04 & 131 & 60 & 16.5 \\
2021 Aug 24 & 23 & 60 & 16.5 & 2022 Aug 19 & 77 & 60 & 16.5 & 2023 May 05 & 175 & 60 & 16.5 
\enddata
\tablecomments{$N_{\rm exp}$ is the number of the exposure. $T_{\rm exp}$ is the exposure time. $T_{\rm ro}$ is the typical readout time.}
\label{tab:obslog:2}
\end{deluxetable*}

\clearpage
\bibliography{namekata_EKDra_paper2_ver1}{}
\bibliographystyle{aasjournalv7}



\end{document}